\documentclass[online,nolinenum=true]{jfp}

\usepackage{newtxmath}
\usepackage{amssymb}
\usepackage{amsfonts}
\usepackage{mathtools}
\usepackage{stmaryrd}
\usepackage[frozencache,cachedir=./minted]{minted}
\usepackage{mathpartir}
\usepackage{xspace}
\usepackage[shortcuts]{extdash}
\usepackage{enumitem}
\usepackage{caption}
\usepackage{subcaption}
\usepackage[title]{appendix}
\usepackage{proof}
\usepackage{suffix}
\usepackage{ifthen}
\usepackage{multicol}
\usepackage{stackengine}
\usepackage[skip=5pt, belowskip=-6pt]{caption}
\usepackage{dirtree}
\usepackage{siunitx}
\usepackage{pifont}
\usepackage{pstc}

\def\sectionautorefname{Section}

\newtheorem{theorem}{Theorem}[section]
\newtheorem{lemma}[theorem]{Lemma}
\newtheorem{definition}[theorem]{Definition}
\newtheorem{conjecture}[theorem]{Conjecture}
\newtheorem{corollary}[theorem]{Corollary}

\renewenvironment{proof}[1][]{\par\addvspace{6pt}\noindent\textbf{Proof{\normalfont\xspace #1}.}\hskip5.5pt}{\par\addvspace{6pt}}

\newmintinline[coqinline]{coq}{escapeinside=<>,mathescape=true}

\begin{document}

\journaltitle{JFP}
\cpr{Cambridge University Press}
\doival{10.1017/xxxxx}
\totalpg{\pageref{lastpage}}
\jnlDoiYr{2021}

\title{Is Sized Typing for Coq Practical?}

\begin{authgrp}
  \author{Jonathan Chan}
  \affiliation{
    University of British Columbia\\
    \emaillink{jcxz@cs.ubc.ca}
  }
  \author{Yufeng Li}
  \affiliation{
    University of Waterloo\\
    \emaillink{yufeng.li@uwaterloo.ca}
  }
  \author{William J. Bowman}
  \affiliation{
    University of British Columbia\\
    \emaillink{wjb@williamjbowman.com}
  }
\end{authgrp}

\begin{abstract}
Contemporary proof assistants such as Coq require that recursive functions be terminating and corecursive functions be productive to maintain logical consistency of their type theories,
and some ensure these properties using syntactic checks.
However, being syntactic, they are inherently delicate and restrictive,
preventing users from easily writing obviously terminating or productive functions at their whim.

Meanwhile, there exist many \emph{sized type theories} that perform type-based termination and productivity checking,
including theories based on the Calculus of (Co)Inductive Constructions (CIC), the core calculus underlying Coq.
These theories are more robust and compositional in comparison.
So why haven't they been adapted to Coq?

In this paper, we venture to answer this question with \lang, a sized type theory based on CIC.
It extends past work on sized types in CIC with additional Coq features such as global and local definitions.
We also present a corresponding size inference algorithm and implement it within Coq's kernel;
for maximal backward compatibility with existing Coq developments,
it requires no additional annotations from the user.

In our evaluation of the implementation, we find a severe performance degradation when compiling parts of the Coq standard library, inherent to the algorithm itself.
We conclude that if we wish to maintain backward compatibility,
using size inference as a replacement for syntactic checking is wildly impractical in terms of performance.
\end{abstract}


\maketitle

\section{Introduction}\label{sec:intro}

Proof assistants based on dependent type theory rely on the termination of recursive functions and the productivity of corecursive functions to ensure two important properties: logical consistency, so that it is not possible to prove false propositions; and decidability of type checking, so that checking that a program proves a given proposition is decidable.

In proof assistants such as Coq, termination and productivity are enforced by a \emph{guard predicate} on fixpoints and cofixpoints respectively.
For fixpoints, recursive calls must be \emph{guarded by destructors}; that is, they must be performed on structurally smaller arguments.
For cofixpoints, corecursive calls must be \emph{guarded by constructors}; that is, they must be the structural arguments of a constructor.
The following examples illustrate these structural conditions.

\begin{minted}{coq}
Fixpoint plus n m : nat :=
  match n with
  | O => m
  | S p => S (plus p m)
  end.
CoFixpoint const {A} a : Stream A := Cons a (const a).
\end{minted}

In the recursive call to \coqinline{plus}, the first argument \coqinline{p} is structurally smaller than \coqinline{S p}, which is the form of the original first argument \coqinline{n}. Similarly, in \coqinline{const}, the constructor \coqinline{Cons} is applied to the corecursive call.

The actual implementation of the guard predicate extends beyond the \guardedbydestructors and \guardedbyconstructors conditions
to accept a larger set of terminating and productive functions.
In particular, function calls will be unfolded (\ie inlined) in the bodies of \cofixpoints
and reduction will be performed as needed before checking the guard predicate.
This has a few disadvantages:
firstly, the bodies of these functions are required, which hinders modular design;
and secondly, as aptly summarized by \citet{coqterm},

\begin{quote}
  ... unfold[ing] all the definitions used in the body of the function, do[ing] reductions, e.t.c.
  ... makes typechecking extremely slow at times.
  Also, the unfoldings can cause the code to bloat by orders of magnitude and become impossible to debug.
\end{quote}

Furthermore, changes in the structural form of functions used in \cofixpoints can cause the guard predicate to reject the program even if the functions still behave the same.
The following simple example, while artificial, illustrates this structural fragility.
\begin{minted}{coq}
Fixpoint minus n m : nat :=
  match n, m with
  | O, _ => n
  | _, O => n
  | S n', S m' => minus n' m'
  end.
\end{minted}
\begin{minted}{coq}
Fixpoint div n m : nat :=
  match n with
  | O => O
  | S n' => S (div (minus n' m) m)
  end.
\end{minted}

If we replace \coqinline{| O, _ => n} with \coqinline{| O, _ => O} in \coqinline{minus}, the behaviour doesn't change, but \coqinline{O} is not a structurally smaller term of \coqinline{n} in the recursive call to \coqinline{div}, so \coqinline{div} no longer satisfies the guard predicate.
The acceptance of \coqinline{div} then depends a separate definition independent of \coqinline{div}.
While the difference is easy to spot here, for larger programs or programs that use many imported library definitions,
this behaviour can make debugging much more difficult.
Furthermore, the guard predicate is unaware of the obvious fact that \coqinline{minus} never returns a \coqinline{nat} larger than its first argument, which the user would have to prove in order for \coqinline{div} to be accepted with our alternate definition of \coqinline{minus}.

In short, the extended syntactic guard condition long used by Coq is anti-modular, anti-compositional, has poor performance characteristics, and requires the programmer to either avoid certain algorithms or pay a large cost in proof burden.

\paragraph*{} This situation is particularly unfortunate, as there exists a non-syntactic termimation- and productivity-checking method that overcomes these issues,
whose theory is nearly as old as the guard condition itself: sized types.

In essence, the \coinductive type of a construction is annotated with a size annotation, which provides some information about the size of the construction.
In this paper, we consider a simple size algebra: \mbox{$s \coloneqq \upsilon \mid \hat{s} \mid \infty$}, where $\upsilon$ ranges over size variables.
If the argument to a constructor has size $s$, then the fully-applied constructor would have a successor size $\hat{s}$.
For instance, the constructors for the naturals follow the below rules:

\vspace{-2ex}
\begin{mathpar}
  \inferrule*
    {~}
    {\Gamma \vdash \Zero : \Nat^{\hat{s}}}
  \and
  \inferrule*
    {\Gamma \vdash n : \Nat^s}
    {\Gamma \vdash \app{\Succ}{n} : \Nat^{\hat{s}}}
\end{mathpar}

Termination and productivity checking is then \emph{just} a type checking rule that uses size information.
For termination, the recursive call must be done on a construction with a smaller size, so when typing the body of the fixpoint, the reference to itself in the typing context must have a smaller size.
For productivity, the returned construction must have a larger size than that of the corecursive call, so the type of the body of the cofixpoint must be larger than the type of the reference to itself in the typing context.
In short, they both follow the following (simplified) typing rule, where $\upsilon$ is an arbitrary fresh size variable annotated on the \coinductive types, and $s$ is an arbitrary size expression as needed.

\vspace{-2ex}
\begin{mathpar}
  \inferrule*[]
    {\Gamma (\assm{f}{t^\upsilon}) \vdash e: t^{\hat{\upsilon}}}
    {\Gamma \vdash \kw{(co)fix}\ f \mathbin{:} t \mathbin{\coloneqq} e : t^s}
\end{mathpar}

We can then assign \coqinline{minus} the type $\text{Nat}^\upsilon \to \text{Nat} \to \text{Nat}^\upsilon$.
The fact that we can assign it a type indicates that it will terminate,
and the $\upsilon$ annotations indicate that the function preserves the size of its first argument.
Then \coqinline{div} uses only the type of \coqinline{minus} to successfully type check, not requiring its body.
Furthermore, being type-based and not syntax-based, replacing \coqinline{| O, _ => n} with \coqinline{| O, _ => O}
doesn't affect the type of \coqinline{minus} or the typeability of \coqinline{div}.
Similarly, some other \cofixpoints that preserve the size of arguments in ways that aren't syntactically obvious may be typed to be size preserving,
expanding the set of terminating and productive functions that can be accepted.
Finally, if additional expressivity is needed, rather than using syntactic hacks like inlining, we could take the semantic approach of enriching the size algebra.

\paragraph*{} It seems perfect; so why doesn't Coq \emph{just} use sized types?

\noindent That is the question we seek to answer in this paper.

Unfortunately, past work on sized types~\citep{cic-hat, cic-hat-minus,cc-hat-omega} for the Calculus of (Co)\-Inductive Constructions (CIC), Coq's underlying calculus, have some practical issues:

\begin{itemize}
    \item They require nontrivial backward-incompatible additions to the surface language,
      such as size annotations on \cofixpoint types and polarity annotations on \coinductive definitions.
    \item They are missing important features found in Coq such as global and local definitions,
      universe cumulativity, and an impredicative $\Prop$.
    \item They restrict size variables from appearing in terms, which precludes, for instance,
      defining type aliases for sized types.
\end{itemize}

To resolve these issues, we extend \CIChat~\citep{cic-hat}, \CIChatminus~\citep{cic-hat-minus-nat,cic-hat-minus}, and \CChatomega~\citep{cc-hat-omega} in our calculus \textbf{\lang} (``\emph{CIC-star-hat}''),
and design a size inference algorithm from CIC to \lang,
borrowing from the algorithms in past work~\citep{f-hat, cic-hat, cc-hat-omega}.

For \lang we prove confluence and subject reduction.
However, new difficulties arise when attempting to prove strong normalization and consistency.
Proof techniques from past work, especially in \citet{cic-hat-minus}, don't readily adapt to our modifications,
in particular to impredicative $\Prop$, universe cumulativity, and unrestricted size variables.
On the other hand, set-theoretic semantics of type theories that do have these features don't readily adapt to the interpretation of sizes, either,
with additional difficulties due to untyped conversion.
We detail a proof attempt on a variant of \lang and discuss its shortcomings.

Even supposing that the metatheoretical problems can be solved and strong normalization and consistency proven,
is an implementation of this sytem practical?
Seeking to answer this question, we have forked Coq~\citep{impl}, implemented the size inference algorithm within its kernel,
and opened a draft pull request to the Coq repository%
\footnote{\url{https://github.com/coq/coq/pull/12426/} (now closed).}.
To maximize backward compatibility, the surface language is completely unchanged,
and sized typing can be enabled by a flag that is off by default.
This flag can be used in conjuction with or without the existing guard checking flag enabled.

While sized typing enables many of our goals, namely increased expressivity with modular and compositional typing for \cofixpoints, the performance cost is unacceptable.
We measure at least a $5.5 \times$ increase in compilation time in some standard libraries.
Much of the performance cost is intrinsic to the size inference algorithm, and thus intrinsic to attempting to maintain backward compatibility.
We analyze the performance of our size inference algorithm and our implementation in detail.

So why doesn't Coq \emph{just} use sized types?
Because it seems it must either sacrifice backward compatibility or compile-time performance,
and the lack of a proof of consistency may be a threat to Coq's trusted core.
While nothing yet leads us to believe that \lang is inconsistent,
the performance sacrifice required for compatibility makes our approach seem wildly impractical.

\paragraph*{} The remainder of this paper is organized as follows.
We begin in \autoref{sec:overview} with a high-level overview of the design of \lang as compared to past work and Coq,
and the consequences resulting from these design decisions.
We formalize the calculus \lang in \autoref{sec:typing},
and discuss the desired metatheoretical properties in \autoref{sec:metatheory}.
In \autoref{sec:algorithm}, we present the size inference algorithm from unsized terms to sized \lang terms,
and evaluate an implementation in our fork in \autoref{sec:impl}.
Finally, we take a look at all of the past work on sized types leading up to \lang in \autoref{sec:related}, and conclude in \autoref{sec:conclusion}.


\section{Overview}\label{sec:overview}

\newcommand{\cmark}{\ding{51}}
\newcommand{\xmark}{\ding{55}}
\begin{table}
\centering
\begin{tabular}{| l | c | c | c | c |}
\hline
\textbf{Feature} & \textit{\CIChat} & \textit{\CIChatminus} & \textit{Coq} & \textit{\lang} \\
\hline
Universe cumulativity & \xmark & \xmark & \cmark & \cmark \\
Impredicative $\Prop$ & \xmark & \xmark & \cmark & \cmark \\
Definitions           & \xmark & \xmark & \cmark & \cmark \\
Parameter polarities  & \cmark & \cmark & \xmark & \xmark \\
Nested \coinductives  & \cmark & \cmark & \cmark & \xmark \\
Normalization proven? & \xmark & \cmark & ?      & \xmark \\
Size inference algorithm & \cmark & \xmark & N/A & \cmark \\
\hline
\end{tabular}
\caption{Comparison of the features in \CIChat, \CIChatminus, Coq, and \lang}
\label{table:CICs}
\end{table}

Past work \CIChat, \CIChatminus, and \CChatomega add sized types to CIC with the explicit philosophy of requiring no size annotations:
a user would write bare CIC code, and the type checker would have the simultaneous task of synthesizing and checking types,
while also inferring all the size annotations.
However, Coq's core calculus extends quite a bit beyond merely CIC,
and the presentation of various analogous features differ subtly but nontrivially.
The goal of \lang is to bring sized types in CIC a few steps closer to Coq,
while keeping with the original philosphy.
In the process of conforming to Coq's calculus, to minimize the changes required to it so that a prototype implementation is
viable---for Coq's codebase itself is old, large, and intricate---we
must also discard some features from past work.
In this section, we discuss what \lang has added or removed relative to past work and to Coq,
and their implications on the metatheory and the implementation.
These features and their presence or absence in the relevant works are summarized in \autoref{table:CICs}.

\subsection{Comparison with Other Work}\label{sec:overview:comparison}

\CIChat and \CIChatminus are extensions of CIC, with dependent functions, a universe hierarchy, inductive definitions, case expressions, and fixpoints.
\CChatomega dually has coinductive streams and cofixpoints instead.
In terms of sized types, they add size expressions as annotations to \coinductive types,
and a \cofixpoint $f$ is well typed if all \corecursive calls in its body occurs on an argument of a smaller sizes,
as we have seen with the simplified typing rule in \autoref{sec:intro}.
Constructors construct constructions of a larger size than their \corecursive functions.

\CIChat and \CIChatminus differ in that \CIChat includes a size inference algorithm but no proof of strong normalization,
while \CIChatminus is proven to be strongly normalizing, with no size inference algorithm explicitly given.
\CIChatminus also restricts where size variables may appear in terms.
Since \lang doesn't have such restrictions, it can be thought of as an extension of \CIChat combined with \CChatomega,
featuring sized (mutual) \coinductive types and (mutual) \cofixpoints,
and further adding universe cumulativity and an impredicative $\Prop$ universe,
which are existing features in Coq.
Cumulativity and impredicativity complicate the set-theoretic model by \citet{cic-hat-minus}, but these issues are known: see, for instance, \citet{not-so-simple-cc}.

The size inference algorithms for \CIChat and \lang both take terms without size annotations,
which are in essence plain CIC terms, along with a local environment
and a set of size constraints for the environment,
and return an annotated term, its annotated size, and a new set of size constraints that must be satisfied.
Because \lang includes global definitions, its algorithm also takes a global environment.

\subsection{Definitions}

Coq's core calculus contains two kinds of variables:%
\footnote{It also has a third type of variable for section-level bindings;
this is beyond the scope of \lang.}
one for local bindings from functions, function types, and let expressions,
and one for global bindings from vernacular declarations such as \coqinline{Definition} and \coqinline{Parameter} (which we call \textit{constants}).
\lang adds let expressions and global declarations to \CIChat,
with separate local and global environments,
and definitions in the environments in the style of a PTS with definitions~\citep{pts}.

Global definitions and let expressions let us define aliases for types for concision and organization of code,
which necessitates a form of size polymorphism if we want the aliases to behave as we expect.
For instance, if we globally define $\Defn{N}{\Type{}}{\Nat^\upsilon}$,
and later want to define an addition function with type $N \rightarrow N \rightarrow N$,
it would \emph{not} be correct to perform the na\"ive substitution to get $\Nat^\upsilon \rightarrow \Nat^\upsilon \rightarrow \Nat^\upsilon$:
addition intuitively does not always return something of the same size.

What we want instead is to allow a different size for each use of $N$,
so that the above type reduces to $\Nat^{\upsilon_1} \rightarrow \Nat^{\upsilon_2} \rightarrow \Nat^{\upsilon_3}$.
This means $N$ must be polymorphic in the sizes involved in its definition,
the same kind of rank-1 or prenex polymorphism in ML-style let polymorphism for type variables.
To retain backward compatibility, there is no explicit size quantification or application ---
every definition and let binding is implicitly quantified over \emph{all} size variables involved.
The corresponding notion of size instantiation comes in the form of size substitution annotations on variables and constants, so that $N^{\set{\upsilon \mapsto s}}$ for instance reduces to $\Nat^s$.
These and all other size annotations are inferrable, as detailed in \autoref{sec:algorithm}.

Having definitions and annotated variables and constants also means we need to now allow sizes to appear
not only in the bodies of let expressions but also in the bodies of functions and in the branches of case expressions,
in contrast to the restrictions of \CIChatminus.

\paragraph*{} Recall that the inference algorithm will return a size constraint set.
To handle inference of global declarations, after the size inference of one declaration,
we now have two options:
pass the resulting constraints along to inference of subsequent declarations,
or ``solve'' the constraints by reassigning size variables with sizes that satisfy those constraints.
Global definitions in Coq are independent of one another in terms of type checking,
so we choose the second option for its modularity:
constraints derived from previous definitions should not interfere with the size inference of the current definition.
However, solving constraints introduces a nontrivial overhead,
so we stick with the first option for local definitions.

Unfortunately, even when only solving constraints for global declarations,
our implementation of the size inference algorithm in the Coq codebase takes a rather significant performance hit.
This is in part due to the proliferation of sizes in which the definitions are polymorphic,
as the worst-case time complexity of solving is at least quadratic in the number of size variables,
and every usage of each definition requires fresh size variables for it.
We analyze the performance of our implementation and discuss possible solutions in \autoref{sec:impl}.

\subsection{Polarities}

\Coinductive definitions in \CIChatminus are also annotated with polarities for each of its parameters to augment the subtyping relation.
For example, if the type parameter of the usual $\List$ type is given a positive polarity,
then $\app{\List^r}{\Nat^s} \leq \app{\List^r}{\Nat^{\hat{s}}}$ holds
because $\Nat^s \leq \Nat^{\hat{s}}$ holds,
which in turn holds because $\hat{s}$ is a larger size than $s$.
Similarly, a negative polarity reverses the subtyping relation,
while an invariant polarity induces no subtyping relation from the parameters at all.
It's not known whether these polarity annotations are inferrable from the \coinductive definitions alone,
so again in the name of backward compatibility, \lang doesn't have these annotations,
and treats all parameters as invariant.
This aligns with Coq's current behaviour, where \coqinline{list Set} is not a subtype of \coqinline{list Type}
despite the presence of cumulativity where \coqinline{Set} is a subtype of \coqinline{Type}.

Unfortunately, the invariance of parameters and subtyping of sized \coinductive types interferes with nested \coinductive types,
where the type itself may appear as a parameter to another type in the type of its constructors.
Subject reduction is violated: it becomes possible to have a well-typed term that becomes no longer well typed after a reduction step.
We give an example in \autoref{sec:metatheory:sub-red} and discuss some solutions,
but the approach \lang takes is to disallow nested \coinductives,
removing them from \CIChat.

\subsection{Implementation}

Whereas \lang can be seen as an extension of \CIChat and \CChatomega,
its implementation is an extension of Coq:
all features of Coq orthogonal to sized types remain untouched,
such as universe polymorphism, strict $\Prop$, various primitives, modules, and so on.
The implementation also retains Coq's nested \coinductives,
especially as it doesn't appear possible for size inference to produce the kind of annotations that break subject reduction.


\section{\titlelang}\label{sec:typing}
In this section, we introduce the syntax and judgements of \lang,
culminating in the typing and well-formedness judgements.
Note that this is the core calculus, which is produced from plain CIC by the inference algorithm,
introduced in \autoref{sec:algorithm}.

\subsection{Syntax}

\begin{figure}
\centering

\begin{align*}
m, n, i, j, k, \ell &\Coloneqq \meta{positive naturals} \\
f, g, h, x, y, z &\Coloneqq \meta{term variables} &
\tau, \upsilon &\Coloneqq \meta{size variables} \\
I &\Coloneqq \meta{(co)inductive type names} &
c &\Coloneqq \meta{constructor names} \\
r, s &\Coloneqq \new{\upsilon \mid \hat{s} \mid \infty} \qquad
\rho \Coloneqq \new{\set{\vec{\upsilon \mapsto s}}} &
U &\Coloneqq \Prop \mid \Set \mid \Type{n} \\
e, a, b, p, q, t, u, v, P &\Coloneqq
  \mathrlap{x
  \mid \new{x^{\rho}}
  \mid U
  \mid \prod{x}{t}{t}
  \mid \abs{x}{\new{t^\circ}}{e}
  \mid \app{e}{e}
  \mid \letin{x}{\new{t^\circ}}{e}{e}
  \mid \new{I^s}
  \mid c} \\
&\mathrlap{\mid \caseof{\new{P^\circ}}{e}{c}{e}
  \mid \fix{m}{f^n}{\new{t^*}}{e}
  \mid \cofix{m}{f}{\new{t^*}}{e}}
\end{align*}

\begin{align*}
\Delta &\Coloneqq \mt \mid \Delta(\assm{x}{e}) &\textit{telescopes} \\
\Gamma &\Coloneqq \mt \mid \Gamma(\assm{x}{e}) \mid \Gamma(\defn{x}{t}{t}) &\textit{local environments} \\
\Gamma_G &\Coloneqq \mt \mid \Gamma_G(\Assm{x}{e}) \mid \Gamma_G(\Defn{x}{t}{t}) &\textit{global environments} \\
\Sigma &\Coloneqq \mt \mid \Sigma(\seq{\vec{\assm{\app{I_i}{\Delta_p}}{\prodctx{\new{\Delta^\infty_i}}{U_i}}}} \coloneqq \seq{\vec{\assm{c_j}{\prodctx{\new{\Delta^\infty_j}}{\app{I_j}{\dom{\Delta_p}}{\new{\overline{t^\infty}_j}}}}}}) &\textit{signatures} \\
\end{align*}

\caption{Syntax of \lang terms, environments, and signatures}
\label{fig:syntax}
\end{figure}

\label{sec:typing:syntax}

The syntax of \lang, environments, and signatures are described in \autoref{fig:syntax}.
It is a standard CIC with expressions (or terms) consisting of cumulative universes, dependent functions, definitions, \coinductives, case expressions, and mutual \cofixpoints.
Additions relevant to sized types are highlighted in grey,
which we explain in detail shortly.
Notation such as syntactic sugar or metafunctions and metarelations will also be highlighted in grey
where they are first introduced in the prose.

The overline $\new{\vec{\ph}}$ denotes a sequence of syntactic constructions.
We use 1-based indexing for sequences using subscripts;
sequences only range over a single index unless otherwise specified.
Ellipses may be used in place of the overline where it is clearer;
for instance, the branches of a case expression are written as
$\langle \vec{c_j \Rightarrow e_j} \rangle$ or
$\langle c_1 \Rightarrow e_1, \dots, c_j \Rightarrow e_j, \dots \rangle$,
and $e_j$ is the $j$th branch expression in the sequence.
Additionally, $\new{\app{e}{\vec{a}}}$ is syntactic sugar for application of $e$ to the terms in $\vec{a}$.

\subsubsection{Size Annotations and Substitutions}

As we have seen, \coinductive types are annotated with a size expression representing its size.
A \coinductive with an \textit{infinite} $\infty$ size annotation is said to be a \textit{full type},
representing \coinductives of all sizes.
Otherwise, an inductive with a \textit{noninfinite} size annotation $s$ represents inductives of size $s$ or \emph{smaller},
while a coinductive with annotation $s$ represents coinductives of size $s$ or \emph{larger}.
This captures the idea that a construction of an inductive type has some amount of content to be consumed,
while one of a coinductive type must produce some amount of content.

As a concrete example, a list with $s$ elements has type $\app{\List^{s}}{t}$, because it has at most $s$ elements,
but it also has type $\app{\List^{\hat{s}}}{t}$, necessarily having at most $\hat{s}$ elements as well.
On the other hand, a stream producing at least $\hat{s}$ elements has type $\app{\Stream^{\hat{s}}}{t}$,
and also has type $\app{\Stream^{s}}{t}$ since it necessarily produces at least $s$ elements as well.
These ideas are formalized in the subtyping rules in an upcoming subsection.

Variables bound by local definitions (introduced by let expressions) and constants bound by global definitions (introduced in global environments)
are annotated with a \textit{size substitution} that maps size variables to size expressions.
The substitutions are performed during their reduction.
As mentioned in the previous section, this makes definitions size polymorphic.

In the type annotations of functions and let expressions, as well as the motive of case expressions,
rather than ordinary \textit{sized terms}, we instead have \textit{bare terms} $t^\circ$.
This denotes terms where size annotations are removed.
These terms are required to be bare in order to preserve subject reduction without requiring explicit size applications or typed reduction,
both of which would violate backward compatibility with Coq.
We give an example of the loss of subject reduction when type annotations aren't bare in \autoref{sec:metatheory:sr:bare}
Similarly, we use $t^\infty$ to denote \textit{full terms}, which appear in signatures.

\subsubsection{Fixpoints and Cofixpoints}

In contrast to \CIChat and \CIChatminus, \lang has mutual \cofixpoints.
In a mutual fixpoint $\fix{m}{f_k^{n_k}}{t_k^*}{e_k}$, each $\defn{f_k^{n_k}}{t_k^*}{e_k}$ is one fixpoint definition.
$n_k$ is the index of the recursive argument of $f_k$, and $\kw{fix}_m$ means that the $m$th fixpoint definition is selected.
Instead of bare terms, fixpoint type annotations are \textit{position terms} $t^*$,
where size annotations are either removed or replaced by a \textit{position annotation} $\ast$.
They occur on the inductive type of the recursive argument, as well as the return type if it is an inductive with the same or smaller size.
For instance (using $\new{t \rightarrow u}$ as syntactic sugar for $\prod{\any}{t}{u}$),
the recursive function $\fix*{1}{\mathit{minus}^1 : \Nat^* \rightarrow \Nat \rightarrow \Nat^* \coloneqq \dots}$
has a position-annotated return type since the return value won't be any larger than that of the first argument.

Mutual cofixpoints $\cofix{m}{f_k}{t_k^*}{e_k}$ are similar, except cofixpoint definitions don't need $n_k$,
as cofixpoints corecursively \emph{produce} a coinductive rather than recursively \emph{consuming} an inductive.
Position annotations occur on the coinductive return type as well as any coinductive argument types with the same size or smaller.
As an example, $\cofix*{1}{\mathit{dup} : \prod{A}{\Set}{\app{\Stream^*}{A} \rightarrow \app{\Stream^*}{A}} \coloneqq \dots}$,
a corecursive function that duplicates each element of a stream,
has a position-annotated argument type since it returns a larger stream.

Position annotations mark the size annotation locations in the type of the \cofixpoint where we are allowed to assign the \emph{same} size expression.
This is why we can give the $\mathit{minus}$ fixpoint the type $\Nat^{\hat{\upsilon}} \rightarrow \Nat^\infty \rightarrow \Nat^{\hat{\upsilon}}$, for instance.
In general, if a \cofixpoint has a position annotation on an argument type and the return type,
we say that it is \textit{size preserving} in that argument.
Inuitively, $f$ is size preserving over an argument $e$ if using $\app{f}{e}$ in place of $e$ should be allowed, size-wise.

\subsubsection{Environments and Signatures}

We divide environments into local and global ones.
They consist of \textit{declarations}, which can be either \textit{assumptions} or \textit{definitions}.
While local environments represent bindings introduced by functions and let expressions,
global environments represent top-level declarations corresponding to Coq vernacular.
We may also refer to global environments alone as \textit{programs}.
Telescopes (that is, environments consisting only of local assumptions) are used in syntactic sugar:
given $\Delta = \vec{(\assm{x_i}{t_i})}$, $\new{\prodctx{\Delta}{t}}$ is sugar for $\prod{x_1}{t_1}{\dots \prod{x_i}{t_i}{\dots t}}$, while $\new{\dom{\Delta}}$ is the sequence $\vec{x_i}$.
Additionally, $\Delta^\infty$ denotes telescopes containing only full terms.

We use $\new{x \in \Gamma}$, $\new{(\assm{x}{t}) \in \Gamma}$, and $\new{(\defn{x}{t}{e}) \in \Gamma}$
to represent the presence of some declaration binding $x$, the given assumption, and the given definition in $\Gamma$, respectively,
and similarly for $\Gamma_G$ and $\Delta$.

Signatures consist of mutual \coinductive definitions.
For simplicity, throughout the judgements in this paper, we assume some fixed, implicit signature $\Sigma$.
Global environments and signatures should be easily extendible to an interleaving of declarations and \coinductive definitions,
which would be more representative of a real program.
A mutual \coinductive definition
$\seq{\vec{\assm{\app{I_i}{\Delta_p}}{\prodctx{\Delta^\infty_i}{U_i}}}} \coloneqq
\seq{\vec{\assm{c_j}{\prodctx{\Delta^\infty_j}{\app{I_j}{\dom{\Delta_p}}{\vec{t^\infty}_j}}}}}$
consists of the following:
\begin{itemize}
  \item $I_i$, the names of the defined \coinductive types;
  \item $\Delta_p$, the \textit{parameters} common to all $I_i$;
  \item $\Delta^\infty_i$, the \textit{indices} of each $I_i$;
  \item $U_i$, the universe to which $I_i$ belongs;
  \item $c_j$, the names of the defined constructors;
  \item $\Delta^\infty_j$, the arguments of each $c_j$;
  \item $I_j$, the \coinductive type to which $c_j$ belongs; and
  \item $\vec{t^\infty}_j$, the indices to $I_j$.
\end{itemize}

We require that the index and argument types be \emph{full} types and terms.
Note also that $I_j$ is the \coinductive type of the $j$th constructor, \emph{not} the $j$th \coinductive in the sequence $\vec{I_i}$.
We forgo the more precise notation $I_{i_j}$ for brevity.

As a concrete example, the usual $\Vector$ type (using $\new{(\assm{x}{t}) \rightarrow u}$ as syntactic sugar for $\prod{x}{t}{u}$) would be defined as:
\begin{align*}
  \seq{\assm{\app{\Vector}{(A : \Type{})}&}{\Nat \to \Type{}}} \coloneqq \\
      \seq{\assm{\VNil&}{\app{\Vector}{A}{\Zero}}, \\
      \assm{\VCons&}{(\assm{n}{\Nat}) \to A \to \app{\Vector}{A}{(\app{\Succ}{n}))}}}.
\end{align*}

As mentioned in the previous section, unlike \CIChat and \CIChatminus, our \coinductive definitions do not have parameter polarity annotations.
In those languages, for $\Vector$'s parameter for instance, they might write $(\assm{A^+}{\Type{}})$, giving it positive polarity, so that 
$\app{\Vector^\infty}{\Nat^s}{n}$ is a subtype of $\app{\Vector^\infty}{\Nat^{\hat{s}}}{n}$.

As is standard, the well-formedness of \coinductive definitions depends not only on the well-typedness of its types but also on syntactic positivity conditions.
We reproduce the \textit{strict positivity} conditions in \autoref{sec:wf-ind}, and refer the reader to clauses I1--I9 in \citet{cic-hat-minus}, clauses 1--7 in \mbox{\citet{cic-hat}}, and the Coq Manual~\citep{coq} for further details.
As \lang doesn't support nested \coinductives,
we don't need the corresponding notion of \textit{nested positivity}.
Furthermore, we assume that our fixed, implicit signature is well-formed.

\subsection{Reduction and Convertibility}

The reduction rules listed in \autoref{fig:reduction} are the usual ones for CIC with definitions:
$\beta$\=/reduction (function application),
$\zeta$\=/reduction (let expression evaluation),
$\iota$\=/reduction (case expressions),
$\mu$\=/reduction (fixpoint expressions),
$\nu$\=/reduction (cofixpoint expressions),
$\delta$\=/reduction (local definitions), and
$\Delta$\=/reduction (global definitions).

\begin{figure}
\fbox{$\gg \vdash e \reduce_{\beta\zeta\delta\Delta\iota\mu\nu} e$} \hfill
\begin{align*}
\gg \vdash x^\rho & \reduce_\delta \rho e \qquad \textit{where} ~ (x : t \coloneqq e) \in \Gamma \\
\gg \vdash x^\rho & \reduce_\Delta \rho e \qquad \textit{where} ~ (\Defn{x}{t}{e}) \in \Gamma_G \\
\gg \vdash \app{(\abs{x}{t^\circ}{e_1})}{e_2} & \reduce_\beta \subst{e_1}{x}{e_2} \\
\gg \vdash \letin{x}{t^\circ}{e_1}{e_2} & \reduce_\zeta \subst{e_2}{x}{e_1} \\
\gg \vdash \caseof{P^\circ}{(\app{c_\ell}{\vec{p}}{\vec{a}})}{c_j}{e_j} & \reduce_\iota \app{e_\ell}{\vec{a}} \\
\gg \vdash \app{q_m}{\vec{b}}{(\app{c_\ell}{\vec{p}}{\vec{a}})}
  & \reduce_\mu \app{\substvec{e_m}{f_k}{q_k}}{\vec{b}}{(\app{c_\ell}{\vec{p}}{\vec{a}})} \\
  \textit{where} ~ & q_i \equiv \fix{i}{f_k^{n_k}}{t_k}{e_k}, \norm{\vec{b}} = n_m - 1 \\
\gg \vdash \caseof{P^\circ}{(\app{q_m}{\vec{b}})}{c_j}{a_j}
  & \reduce_\nu \caseof{P^\circ}{(\app{\substvec{e_m}{f_k}{q_k}}{\vec{b}})}{c_j}{a_j} \\
  \textit{where} ~ & q_i \equiv \cofix{i}{f_k}{t_k}{e_k}
\end{align*}
\caption{Reduction rules}
\label{fig:reduction}
\end{figure}

\begin{figure}
\fbox{$\gg \vdash e \reduce^* e$} \hfill
\vspace{-2ex}

\begin{mathpar}
  \inferrule[\defrule{red-refl}]{~}{
    \gg \vdash e \reduce^* e
  }
  \and
  \inferrule[\defrule{red-trans}]{
    \gg \vdash e_1 \reduce e_2 \and
    \gg \vdash e_2 \reduce^* e_3
  }{
    \gg \vdash e_1 \reduce^* e_3
  }
\end{mathpar}
\caption{Multi-step reduction rules}
\label{fig:reductions}
\end{figure}

\begin{figure}
\fbox{$\gg \vdash e \conv* e$} \hfill

\begin{mathpar}
  \inferrule[\defrule{conv-red}]
    {\gg \vdash e_1 \reduce^* e'_1 \\\\
      \gg \vdash e_2 \reduce^* e'_2 \\\\
      \gg \vdash e'_1 \conv e'_2}
    {\gg \vdash e_1 \conv* e_2}
  \and
  \inferrule[\defrule{conv-cong}]
    {\text{For every $i$:} \\
      \gg \vdash a_i \conv b_i}
    {\gg \vdash \substvec{e}{x_i}{a_i} \conv \substvec{e}{x_i}{b_i}}
  \\
  \inferrule[\defnamerule{conv-eta-r}{conv-$\eta$-l}]
    {\gg \vdash e_1 \reduce^* \abs{x}{\erase{t}}{e} \\\\
      \gg (x:t) \vdash e \conv* e_2 ~ x}
    {\gg \vdash e_1 \conv e_2}
  \and
  \inferrule[\defnamerule{conv-eta-1}{conv-$\eta$-r}]
    {\gg \vdash e_2 \reduce^* \abs{x}{\erase{t}}{e} \\\\
      \gg (x:t) \vdash e'_1 ~ x \conv* e}
    {\gg \vdash e_1 \conv e_2}
\end{mathpar}
\caption{Convertibility rules}
\label{fig:convertibility}
\end{figure}


In the case of \deltaDeltareduction, where the variable or constant has a size substitution annotation, we modify the usual rules.
These reduction rules are important for supporting size inference with definitions.
If the definition body contains \coinductive types (or other defined variables and constants), we can assign them fresh size variables for each distinct usage of the defined variable.
Further details are discussed in \autoref{sec:algorithm}.

Much of the reduction behaviour is expressed in terms of term and size substitution.
Capture-avoiding substitution is denoted with $\new{\subst{e}{x}{e'}}$,
and simultaneous substitution with $\new{\substvec{e}{x_i}{e_i}}$.
$\new{\rho e}$ denotes applying the substitutions $\substvec{e}{\upsilon_i}{s_i}$ for every $\upsilon_i \mapsto s_i$ in $\rho$,
and similarly for $\new{\rho s}$.

This leaves applications of size substitutions to environments,
and to size substitutions themselves when they appear as annotations on variables and constants.
A variable $x^{\set{\vec{\upsilon \mapsto s}}}$ bound to $\defn{x}{t}{e}$ in the environment, for instance,
can be thought of as a delayed application of the sizes $\vec{s}$,
with the definition implicitly abstracting over all size variables $\vec{\upsilon}$.
Therefore, the ``free size variables'' of the annotated variable are those in $\vec{s}$,
and given some size substitution $\rho$,
$\rho x^{\set{\vec{\upsilon \mapsto s}}} = x^{\set{\vec{\upsilon \mapsto \rho s}}}$.
Meanwhile, we treat all $\vec{\upsilon}$ in the definition as \emph{bound},
so that $\rho(\Gamma_1 \defn{x}{t}{e} \Gamma_2) = (\rho\Gamma_1)(\defn{x}{t}{e})(\rho\Gamma_2)$,
skipping over all definitions, and similarly for global environments.

Finally, $\new{\ph \equiv \ph}$ is syntactic equality up to $\alpha$-equivalence (renaming),
and $\new{\norm{\ph}}$ yields the cardinality of its argument (\eg sequence length, set size, \etc).

We define reduction ($\reduce$) as the congruent closure of the reductions,
multi-step reduction ($\rhd^*$) in \autoref{fig:reductions} as the reflexive--transitive closure of $\rhd$,
and convertibility ($\conv*$) in \autoref{fig:convertibility}.
The latter also includes $\eta$-convertibility,
which is presented informally in the Coq manual~\citep{coq} and formally (but part of typed conversion) in \citet{conversion}.
Note that there are no explicit rules for symmetry and transitivity of convertibility
because these properties are derivable, as proven in\opcitt{conversion}

\subsection{Subtyping and Positivity}\label{subsec:typing:subtyping}

\begin{figure}
\fbox{$s \sqsubseteq s$} \hfill
\begin{mathpar}
  \inferrule[\defrule{ss-infty}]{~}
    {s \sqsubseteq \infty}
  \and
  \inferrule[\defrule{ss-refl}]{~}
    {s \sqsubseteq s}
  \and
  \inferrule[\defrule{ss-succ}]{~}
    {s \sqsubseteq \hat{s}}
  \and
  \inferrule[\defrule{ss-trans}]
    {s_1 \sqsubseteq s_2 \and s_2 \sqsubseteq s_3}
    {s_1 \sqsubseteq s_3}
\end{mathpar}
\caption{Subsizing rules}
\label{fig:subsizing}
\end{figure}


First, we define the subsizing relation in \autoref{fig:subsizing}.
Subsizing is straightforward since our size algebra is simple.
Notice that both $\infty \sqsubseteq \succ{\infty}$ and $\succ{\infty} \sqsubseteq \infty$ hold.

\begin{figure}
\begin{flushleft}
  \fbox{$\gg \vdash t \leq t$}
\end{flushleft}
\begin{mathpar}
  \inferrule[\defrule{st-cumul}]{~}
    {\gg \vdash \Prop \leq \Set \leq \Type{1}}
  \quad
  \inferrule{~}
    {\gg \vdash \Type{i} \leq \Type{i+1}}
  \and
  \inferrule[\defrule{st-conv}]
    {\gg \vdash t \conv* u}
    {\gg \vdash t \leq u}
  \and
  \inferrule[\defrule{st-trans}]
    {\gg \vdash t \leq u \\\\ \gg \vdash u \leq v}
    {\gg \vdash t \leq v}
  \and
  \inferrule[\defrule{st-prod}]
    {\gg \vdash t_1 \conv* t_2 \\\\ \gg(\assm{y}{t_2}) \vdash \subst{u_1}{x}{y} \leq u_2} 
    {\gg \vdash \prod{x}{t_1}{u_1} \leq \prod{y}{t_2}{u_2}}
  \and
  \inferrule[\defrule{st-app}]
    {\gg \vdash t_1 \leq t_2 \\\\ \gg \vdash u_1 \conv* u_2} 
    {\gg \vdash t_1 ~ u_1 \leq t_2 ~ u_2}
  \and
  \inferrule[\defrule{st-ind}]
    {I \textrm{ inductive } \and s \sqsubseteq s'}
    {\gg \vdash I^s \leq I^{s'}}
  \and
  \inferrule[\defrule{st-coind}]
    {I \textrm{ coinductive } \and s' \sqsubseteq s}
    {\gg \vdash I^s \leq I^{s'}}
\end{mathpar}
\caption{Subtyping rules}
\label{fig:subtyping}
\end{figure}


The subtyping rules in \autoref{fig:subtyping} extend those of cumulative CIC with rules for sized \coinductive types.
In other words, they extend those of \CIChat, \CIChatminus, and \CChatomega with universe cumulativity (and a $\Prop$ universe).
Inductive types are \emph{covariant} in their size annotations with respect to subsizing (\refrule{st-ind}),
while coinductive types are \emph{contravariant} (\refrule{st-coind}).
Coming back to the examples in the previous section, this means that
$\app{\List^{s}}{t} \leq \app{\List^{\hat{s}}}{t}$ holds as we expect,
since a list with $s$ elements has no more than $\hat{s}$ elements;
dually, $\app{\Stream^{\hat{s}}}{t} \leq \app{\Stream^{s}}{t}$ holds as well,
since a stream producing $\hat{s}$ elements also produces no fewer than $s$ elements.

Rules \refnorule{st-prod} and \refnorule{st-app} differ from past work in their variance, but correspond to those in Coq.
As \coinductive definitions have no polarity annotations,
we treat all parameters as ordinary, invariant function arguments.
The remaining rules are otherwise standard.

\begin{figure}
\centering
\begin{flushleft}
  \fbox{$\gg \vdash \upsilon \pos t$} \quad
  \fbox{$\gg \vdash \upsilon \neg t$}
\end{flushleft}
\begin{mathpar}
  \inferrule[\defnamerule{pos-neg-notin}{pos-neg-$\notin$}]
    {\gg \vdash \upsilon \notin \SV{t}}
    {\gg \vdash \upsilon \pos t \\\\ \gg \vdash \upsilon \neg t}

  \inferrule[\defrule{pos-conv}]
    {\gg \vdash t \conv* t' \\\\
      \gg \vdash \upsilon \pos t'}
    {\gg \vdash \upsilon \pos t}

  \inferrule[\defrule{neg-conv}]
    {\gg \vdash t \conv* t' \\\\
      \gg \vdash \upsilon \neg t'}
    {\gg \vdash \upsilon \neg t}
  
  \inferrule[\defnamerule{pos-prod}{pos-$\Pi$}]
    {\upsilon \pos u \\\\
      \gg \vdash \upsilon \notin \SV{t}}
    {\gg \vdash \upsilon \pos \prod{x}{t}{u}}

  \inferrule[\defnamerule{neg-prod}{neg-$\Pi$}]
    {\upsilon \neg u \\\\
      \gg \vdash \upsilon \notin \SV{t}}
    {\gg \vdash \upsilon \neg \prod{x}{t}{u}}

  \inferrule[\defrule{pos-ind}]
    {I ~ \text{inductive} \\\\
      \gg \vdash \upsilon \notin \SV{\overline{a}}}
    {\gg \vdash \upsilon \pos \app{I^s}{\overline{a}}}

  \inferrule[\defrule{neg-coind}]
    {I ~ \text{coinductive} \\\\
      \gg \vdash \upsilon \notin \SV{\overline{a}}}
    {\gg \vdash \upsilon \neg \app{I^s}{\overline{a}}}
\end{mathpar}
\caption{Positivity/negativity of size variables in terms}
\label{fig:posneg}
\end{figure}


In addition to subtyping, we define a \textit{positivity} and \textit{negativity} judgements like in past work.
They are syntactic approximations of monotonicity properties of subtyping with respect to size variables;
we have that
$\upsilon \pos t \Leftrightarrow t \leq \subst{t}{\upsilon}{\hat{\upsilon}}$ and
$\upsilon \neg t \Leftrightarrow \subst{t}{\upsilon}{\hat{\upsilon}} \leq t$ hold.
Positivity and negativity are then used to indicate where position annotations are allowed to appear in the types of \cofixpoints,
as we will see in the typing rules.

\subsection{Typing and Well-Formedness}\label{sec:typing:rules}

\begin{figure}
\begin{flushleft}
  \fbox{$\WF{\gg}$}
\end{flushleft}
\begin{mathpar}
  \inferrule[\defrule{wf-nil}]{~}
    {\WF{\mt, \mt}}
  \and
  \inferrule[\defrule{wf-local-assum}]
    {\gg \vdash t: U \and x \notin \Gamma}
    {\WF{\Gamma_G, \Gamma (x:t)}}
  \and
  \inferrule[\defrule{wf-local-def}]
    {\gg \vdash e: t \and x \notin \Gamma}
    {\WF{\Gamma_G, \Gamma (x:t \coloneqq e)}}
  \and
  \inferrule[\defrule{wf-global-assum}]
    {\Gamma, \mt \vdash t: U \and x \notin \Gamma_G}
    {\WF{\Gamma_G (\Assm{x}{t}), \mt}}
  \and
  \inferrule[\defrule{wf-global-def}]
    {\Gamma, \mt \vdash e: t \and x \notin \Gamma_G}
    {\WF{\Gamma_G (\Defn{x}{t}{e}), \mt}}
\end{mathpar}
\caption{Well-formedness of environments}
\label{fig:wf}
\end{figure}


We begin with the rules for well-formedness of local and global environments, presented in \autoref{fig:wf}.
As mentioned, this and the typing judgements implicitly contain a signature $\Sigma$, whose well-formedness is assumed.
Additionally, we use $\any$ to omit irrelevant constructions for readability.

\begin{figure}
\centering
\begin{align*}
\Axioms
    &= \set{(\Prop, \Type{1}), (\Set, \Type{1}), (\Type{i}, \Type{i+1})} \\
\Rules
    &= \set{(U, \Prop, \Prop)}
    \cup \set{(U, \Set, \Set) : U \in \set{\Prop, \Set}} \\
    &\cup \set{(\Type{i}, \Type{j}, \Type{k}) : k = \meta{max}(i, j)} \\
\Elims
    &= \set{(U_i, U, I_i) : U_i \in \set{\Set, \Type{}}}
    \cup \set{(\Prop, \Prop, I_i)} \\
    &\cup \set{(\Prop, U, I_i) : \text{$I_i$ empty or singleton}}
\end{align*}
\caption{Universe relations: Axioms, Rules, and Eliminations}
\label{fig:axruel}
\end{figure}


\begin{figure}
\centering

\begin{align*}
    \indtype{I_i} &=
        \prodctx{\Delta_p}{\prodctx{\Delta^\infty_i}{U_i}} \\
    \constrtype{c_j}{s} &=
        \prodctx{\Delta_p}{\prodctx{\Delta^\infty_j [I_j^\infty \coloneqq I_j^s]}{I_{j}^{\hat{s}} ~ \dom{\Delta_p} ~ \overline{t^\infty}_j}} \\
    \motivetype{\overline{p}}{U}{I_i^s} &=
        \prodctx{\Delta^\infty_i[\dom{\Delta_p} \coloneqq \overline{p}]}{\prod{\any}{I_i^s ~ \overline{p} ~ \dom{\Delta^\infty_i}}{U}} \\
    \branchtype{\overline{p}}{c_j}{s}{P} &=
        \prodctx{\Delta^\infty_j [I_j^\infty \coloneqq I_j^s][\dom{\Delta_p} \coloneqq \overline{p}]}{P ~ \overline{t^\infty}_j[\dom{\Delta_p}\coloneqq\overline{p}] ~ (c_j ~ \overline{p} ~ \dom{\Delta^\infty_j})}
\end{align*}
\begin{displaymath}
    \textit{where}\;\:
    \big(\seq{\vec{\assm{\app{I_i}{\Delta_p}}{\prodctx{\Delta^\infty_i}{U_i}}}} \coloneqq \seq{\vec{\assm{c_j}{\prodctx{\Delta^\infty_j}{\app{I_j}{\any}{\overline{t^\infty}_j}}}}}\big) \in \Sigma
\end{displaymath}

\caption{Metafunctions for typing rules}
\label{fig:metafunctions}
\end{figure}


\begin{figure*}[!tp]
\centering

\begin{flushleft}
\fbox{$\gg \vdash e : t$}
\end{flushleft}

\vspace{-2ex}

\begin{mathpar}
  \inferrule[\defrule{var-assum}]
    {\WF{\gg} \\\\ (x:t) \in \Gamma}
    {\gg \vdash x:t}

  \inferrule[\defrule{var-def}]
    {\WF{\gg} \\\\
      \Gamma \equiv \Gamma_1 (x:t \coloneqq e) \Gamma_2 \and
      \Gamma_G, \Gamma_1 \vdash \rho e : \rho t}
    {\gg \vdash x^\rho : \rho t}

  \\

  \inferrule[\defrule{const-assum}]
    {\WF{\gg} \\\\ (\Assm{x}{t}) \in \Gamma_G}
    {\gg \vdash x:t}

  \inferrule[\defrule{const-def}]
    {\WF{\gg} \\\\
      \Gamma_G \equiv \Gamma_{G1}(\Defn{x}{t}{e})\Gamma_{G2} \and
      \Gamma_{G1}, \mt \vdash \rho e : \rho t}
    {\gg \vdash x^\rho : \rho t}

  \inferrule[\defrule{univ}]
    {\WF{\gg} \and (U_1, U_2) \in \Axioms}
    {\gg \vdash U_1 : U_2}

  \inferrule[\defrule{conv}]
    {\gg \vdash e:t \and \gg \vdash u:U \and \gg \vdash t \leq u}
    {\gg \vdash e:u}

  \inferrule[\defrule{pi}]
    {(U_1, U_2, U_3) \in \Rules \\\\
      \gg \vdash t : U_1 \and
      \gg (x:t) \vdash u : U_2}
    {\gg \vdash \prod{x}{t}{u} : U_3}

  \inferrule[\defrule{lam}]
    {\gg \vdash t : U \and \gg (x:t) \vdash e:u}
    {\gg \vdash \abs{x}{\erase{t}}{e} : \prod{x}{t}{u}}

  \inferrule[\defrule{app}]
    {\gg \vdash e_1 : \prod{x}{t}{u} \and \gg \vdash e_2 : t}
    {\gg \vdash \app{e_1}{e_2} : u[x \coloneqq e_2]}

  \inferrule[\defrule{let}]
    {\gg \vdash e_1:t \and \gg (x:t \coloneqq e_1) \vdash e_2:u}
    {\gg \vdash \letin{x}{\erase{t}}{e_1}{e_2} : u[x \coloneqq e_1]}
  \\
  \inferrule[\defrule{ind}]
    {\WF{\gg}}
    {\gg \vdash I^s : \indtype{I}}

  \inferrule[\defrule{constr}]
    {\WF{\gg}}
    {\gg \vdash c : \constrtype{c}{s}}

  \inferrule[\defrule{case}]
    {\gg \vdash e : \app{I^{\hat{s}}}{\overline{p}}{\overline{a}} \and
      \indtype{I} = \prodctx{\any}{\prodctx{\any}{U'}} \\
      (U', U, I) \in \Elims \and
      \gg \vdash P : \motivetype{\overline{p}}{U}{I^{\hat{s}}} \\
      \text{For each $j$:} \and
      \gg \vdash e_j : \branchtype{\overline{p}}{c_j}{s}{P}}
    {\gg \vdash \caseof{\erase{P}}{e}{c_j}{e_j} : \app{P}{\overline{a}}{e}}

  \inferrule[\defrule{fix}]
    {\text{For each $k$:} \and
      \gg \vdash t_k \conv* \prodctx{\Delta_{k}}{\prod{x_k}{\app{I_k^{\upsilon_k}}{\overline{a}_k}}{u_k}} \and
      \norm{\Delta_{k}} = n_k - 1 \and
      \gg \vdash \upsilon_k \pos u_k \\
      \upsilon_k \notin \SV{\Gamma, \overline{a}_k, e_k, \Delta_k} \and
      \gg \vdash t_k : U_k \and
      \gg \overline{(f_k : t_k)} \vdash e_k : t_k[\upsilon_k \coloneqq \hat{\upsilon}_k]}
    {\gg \vdash \fix{m}{f_k^{n_k}}{\erase{t_k}^{\upsilon_k}}{e_k} : t_m[\upsilon_m \coloneqq s]}

  \inferrule[\defrule{cofix}]
    {\text{For each $k$:} \and
      \gg \vdash t_k \conv* \prodctx{\Delta_k}{\app{I_k^{\upsilon_k}}{\overline{a}_k}} \and
      \gg \vdash \upsilon_k \neg \Delta_k \\
      \upsilon_k \notin \SV{\Gamma, \overline{a}_k, e_k} \and
      \gg \vdash t_k : U_k \and
      \gg \overline{(f_k : t_k)} \vdash e_k : t_k[\upsilon_k \coloneqq \hat{\upsilon}_k]}
    {\gg \vdash \cofix{m}{f_k}{\erase{t_k}^{\upsilon_k}}{e_k} : t_m[\upsilon_m \coloneqq s]}
\end{mathpar}

\caption{Typing rules}
\label{fig:typing}
\end{figure*}


The typing rules for sized terms are given in \autoref{fig:typing}. As in CIC, we define the three sets \Axioms, \Rules, and \Elims in \autoref{fig:axruel}, which describe how universes are typed, how products are typed, and what eliminations are allowed in case expressions, respectively.
Metafunctions that construct important function types for inductive types, constructors, and case expressions are listed in \autoref{fig:metafunctions}; they are also used by the inference algorithm in \autoref{sec:algorithm}.

Rules \refnorule{var-assum}, \refnorule{const-assum}, \refnorule{univ}, \refnorule{conv} \refnorule{pi}, and \refnorule{app} are essentially unchanged from CIC.
Rules \refnorule{lam} and \refnorule{let} differ only in that type annotations need to be bare to preserve subject reduction,
and the erasure metafunction $\new{\erase{\ph}}$ removes all size annotations from a sized term.

The first significant usage of size annotations are in Rules \refnorule{var-def} and \refnorule{const-def}.
If a variable or a constant is bound to a term in the local or global environment, it is annotated with a size substitution such that the term is well typed after performing the substitution, allowing for proper $\delta$-/$\Delta$-reduction of variables and constants.
Notably, each usage of a variable or a constant doesn't have to have the same size annotations.

Inductive types and constructors are typed mostly similar to CIC,
with their types specified by \indtype and \constrtype.
In \refrule{ind}, the \coinductive type itself holds a single size annotation.
In \refrule{constr}, size annotations appear in two places:
\begin{itemize}
    \item In the argument types of the constructor.
      We annotate each occurrence of $I_j$ in the arguments $\Delta^\infty_j$ with a size expression $s$.
    \item On the \coinductive type of the fully-applied constructor,
      which is annotated with the size expression $\hat{s}$.
      Using the successor guarantees that the constructor always constructs a construction that is \textit{larger} than any of its arguments of the same type.
\end{itemize}
As an example, consider a possible typing of \text{VCons}:
\begin{align*}
\mt, \mt \vdash \VCons &: \underbrace{(A: \Type{})}_{\text{parameter}} \to \underbrace{(n:\Nat^\infty) \to A \to \Vector^s ~ A ~ n}_{\text{arguments}} \to \underbrace{\Vector^{\hat{s}} ~ A ~ (\Succ ~ n)}_{\text{return type}}
\end{align*}
It has a single parameter $A$ and $\Succ ~ n$ corresponds to the index $\vec{t^\infty}_j$ of the constructor's inductive type.
The input $\Vector$ has size $s$, while the output $\Vector$ has size $\hat{s}$.

In \refrule{case}, a case expression has three parts:
\begin{itemize}
    \item The \textbf{target} $e$ that is being destructed.
      It must have a \coinductive type $I$ with a successor size annotation $\hat{s}_k$ so that recursive constructor arguments have the predecessor size annotation.

    \item The \textbf{motive} $P$, which yields the return type of the case expression.
      While it ranges over the \coinductive's indices,
      the parameter variables $\dom{\Delta_p}$ in the indices' types are bound to the parameters $\vec{p}$ of the target type.
      As usual, the universe of the motive $U$ is restricted by the universe of the \coinductive $U'$ according to \Elims.

      (This presentation follows that of Coq, but differs from that by~\citet{cic-hat-minus, cic-hat-l, cc-hat-omega}, where the case expression contains a return type in which the index and target variables are free and explicitly stated, in the syntactic form $\vec{y}.x.P$.)

    \item The \textbf{branches} $e_j$, one for each constructor $c_j$.
      Again, the parameters of its type are fixed to $\vec{p}$, while ranging over the constructor arguments.
      Note that like in the type of constructors, we annotate each occurrence of $c_j$'s \coinductive type $I$ in $\Delta_j$ with the size expression $s$.
\end{itemize}

Finally, we have the typing of mutual \cofixpoints in rules \refnorule{fix} and \refnorule{cofix}.
We take the annotated type $t_k$ of the $k$th \cofixpoint definition to be convertible to a function type containing a \coinductive type, as usual.
However, instead of the guard condition, we ensure termination/productivity using size expressions.

The main complexity in these rules is supporting size-preserving \cofixpoints.
We must restrict how the size variable $v_k$ appears in the type of the \cofixpoints using the positivity and negativity judgments.
For fixpoints, the type of the $n_k$th argument, the recursive argument, is an inductive type annotated with a size variable $v_k$.
For cofixpoints, the return type is a coinductive type annotated with $v_k$.
The positivity or negativity of $v_k$ in the rest of $t_k$ indicate where $v_k$ may occur other than in the \corecursive position.
For instance, supposing that $n = 1$,
$\app{\List^\upsilon}{\Nat} \to \app{\List}{\Nat} \to \app{\List^\upsilon}{\Nat}$
is a valid fixpoint type with respect to $\upsilon$, while
$\app{\List^\upsilon}{\Nat} \to \app{\List}{\Nat^\upsilon} \to \app{\Stream^\upsilon}{\Nat}$
is not, since $\upsilon$ illegally appears negatively in $\Stream$ and must not appear at all in the parameter of the second $\List$ argument type.
This restriction ensures the aforementioned monotonicity property of subtyping for the \cofixpoints' types,
so that $u_k \leq \subst{u_k}{\upsilon_k}{\hat{\upsilon}_k}$ holds for fixpoints,
and that $\subst{u}{\upsilon_k}{\hat{\upsilon}_k} \leq u$ for each type $u$ in $\Delta_k$ holds for cofixpoints.

As in \refrule{lam}, to maintain subject reduction, we cannot keep the size annotations, instead replacing them with position variables.
The metafunction $\new{\erase{\ph}^\upsilon}$ replaces $\upsilon$ annotations with the position annotation $\ast$ and erases all other size annotations.

Checking termination and productivity is then relatively straightforward.
If $t_k$ are well typed, then the \cofixpoint bodies should have type $t_k$ with a successor size when $(\assm{f_k}{t_k})$ are in the local environment.
This tells us that the recursive calls to $f_k$ in fixpoint bodies are on smaller-sized arguments, and that corecursive bodies produce constructions with size larger than those from the corecursive call to $f_k$.
The type of the $m$th \cofixpoint is then the $m$th type $t_m$ with some size expression $s$ substituted for the size variable $v_m$.

In Coq, the indices of the recursive elements are often elided, and there are no user-provided position annotations at all.
We show how indices and position annotations can be computed during size inference in \autoref{sec:algorithm}.


\section{Metatheoretical Results}
\label{sec:metatheory}

In this section, we describe the metatheory of \lang.
Some of the metatheory is inherited or essentially similar to past work~\citep{cic-hat-minus,cc-hat-omega,cic-hat},
although we must adapt key proofs to account for differences in subtyping and definitions.
Complete proofs for a language like \lang are too involved to present in full,
so we provide only key lemmas and proof sketches.

In short, \lang satisfies confluence and subject reduction, with the same caveats as in CIC for cofixpoints.
While strong normalization and logical consistency have been proven for a variant of \lang
with features that violate backward compatibility,
proofs for \lang itself remain future work.

\subsection{Confluence}

Recall that we define $\rhd$ as the congruent closure of \reduction and $\rhd^*$ as the reflexive--transitive closure of $\rhd$.

\begin{theorem}[Confluence]
\label{thm:metatheory:confluence}
  If $\gg \vdash e \rhd^* e_1$ and $\gg \vdash e \rhd^* e_2$,
  then there is some term $e'$ such that $\gg \vdash e_1 \rhd^* e'$ and $\gg \vdash e_2 \rhd^* e'$.
\end{theorem}

\begin{proof}[{[sketch]}]
  We use the Takahashi translation technique due to \citet{takahashitrans},
  which is a simplification of the standard parallel reduction technique.
  It uses the Takahashi translation $e^\dagger$ of terms $e$,
  defined as the simultaneous single-step reduction of all
  $\beta\zeta\delta\Delta\iota\mu\nu$-redexes of $e$ in left-most inner-most order.
  The proof is relatively standard thanks to reduction being purely syntactic and untyped.
\end{proof}

\subsection{Subject Reduction}
\label{sec:metatheory:sub-red}

Suject reduction does not hold in \lang or in Coq due to the way coinductives are presented.
This is a well-known problem, discussed previously in a sized-types setting by \citet{cc-hat-omega},
on which our presentation of coinductives is based,
as well as by the Coq developers%
\footnote{The discussion of the problem and suggested solutions can be found here: \url{https://github.com/coq/coq/issues/5288/}.}.

In brief, the current presentation of coinductives requires that cofixpoint reduction be \textit{restricted},
\ie occurring only when it is the target of a case expression.
This allows for strong normalization of cofixpoints in much the same way restricting fixpoint reduction to when the recursive argument is syntactically a fully-applied constructor does.
One way this can break subject reduction is by making the type of a case expression not be convertible before and after the cofixpoint reduction.
As a concrete example, consider the following coinductive definition for conaturals.
\begin{displaymath}
  \seq{\assm{\Conat}{\Type{}}} \coloneqq {\seq{\assm{\Succ}{\Conat \to \Conat}}}
\end{displaymath}
For some motive $P$ and branch $e$, we have the following $\nu$-reduction.
\begin{align*}
  &\caseof*{\erase{P}}{\cofix*{1}{\defn{\omega}{\Conat}{\app{\Succ}{\omega}}}}{\seq{\Succ \Rightarrow e}} \rhd_\nu \\
  &\caseof*{\erase{P}}{\app{\Succ}{(\cofix*{1}{\defn{\omega}{\Conat}{\app{\Succ}{\omega}}})}}{\seq{\Succ \Rightarrow e}}
\end{align*}
Assuming both terms are well typed, the former has type $\app{P}{(\cofix*{1}{\defn{\omega}{\Conat}{\app{\Succ}{\omega}}})}$ while the latter has type $\app{P}{(\app{\Succ}{(\cofix*{1}{\defn{\omega}{\Conat}{\app{\Succ}{\omega}})}})}$, but for an arbitrary $P$ these aren't convertible without allowing cofixpoints to reduce arbitrarily.

\begin{figure}
  \fbox{$\gg \vdash e \reduce_{\beta\zeta\delta\Delta\iota\mu\nu'} e$} \hfill
  \vspace{-3ex}
  \begin{align*}
    \dots & \\
    \gg \vdash q_m & \reduce_{\nu'} \substvec{e_m}{f_k}{q_k} \\
    \textit{where} ~ & \forall i \in \vec{k}, q_i \equiv \cofix{i}{f_k}{t_k}{e_k}
  \end{align*}
  \caption{Reduction rules with unrestricted cofixpoint reduction}
  \label{fig:reduction-alt}
\end{figure}

On the other hand, if we do allow unrestricted $\nu'$-reduction as in \autoref{fig:reduction-alt}, subject reduction does hold,
at the expense of normalization,
as a cofixpoint on its own could reduce indefinitely.

\begin{theorem}[Subject Reduction]
  \label{thm:metatheory:sr}
  Let $\Sigma$ be a well-formed signature.
  Suppose $\rhd$ includes unrestricted $\nu'$-reduction of cofixpoints.
  Then $\gg \vdash e : t$ and $e \rhd e'$ implies $\gg \vdash e' : t$.
\end{theorem}

\begin{proof}[{[sketch]}]
  By induction on $\gg \vdash e : t$.  Most cases are straightforward,
  making use of confluence when necessary, such as for a lemma of
  $\Pi$-injectivity to handle $\beta$-reduction in \refrule{app}.
  The case for \refrule{case} where $e \rhd e'$ by $\iota$-reduction relies on the fact that
  if $x$ is the name of a \coinductive type and appears strictly positively in $t$,
  then $x$ appears covariantly in $t$.
  (This is only true without nested \coinductive types, which \lang disallows in well-formed signatures.)

  The case for \refrule{case} and $e$ (guarded) $\nu$-reduces to $e'$ requires an unrestricted $\nu$-reduction.
  After guarded $\nu$-reduction, the target (a cofixpoint) appears in the motive unguarded by a case expression, but must be unfolded to re-establish typing the type $t$.
\end{proof}

\subsubsection{The Problem with Nested Inductives}

\newcommand{\nat}{\const{N}}

Recall from \autoref{sec:typing} that we disallow nested \coinductive types.
This means that when defining a \coinductive type, it cannot recursively appear as the parameter of another type.
For instance, the following definition $\nat$, while equivalent to $\Nat$,
is disallowed due to the appearance of $\nat$ as a parameter of $\Box$.
\begin{align*}
  \seq{\assm{\app{\Box}{(\assm{A}{\Type{}})}}{\Type{}}} &\coloneqq \seq{\assm{\MkBox}{A \rightarrow \app{\Box}{A}}} \\
  \seq{\assm{\nat}{\Type{}}} &\coloneqq \seq{\assm{\Zero}{\nat}, \assm{\Succ}{\app{\Box}{\nat} \to \nat}}
\end{align*}
Notice that we have the subtyping relation $\nat^\upsilon \leq \nat^{\hat{\upsilon}}$,
but as all parameters are invariant for backward compatibility and need to be convertible,
we do \emph{not} have $\app{\Box^\infty}{\nat^\upsilon} \leq \app{\Box^\infty}{\nat^{\hat{\upsilon}}}$.
But because case expressions on some target $\nat^{\hat{s}}$ force recursive arguments to have size $s$ exactly,
and the target also has type $\nat^{\hat{\hat{s}}}$ by cumulativity,
the argument of $\Succ$ could have both type $\app{\Box^\infty}{\nat^s}$ and $\app{\Box^\infty}{\nat^{\hat{s}}}$, violating convertibility.
We exploit this fact and break subject reduction explicitly with the following counterexample term.
\begin{displaymath}
\begin{array}{l}
  \caseof*{\erase{\abs{\any}{\nat}{\nat^\infty}}}{\app{\Succ}{(\app{\MkBox}{\nat^{\hat{\upsilon}}}{\Zero})}}{\\
  \seq{\Zero \Rightarrow \Zero,\\
  \phantom{\langle} \Succ \Rightarrow \app{(\abs{A}{\Type{}}{\abs{x}{A}{\Zero}})}{(\app{\Box^\infty}{\nat^{\hat{\hat{\upsilon}}}})}}}
\end{array}
\end{displaymath}
By cumulativity, the target can be typed as $\nat^{\hat{\upsilon}^{3}}$ (that is, with size $\hat{\hat{\hat{\upsilon}}}$).
By \refrule{case}, the second branch must then have type $\prod{x}{\app{\Box}{\nat^{\hat{\hat{\upsilon}}}}}{\nat^\infty}$ (and so it does).
Then the case expression is well typed with type $\nat^\infty$.
However, once we reduce the the case expression, we end up with a term that is no longer well typed.
\begin{displaymath}
  \app{(\abs{A}{\Type{}}{\abs{x}{A}{\Zero}})}
    {(\app{\Box^\infty}{\nat^{\hat{\hat{\upsilon}}}})}
    {(\app{\MkBox}{\nat^{\hat{\upsilon}}}{\Zero})}
\end{displaymath}
By \refrule{app}, the second argument should have type $\app{\Box^\infty}{\nat^{\hat{\hat{\upsilon}}}}$ (or a subtype thereof), but it cannot:
the only type the second argument can have is $\app{\Box^\infty}{\nat^{\hat{\upsilon}}}$.

There are several possible solutions, all threats to backward compatibility.
\CIChat's solution is to require that constructors be fully-applied and that their parameters be bare terms,
so that we are forced to write $\app{\MkBox}{\nat}{\Zero}$.
The problem with this is that Coq treats constructors essentially like functions,
and assuring that they are fully applied with bare parameters would require either reworking how they are represented internally
or adding an intermediate step to elaborate partially-applied constructors into functions whose bodies are fully-applied constructors.
The other solution, as mentioned, is to add polarities back in, so that $\Box$ with positive polarity in its parameter yields the subtyping relation $\app{\Box^\infty}{\nat^{\hat{\upsilon}}} \leq \app{\Box^\infty}{\nat^{\hat{\hat{\upsilon}}}}$.

Interestingly, because the implementation infers all size annotations from a completely bare program,
our counterexample and similar ones exploiting explicit size annotations aren't directly expressible,
and don't appear to be generated by the algorithm, which would solve for the smallest size annotations.
For the counterexample, in the second branch, the size annotation would be (a size constrained to be equal to) $\hat{\upsilon}$.
We conjecture that the terms synthesized by the inference algorithm do indeed satisfy subject reduction even in the presence of nested \coinductives
by being a strict subset of all well-typed terms that excludes counterexamples like the above.

\subsubsection{Bareness of Type Annotations}\label{sec:metatheory:sr:bare}

As mentioned in \autoref{sec:typing:syntax}, type annotations on functions and let expressions
as well as case expression motives and \cofixpoint types
need to be bare terms (or position terms, for the latter) to maintain subject reduction.
To see why, suppose they were not bare, and consider the term
$\app{(\fix*{1}{\defn{f^1}{\Nat^\tau \rightarrow \Nat^\tau}{\abs{n}{\Nat^{\hat{\tau}}}{n}}})}{(\app{\Succ}{\Zero})}$.
Under empty environments, the fixpoint argument is well typed with type $\Nat^{\hat{\hat{s}}}$ for any size expression $s$,
while the fixpoint itself is well typed with type $\Nat^{r} \rightarrow \Nat^{r}$ for any size expression $r$.
For the application to be well typed, it must be that $r$ is $\hat{\hat{s}}$,
and the entire term has type $\Nat^{\hat{\hat{s}}}$.

By the $\mu$-reduction rule, this steps to the term $\app{(\abs{n}{\Nat^{\hat{\tau}}}{n})}{(\app{\Succ}{\Zero})}$.
Unfortunately, the term is no longer well typed, as $\app{\Succ}{\Zero}$ cannot be typed with type $\Nat^{\hat{\tau}}$ as is required.
By erasing the type annotation of the function,
there is no longer a restriction on what size the function argument must have,
and subject reduction is no longer broken.
An alternate solution is to substitute $\tau$ for $\hat{s}$ during $\mu$-reduction,
but this requires typed reduction to know what the correct size to substitute is,
violating backward compatibility with Coq,
whose reduction and convertibility rules are untyped.

\subsection{Strong Normalization and Logical Consistency}\label{sec:metatheory:sn}

Following strong normalization and logical consistency for \CIChatminus and \CChatomega,
we conjecture that they hold for \lang as well.
We present some details of a model constructed in our a proof attempt;
unfortunately, the model requires changes to \lang that are backward
incompatible with Coq, so we do not pursue it further.
We discuss from where these backward-incompatible changes arise for posterity.

\begin{conjecture}[Strong Normalization]\label{thm:metatheory:sn}
  If $\gg \vdash e : t$ then $e$ contains no infinite
  reduction sequences.
\end{conjecture}

\begin{conjecture}[Logical Consistency]\label{thm:metatheory:lc}
  There is no $e$ such that $\mt, \mt \vdash e : \prod{p}{\Prop}{p}$.
\end{conjecture}

Note that strong normalization is a stricter requirement than Coq, which is only \emph{weakly} normalizing:
every well-typed term has \emph{some} finite reduction sequence.
This relaxation enables more programs to pass the guard predicates while still being consistent.
For instance, the first fixpoint definition is accepted, as \texttt{x} is never used,
but the second is not.

\begin{minted}{coq}
Fixpoint f (u: unit): unit := let x := f u in tt.
Fail Fixpoint f (u: unit): False := let x := f u in x.
\end{minted}

\subsubsection{Proof Attempt and Apparent Requirements for Set-Theoretic Model}

In attempting to prove normalization and consistency, we developed a variant of
\lang called \langAnother which made a series of simplifying assumptions
suggested by the proof attempt:

\begin{itemize}
  \item Reduction, subtyping, and convertibility are typed, as is the case for
    most set-theoretic models.
    That is, each judgement requires the type of the terms,
    and the derivation rules may have typing judgements as premises.
  \item A new size irrelevant typing judgement is needed, similar to that
    introduced by \citet{barras-thesis}. While \lang is probably size
    irrelevant, this is not clear in the model without an explicit judgment.
  \item Fixpoint type annotations require explicit size annotations
    (\ie are no longer merely position terms) and explicitly abstract over a size
    variable, and fixpoints are explicitly applied to a size expression.
    The typing rule no longer erases the type, and the size in the fixpoint type
    is fixed.
    \begin{mathparpagebreakable}
      \inferrule*[right=\defrule{fix-explicit}]
      { \Gamma (f : t) \vdash e : \subst{t}{\upsilon}{\hat{\upsilon}} }
      { \Gamma \vdash \fixE{\upsilon}{f}{t}{e}{s} : \subst{t}{\upsilon}{s}}
    \end{mathparpagebreakable}
    The fixpoint above binds the size variable $\upsilon$ in $t$ and in $e$.
    The reduction rule adds an additional substitution of the predecessor of the size expression,
    in line with how $f$ may only be called in $e$ with a smaller size.
    \begin{mathparpagebreakable}
      \Gamma \vdash \app{\fixE{\upsilon}{f}{t}{e}{\hat{s}}\,}{\vec{b}}{(\app{c_\ell}{\vec{p}}{\vec{a}})}
      \rhd_\mu \app{\subst{\subst{e}{\upsilon}{s}}{f}{\fixE{\upsilon}{f}{t}{e}{s}}}{\vec{b}}{(\app{c_\ell}{\vec{p}}{\vec{a}})}
    \end{mathparpagebreakable}
  \item Rather than inductive definitions in general, only predicative W types are considered.
    W types can be defined as an inductive type:
    \begin{mathparpagebreakable}
      \seq{\app{\W}{(\assm{A}{U})}{(\assm{B}{A \rightarrow U})}} \coloneqq
      \seq{\assm{\Sup}{(\assm{a}{A}) \rightarrow (\assm{b}{\app{B}{a} \rightarrow \app{\W}{A}{B}}) \rightarrow \app{\W}{A}{B}}}
    \end{mathparpagebreakable}
    Predicative W types only allow $U$ to be $\Set$ or $\Type{}$,
    while impredicative W types also allow it to be $\Prop$.
    Including impredicative W types as well poses several technical challenges.
\end{itemize}

Because some of these changes violate backward compatibility, they cannot be
adopted in \lang.

The literature suggests that future work could prove \langAnother and \lang
equivalent to derive that strong normalization (and therefore logical
consistency) of \langAnother implies that they hold in \lang.
More specifically, \cite{conversion} show that a typed and an untyped convertibility in a Martin--L\"of type theory (MLTT) imply each other;
and \citet{w-types, polynomial-functors-w} show that W types in an extensional MLTT
can encode well-formed inductive types, including nested inductive types
(while \citet{hofmann} shows that extensional MLTT is a conservative extension of intensional MLTT).

We leave this line of inquiry as future work\footnote{In fact, ongoing work by
  the second author, Yufeng Li, in collaboration with Bruno Barras has reportedly
  finished the strong normalization proof of \langAnother using realisability
  candidates based on \citet{barras-thesis}. (Private
  communication, Dec. 2021).}, since we have other reasons to believe backward-incompatible
  changes are necessary in \lang to make sized typing practical.
Nevertheless, we next explain where each these changes originate and why they
seem necessary for the model.

\subsubsection{Typed Reduction}

Recall from \autoref{sec:overview:comparison} that we add an impredicative $\Prop$ universe
and universe cumulativity to the existing universe hierarchy in \CIChatminus.
We follow the set-theoretical model presented by \citet{not-so-simple-cc},
where $\Prop$ is treated proof-irrelevantly:
its set-theoretical interpretation is the set $\set{\emptyset, \set{\emptyset}}$,
and a type in $\Prop$ is either $\set{\emptyset}$ (representing true, inhabitated propositions)
or $\emptyset$ (representing false, uninhabited propositions).

Impredicativity of function types are encoded using a \emph{trace encoding} \citep{aczel-trace}.
First, the \emph{trace} of a (set-theoretical) function $f : A \to B$ is defined as
$$\trace(f) = \set{(a, b) \mid a \in A, b \in f(a)}.$$
Then the interpretation of a function type $\prod{x}{t}{u}$ is defined as
$$\bigg\{\trace(f) \mathbin{\big|} f \in A \times \bigcup_{a \in A} B_a \text{ and } \forall a \in A, f(a) \in B_a\bigg\}$$
where $A$ is the interpretation of $t$ and $B_a$ is the interpretation of $u$ when $x = a$,
while a function $\abs{x}{t}{e}$ is interpreted as
$\set{(a, b) \mid a \in A, b \in y_a}$
where $y_a$ is the interpretation of $e$ when $x = a$.

To see that this definition satisfies impredicativity,
suppose that $u$ is in $\Prop$.
Then $B_a$ is either $\emptyset$ or $\set{\emptyset}$.
If it is $\emptyset$, then there is no possible $f(a)$,
making the interpretation of the function type itself $\emptyset$.
If it is $\set{\emptyset}$, then $f(a) = \emptyset$,
and $\trace(f) = \emptyset$ since there is no $b \in f(a)$,
making the interpretation of the function type itself $\set{\emptyset}$.

Since reduction is untyped, it is perfectly fine for ill-typed terms to reduce.
For instance, we can have the derivation
$\mt, \mt \vdash \app{(\abs{x}{(\prod{p}{\Prop}{p \to p})}{x})}{\Prop} \rhd_\beta \Prop$
even though the left-hand side is not well typed.
However, to justify a convertibility (such as a reduction) in the model,
we need to show that the set-theoretic interpretations of both sides are equal.
For the example above, since $\prod{p}{\Prop}{p \to p}$ is in $\Prop$
and is inhabited by $\abs{p}{\Prop}{\abs{x}{p}{x}}$,
its interpretation must be $\set{\emptyset}$.
Then the interpretation of the function on the left-hand side must be $\set{(\emptyset, \emptyset)}$.
By the definition of the interpretation of application,
since the interpretation of $\Prop$ is not in the domain of the function,
the left-hand side becomes $\emptyset$.
Meanwhile, the right-hand side is $\set{\emptyset, \set{\emptyset}}$,
and the interpretations of the two sides aren't equal.

Ultimately, the set-theoretic interpretations of terms only make sense for well-typed terms,
despite being definable for ill-typed ones as well.
Therefore, to ensure a sensible interpretation,
reduction (and therefore subtyping and convertibility) needs to be typed.

\subsubsection{Size Irrelevance}

In the model, we need to know that functions cannot make computational decisions
based on the value of a size variable, \ie that computation is size irrelevant.
This is necessary to model functions as set-theoretic functions, since sizes are
ordinals and (set-theoretic) functions quantifying over ordinals may be too
large to be proper sets.

In short, while we conjecture that size irrelevance holds in \lang since size
expressions are second class and size variables are implicitly quantified, it is
no longer true in the model, where sizes are modeled as ordinals and size
variables must be explicitly quantified.
As a result, we follow \citet{barras-thesis} in creating two typing modes in
\langAnother (normal and size irrelevant) and two function spaces (normal
and sized) which allow proving that functions respect sizes in necessary
situations in the model.
The sized function space and size irrelevant mode enforce that the size of the
function's domain is irrelevant during typing, and this is use to type check
fixpoints.

In detail, the problem arises as follows.

Given a recursive call $f$ of some fixpoint whose body is $e'$
and two functions $\psi_1, \psi_2$ of the same type as $f$,
if they behave identically, then the model requires that
$\subst{e'}{f}{\psi_1}$ and $\subst{e'}{f}{\psi_2}$ are indistinguishable.
However, this cannot be shown with the current typing rules,
which is why \emph{size irrelevance} is introduced.

Formally, the set-theoretic model interpret terms and types as their natural
set-theoretic counterparts and size expressions as ordinals;
we call these their \emph{valuations}.
Given some environment $\Gamma$ and a term $e$ that is well typed under
$\Gamma$ with size variables $V = \SV(e)$,
letting $\rho$ be the valuations of the term variables of $\Gamma$ and $\pi$ be
the valuations of the size variables in $V$,
the valuation of $e$ is denoted by $\Val(e)_\rho^\pi$.

Consider now the valuation of the following term that is well typed under
$\Gamma$.
$$e = \fixE{\upsilon}{f}{\Nat^\upsilon \rightarrow \Nat^\upsilon}{e'}{\infty}$$
As the fixpoint is evaluated at the infinite size $\infty$,
intuitively the valuation of $e$ must be the fixed point of
$e'$ with respect to $f$.
Then to compute $e$, we take an initial approximation of $e'$ and iterate until
the fixed point has been reached.

For simplicity, suppose that for every ordinal $\alpha$, we have some valuation
$\Val(\Nat^\upsilon)_\rho^{\subst{\pi}{\upsilon}{\alpha}} = \McN^\alpha$,
where given some ordered ordinals $\vec{\alpha}$,
$\vec{\McN^{\alpha}}$ is a $\subseteq$-increasing sequence of sets constant
beyond $\alpha = \omega$.
Let $D_0 = \set{\emptyset}$, the vacuous function space
(representing $\McN^0 \to \McN^0$),
and define the following:
\begin{align*}
  D_\alpha
  &= \Val(\Nat^\upsilon \rightarrow \Nat^\upsilon)_\rho^{\subst{\pi}{\upsilon}{\alpha}}
  = \McN^\alpha \rightarrow \McN^\alpha \\
  \varphi_\alpha(\psi) &= \Val(e')_{\subst{\rho}{f}{\psi}}^{\subst{\pi}{\upsilon}{\alpha}}
  \qquad (\text{where} ~ \psi \in D_\alpha)
\end{align*}

The usual approach to compute $\Val(e)_\rho^\pi$ is to iterate up to the least fixed point of
$\varphi_\alpha$ starting at $\psi_0 = D_0$ and setting
$\psi_{\alpha + 1} = \varphi_\alpha(\psi_\alpha)$.
\refrule{fix-explicit} ensures that $\psi_\alpha \in D_\alpha$;
however, we also need to ensure that the sequence
$\vec{\psi_\alpha}$ eventually converges.

What would be a sufficient condition for convergence?  As
$\vec{\psi_\alpha}$ is obtained by successively improving upon
approximations of the fixed point of (the interpretation of) the
defining body $e'$, we expect that subsequent approximations to use
the results of previous approximations, and so that
\begin{align*}
  \forall x \in \McN^\alpha \subseteq \McN^\beta, \psi_\alpha(x) = \psi_\beta(x).
  \label{eqn:irrel}\tag{\textsc{irrel}}
\end{align*}
This is the formal statement of size irrelevance in the model:
size variables bound by fixpoints merely restrict their domains and don't
affect their computation.
It turns out that size irrelevance ensures that $\psi_\alpha$
converges at $\psi_\omega$,
so it suffices to prove \eqref{eqn:irrel}.

We proceed by induction on $\alpha$ and $\beta$.
Assuming \eqref{eqn:irrel} holds for some $\alpha$ and $\beta$,
unfolding definitions, the goal is to show that
$$\forall x \in \McN^{\alpha+1} \subseteq \McN^{\beta+1},
  \Val(e')_{\subst{\rho}{f}{\psi_\alpha}}^{\subst{\pi}{\upsilon}{\alpha}}(x) =
  \Val(e')_{\subst{\rho}{f}{\psi_\beta}}^{\subst{\pi}{\upsilon}{\beta}}(x).$$
Inductively, $\psi_\alpha$ and $\psi_\beta$ behave identically,
but from \refrule{fix-explicit} we cannot easily conclude that $e'$ cannot tell
them apart.
This is the same problem encountered by \citet{barras-thesis},
who resolves it using a new size irrelevant judgement.
We use a similar judgement for \langAnother,
expanding it to allow recursive references of fixpoints as arguments
to other functions.

\subsubsection{Size-Annotated Fixpoints}

As shown above, the set-theoretic interpretation of a fixpoint evaluated at some size $s$
is the iteration of its corresponding operator up to $\alpha$ times,
where $\alpha$ is the valuation of $s$.
Without explicitly annotating the size, we wouldn't know how many times to iterate,
since $s$ is otherwise only found in the type of the fixpoint and not in the fixpoint term itself.

\section{Size Inference}\label{sec:algorithm}

In this section, we present a size inference algorithm, based on that of \CIChat~\citep{cic-hat}.
Starting with a program (that is, a global environment) consisting of bare global declarations,
we want to obtain a program consisting of the corresponding size-annotated global declarations.
Given bare terms corresponding to terms in CIC
(with no size annotations but with the recursive argument index marked on fixpoints),
the algorithm assigns size annotations while collecting a set of subsizing constraints.
Because the subsizing constraints that must be satisfied are based on the typing rules,
this algorithm is also necessarily a type checking algorithm,
ensuring well-typedness of the size-annotated term.
The algorithm then returns either a size-annotated and well-typed term along with the set of subsizing constraints, or fails.

Before proceeding to the next global declaration,
we solve its constraints by finding an assignment from size variables to size expressions
such that these size expressions satisfy all of the constraints.
Then we perform the substitution of these assignments on the declaration.
This lets us run the inference algorithm on each declaration independently,
without needing to manipulate a constraint set every time a global declaration is used in a subsequent one.

One of the most involved parts of the algorithm is the size inference and type checking of \cofixpoints,
which adapts the \RecCheck algorithm from \Fhat~\citep{f-hat}.
The other notably involved part of the algorithm is the \solve algorithm,
which given a set of constraints produces a valid solution.
Finally, we state soundness and completeness theorems for the algorithm as a whole,
proving only soundness and leaving completeness as a conjecture.

\subsection{Preliminaries}

We first formally define the notions of constraints and solutions,
as well as some additional notation.

\begin{definition}
A \textbf{subsizing constraint set} (or simply \textbf{constraint set}) $C$ is a set of pairs of size expressions $s_1 \sqsubseteq s_2$ (also referred to as a \textbf{constraint})
representing a subsizing relation from $s_1$ to $s_2$ that must be enforced.
(When ambiguous, we will explicitly distinguish between the \emph{constraint} $s_1 \sqsubseteq s_2$ and the \emph{judgement} $s_1 \sqsubseteq s_2$.)

We write $s_1 = s_2$ to mean the two pairs $s_1 \sqsubseteq s_2$ and $s_2 \sqsubseteq s_1$.
Given a set of size variables $V$, we also write $\upsilon \sqsubseteq V$ for the pointwise constraint set $\set{\upsilon \sqsubseteq \upsilon' \mid \upsilon' \in V}$,
and similarly for $V \sqsubseteq \upsilon$.
\end{definition}

This is the natural representation of the constraints:
the algorithm is based on the typing rules,
and they produce constraints representing subsizing judgements that need to hold.
However, in \RecCheck and in \solve we will need to view these constraints as a graph.
We use $C$ to represent either the constraint set or the graph depending on the context.
First, notice that any noninfinite size consists of a size variable and some finite number $n$ of successor ``hats'';
we will write this as $\hat{\upsilon}^n$ so that, for instance, $\hat{\hat{\hat{\upsilon}}}$ is instead $\hat{\upsilon}^3$.

\begin{definition}
A \textbf{subsizing constraint graph} (or simply \textbf{constraint graph}) $C$ of a constraint set is a weighted, directed graph whose vertices are size variables, edges are constraints, and weights are integers.

Given a constraint \mbox{$\hat{\upsilon}_1^{n_1} \sqsubseteq \hat{\upsilon}_2^{n_2}$},
the constraint graph contains an edge from $\upsilon_1$ to $\upsilon_2$ with weight $n_2 - n_1$.
Constraints of the form $s \sqsubseteq \infty$ are trivially true and aren't added to the graph.
Constraints of the form $\infty \sqsubseteq \hat{\upsilon}^n$ correspond to an edge from $\infty$ to $\upsilon$ with weight $0$.
\end{definition}

Given a constraint graph and some set of size variables $V$,
it's useful to know which variables in the graph can be reached from $V$ or will reach $V$.

\begin{definition}[Transitive closures]~\\[-4ex]
\begin{itemize}
  \item Given a set of size variables $V$, the \textbf{upward closure} $\bigsqcup V$ with respect to $C$ is the set of variables that can be reached from $V$ by travelling along the directed edges of $C$.
  That is, $V \subseteq \bigsqcup V$, and if $\upsilon_1 \in \bigsqcup V$ and $\hat{\upsilon}_1^{n_1} \sqsubseteq \hat{\upsilon}_2^{n_2}$, then $\upsilon_2 \in \bigsqcup V$.
  \item Given a set of size variables $V$, the \textbf{downward closure} $\bigsqcap V$ with respect to $C$ is the set of variables that can reach $V$ by travelling along the directed edges of $C$.
  That is, $V \subseteq \bigsqcap V$, and if $\upsilon_2 \in \bigsqcap V$ and $\hat{\upsilon}_1^{n_1} \sqsubseteq \hat{\upsilon}_2^{n_2}$, then $\upsilon_1 \in \bigsqcap V$.
\end{itemize}
\end{definition}

Finally, we can define what it means to be a solution of a constraint set,
as well as some useful related notation.

\begin{definition}[Size substitutions and constraint satisfaction]~\\[-4ex]
\begin{itemize}
  \item A size substitution $\rho$ applied to a set of size variables produces a set of size expressions:
  $\rho V \coloneqq \set{\rho \upsilon \mid \upsilon \in V}$.
  Applying $\rho$ to a constraint set works similarly.

  \item The composition of size substitutions $\rho_1$ and $\rho_2$ is defined as \mbox{$(\rho_1 \circ \rho_2) \upsilon \coloneqq \rho_1(\rho_2 \upsilon)$}.

  \item A size substitution $\rho$ \textbf{satisfies the constraint set} $C$ (or is a \textbf{solution} of $C$), written as $\rho \vDash C$, if for every constraint $s_1 \sqsubseteq s_2$ in $C$, the judgement $\rho s_1 \sqsubseteq \rho s_2$ holds.
  For convenience, we will also require that $\rho$ maps to size expressions whose size variables are fresh,
  and that it doesn't map any size variables not in $C$.
\end{itemize}
\end{definition}

We now define four judgements to represent \emph{algorithmic subtyping}, \emph{checking}, \emph{inference}, and \emph{well-formed\-ness}.
They all use the symbol $\rightsquigarrow$, with inputs on the left and outputs on the right.
\begin{itemize}
  \item $\gg \vdash t \constrain u \rightsquigarrow C$ (algorithmic subtyping) takes environments $\Gamma_G, \Gamma$ and annotated terms $t, u$, and produces a set of constraints $C$ that must be satisfied in order for $t$ to be a subtype of $u$.
  \item $C, \gg \vdash e^\circ \Leftarrow t \rightsquigarrow C', e$ (checking)
  takes a set of constraints $C$, environments $\Gamma_G, \Gamma$,
  a bare term $e^\circ$, and an annotated type $t$,
  and produces the annotated term $e$ with a set of constraints $C'$
  that ensures that the type of $e$ subtypes $t$.
  \item $C, \gg \vdash e^\circ \rightsquigarrow C', e \Rightarrow t$ (inference)
  takes a set of constraints $C$, environments $\Gamma_G, \Gamma$,
  and a bare term $e^\circ$, and produces the annotated term $e$, its annotated type $t$, and a set of constraints $C$.
  \item $\Gamma_G^\circ \rightsquigarrow \Gamma_G$ (well-formedness) takes a global environment with bare declarations and produces a global environment where each declaration has been properly annotated via size inference.
\end{itemize}

In checking and inference, the input constraint set represents the constraints that must be satisfied
for the local environment to be well formed.
Algorithmic subtyping doesn't need this constraint set since subtyping doesn't rely on well-formedness of the environments.
The output constraint set represents \emph{additional} constraints that must be satisfied
for the input term(s) to be well typed.

The algorithm is implicitly parameterized over a fixed signature $\Sigma$,
as well as two mutable sets of size variables $\V, \V^*$, such that $\V^* \subseteq \V$.
Their assignment is denoted with $\coloneqq$ and they are initialized as empty.
The set $\V^*$ contains \textit{position} size variables,
which mark size-preserving types by replacing position annotations,
and we use $\tau$ for these position size variables.
We define two related metafunctions: \PV returns all position size variables in a given term,
while $\erase{\ph}^*$ erases position size variables to position annotations and all other annotations to bare.

Finally, on the right-hand size of inference judgements, we use $e \Rightarrow^* t$ to mean $e \Rightarrow t' \wedge t = \whnf{t'}$.
\whnf reduces a term until it is in \emph{weak head normal form} (WHNF),
which allows us to syntactically match on and take apart the term.
A term is in weak head normal form when no reduction rule can be applied to the entire term,
when it has the form $\app{X}{\vec{a}}$ and $X$ is in WHNF,
when it is a case expression whose target is in WHNF,
or when it is an applied fixpoint whose arguments are in WHNF.
For our purposes, the important thing is that we can tell whether or not a term is
a universe, a function type, or an inductive type.

We define a number of additional metafunctions to translate the side conditions from the typing rules into procedural form.
They are introduced as needed.

\paragraph*{} The entry point of the algorithm is the well-formedness judgement,
which take and produce global environments representing whole programs.
Its rules are defined in \autoref{fig:algorithm-wf} and use the mutually-defined rules of the checking and inference judgements,
defined in \autoref{fig:algorithm-check}, \autoref{fig:algorithm-core}, and \autoref{fig:algorithm-coind} respectively.
We begin with the latter two first in \autoref{sec:algorithm:infer},
followed by a detailed look at \RecCheck in \autoref{sec:algorithm:reccheck}.
Well-formedness is discussed in \autoref{sec:algorithm:wf}.
Finally, we make our way up to soundness and completeness with respect to the typing rules in \autoref{sec:algorithm:metatheory}.

\subsection{Inference Algorithm}\label{sec:algorithm:infer}

Size inference begins with either a bare term or a position term. For the bare terms, even type annotations of \cofixpoints are bare, \ie
  $$e^\circ \Coloneqq \dots
    \mid \fix{m}{f^n}{t^\circ}{e^\circ}
    \mid \cofix{m}{f}{t^\circ}{e^\circ}$$
Notice that fixpoints still have the indices $n$ of the recursive arguments, whereas surface Coq programs generally have no indices.
To produce these indices, we do what Coq currently does: brute-force search.
We try the algorithm on every combination of indices from left to right.
This continues until one combination works, or fails if none do.
Then the type annotations are initially position annotated on as many types as possible,
and this position term itself is passed into the algorithm.
Because of the way the Coq kernel is structured, this may not always be possible in the implementation.
We discuss this and other implementational issues in the next section.

\subsubsection{Checking}

\begin{figure*}
\centering

\begin{flushleft}
\fbox{$C, \gg \vdash e^\circ \Leftarrow t \rightsquigarrow C, e$}
\end{flushleft}

\vspace{-3ex}

\begin{mathpar}
\inferrule[\defrule{a-check}]{
    C, \gg \vdash e^\circ \rightsquigarrow C_1, e \Rightarrow t \\
    \gg \vdash t \constrain u \rightsquigarrow C_2
}{
    C, \gg \vdash e^\circ \Leftarrow u \rightsquigarrow C_1 \cup C_2, e
}
\end{mathpar}
\caption{Size inference algorithm: Checking}
\label{fig:algorithm-check}
\end{figure*}


\begin{figure*}
\centering

\begin{flushleft}
\fbox{$\Gamma_G, \Gamma \vdash t \constrain u \rightsquigarrow C$}
\end{flushleft}

\vspace{-2ex}

\begin{mathpar}
\dots
\and
\inferrule[\defrule{a-st-ind}]
  {I \textrm{ inductive }}
  {\gg \vdash I^s \leq I^{s'} \rightsquigarrow \set{s \sqsubseteq s'}}
\and
\inferrule[\defrule{a-st-coind}]
  {I \textrm{ coinductive }}
  {\gg \vdash I^s \leq I^{s'} \rightsquigarrow \set{s' \sqsubseteq s}}
\end{mathpar}
\caption{Size inference algorithm: Algorithmic subtyping (excerpt)}
\label{fig:algorithm-subtyping}
\end{figure*}

\paragraph*{} \refrule{a-check} in \autoref{fig:algorithm-check} is the checking component of the algorithm.
It uses algorithmic subtyping in \autoref{fig:algorithm-subtyping} to ensure that the inferred type of the term is a subtype of given type.
This subtyping is defined inductively over the rules of the subtyping judgement in the straightforward manner, taking the union of constraint sets from their premises;
we present only Rules \refnorule{a-st-ind} and \refnorule{a-st-coind},
which shows the concrete subsizing constraints derived from comparing two \coinductive types.
It may also fail if two terms are not subtypes and are inconvertible.

\subsubsection{Inference: Part 1}

\begin{figure*}
\centering

\begin{flushleft}
\fbox{$C, \gg \vdash e^\circ \rightsquigarrow C, e \Rightarrow t$}
\end{flushleft}

\vspace{-4ex}

\begin{mathpar}
\inferrule[\defrule{a-var-assum}]{
    (\assm{x}{t}) \in \Gamma
}{
    C, \gg \vdash x \rightsquigarrow \set{}, x \Rightarrow t
}

\inferrule[\defrule{a-var-def}]{
    (\defn{x}{t}{e}) \in \Gamma \and
    \overline{\upsilon'_i} = \SV(e, t) \setminus \SV(C) \\\\
    \overline{\upsilon_i} = \fresh{\norm{\overline{\upsilon'_i}}} \and
    \rho = \set{\vec{\upsilon'_i \mapsto \upsilon_i}}
}{
    C, \gg \vdash x \rightsquigarrow \set{}, x^\rho \Rightarrow \rho t
}

\inferrule[\defrule{a-const-assum}]{
    (\Assm{x}{t}) \in \Gamma_G
}{
    C, \gg \vdash x \rightsquigarrow \set{}, x \Rightarrow t
}

\inferrule[\defrule{a-const-def}]{
    (\Defn{x}{t}{e}) \in \Gamma_G \and
    \overline{\upsilon'_i} = \SV(e, t) \setminus \SV(C) \\\\
    \overline{\upsilon_i} = \fresh{\norm{\overline{\upsilon'_i}}} \and
    \rho = \set{\vec{\upsilon'_i \mapsto \upsilon_i}}
}{
    C, \gg \vdash x \rightsquigarrow \set{}, x^\rho \Rightarrow \rho t
}

\inferrule[\defrule{a-univ}]{~}{
    C, \gg \vdash U \rightsquigarrow \set{}, U \Rightarrow \axiom{U}
}

\inferrule[\defrule{a-prod}]{
    C, \gg \vdash t^\circ \rightsquigarrow C_1, t \Rightarrow^* U_1 \\\\
    C \cup C_1, \gg (x:t) \vdash u^\circ \rightsquigarrow C_2, u \Rightarrow^* U_2
}{
    C, \gg \vdash \prod{x}{t^\circ\!}{u^\circ} \rightsquigarrow C_1 \cup C_2, \prod{x}{t}{u} \Rightarrow \rules{U_1}{U_2}
}

\inferrule[\defrule{a-abs}]{
    C, \gg \vdash t^\circ \rightsquigarrow C_1, t \Rightarrow^* U \\\\
    C \cup C_1, \gg (x:t) \vdash e^\circ \rightsquigarrow C_2, e \Rightarrow u
}{
    C, \gg \vdash \abs{x}{t^\circ\!}{e^\circ} \rightsquigarrow C_1 \cup C_2, \abs{x}{|t|}{e} \Rightarrow \prod{x}{t}{u}
}

\inferrule[\defrule{a-app}]{
    C, \gg \vdash e_1^\circ \rightsquigarrow C_1, e_1 \Rightarrow^* \prod{x}{t}{u} \\\\
    C, \gg \vdash e_2^\circ \Leftarrow t \rightsquigarrow C_2, e_2
}{
    C, \gg \vdash e_1^\circ ~ e_2^\circ \rightsquigarrow C_1 \cup C_2, e_1 ~ e_2 \Rightarrow u[x \coloneqq e_2]
}

\inferrule[\defrule{a-let-in}]{
    C, \gg \vdash t^\circ \rightsquigarrow C_1, t \Rightarrow^* U \\
    C, \gg \vdash e_1^\circ \Leftarrow t \rightsquigarrow C_2, e_1 \\\\
    C \cup C_1 \cup C_2, \gg (\defn{x}{t}{e_1}) \vdash e_2^\circ \rightsquigarrow C_3, e_2 \Rightarrow u
}{
    C, \gg \vdash \letin{x}{t^\circ}{e_1^\circ}{e_2^\circ} \rightsquigarrow C_1 \cup C_2 \cup C_3, \letin{x}{|t|}{e_1}{e_2} \Rightarrow u[x \coloneqq e_1]
}

\dots

\end{mathpar}
\caption{Size inference algorithm: Inference (1/2)}
\label{fig:algorithm-core}
\end{figure*}


\autoref{fig:algorithm-core} is the first half of the inference component of the algorithm,
presenting the rules for basic language constructs.
In general, when the local environment is extended with some term, we make sure that the input constraint set is extended as well
with the constraints generated from size inference on that term.
The constraints returned are simply the union of all the constraints generated by the premises.
Note that since type annotations need to be bare (to maintain subject reduction, as discussed in \autoref{sec:metatheory:sr:bare}),
they must be erased first before reconstructing the term.

Rules \refnorule{a-var-assum}, \refnorule{a-const-assum}, \refnorule{a-univ}, \refnorule{a-prod}, \refnorule{a-abs}, \refnorule{a-app}, and \refnorule{a-let-in} are all fairly straightforward.
These rules use the metafunctions \axiom, \rules, and \elim, which correspond to the sets \Axioms, \Rules, and \Elims, defined in \autoref{fig:axruel}.
The metafunction \axiom produces the type of a universe; \rules produces the type of a function type given the universes of its argument and return types; and \elim directly checks membership in \Elims and can fail.

In Rules \refnorule{a-var-def} and \refnorule{a-const-def},
we generate a size substitution that freshens the size variables in the associated definition
using $\fresh$, which freshly generates the given number of variables and adds them to $\mathcal{V}$.
By freshening the size variables, we can define let-bound type aliases that can be used like regular types.
For instance, in the term
$$\letin{N}{\Type{}}{\Nat^{\upsilon_1}}{\abs{n}{N \rightarrow N}{n}}$$
the two uses of $N$ need not have the same size:
the type of the function might be inferred as
$N^{\seq{\upsilon_1 \mapsto \upsilon_2}} \rightarrow N^{\seq{\upsilon_1 \mapsto \upsilon_3}}$,
which by $\delta$-reduction is convertible with $\Nat^{\upsilon_2} \rightarrow \Nat^{\upsilon_3}$.

\subsubsection{Inference: Part 2}

\begin{figure*}
\centering

\begin{flushleft}
\fbox{$C, \gg \vdash e^\circ \rightsquigarrow C, e \Rightarrow t$}
\end{flushleft}

\vspace{-4ex}

\begin{mathpar}
\dots

\inferrule[\defrule{a-ind}]{
    \upsilon = \fresh{1}
}{
    C, \gg \vdash I \rightsquigarrow \set{}, I^\upsilon \Rightarrow \indtype{I}
}

\inferrule[\defrule{a-ind-star}]{
    \tau = \fresh{1} \and \mathcal{V}^* \coloneqq \mathcal{V}^* \cup \set{\tau}
}{
    C, \gg \vdash I^* \rightsquigarrow \set{}, I^\tau \Rightarrow \indtype{I}
}

\inferrule[\defrule{a-constr}]{
    \upsilon = \fresh{1}
}{
    C, \gg \vdash c \rightsquigarrow \set{}, c \Rightarrow \constrtype{c}{\upsilon}
}

\inferrule[\defrule{a-case}]{
    C, \gg \vdash e^\circ \rightsquigarrow C_1, e \Rightarrow^* I_k^s ~ \overline{p} ~ \overline{a} \\
    C, \gg \vdash P^\circ \rightsquigarrow C_2, P \Rightarrow t_p \\
    \prodctx{\any}{\prodctx{\Delta_k}{U_k}} = \indtype{I_k} \\
    U = \decompose{t_p}{\norm{\Delta_k} + 1} \\
    \elim{U_k}{U}{I_k} \\
    \upsilon = \fresh{1} \\
    \gg \vdash t_p \constrain \motivetype{\overline{p}}{U}{I_k^{\hat{\upsilon}}} \rightsquigarrow C_3 \\
    \textrm{For each $j$:} \\
    C, \gg \vdash e^\circ_j \Leftarrow \branchtype{\overline{p}}{c_j}{\upsilon}{P} \rightsquigarrow C_{4j}, e_j \\
    C_5 = \casesize{I_k^s}{\hat{\upsilon}} \cup C_1 \cup C_2 \cup C_3 \cup (\textstyle\bigcup_j C_{4j})
}{
    C, \gg \vdash \caseof{P^\circ}{e^\circ}{c_j}{e_j^\circ} \rightsquigarrow C_5, \caseof{|P|}{e}{c_j}{e_j} \Rightarrow P ~ \overline{a} ~ e
}

\inferrule[\defrule{a-fix}]{
    \textrm{For each $k$:} \\
    C, \gg \vdash t_k^\circ \rightsquigarrow \any, \any \Rightarrow \any \\
    C, \gg \vdash \setrecstars{t_k^\circ}{n_k} \rightsquigarrow C_{1k}, t_k \Rightarrow^* U \\
    \prodctx{\Delta_k}{u_k} = \whnf{t_k} \and \prodctx{\Delta_k}{u'_k} = \shift{\prodctx{\Delta_k}{u_k}} \\
    \textstyle\bigcup_k C_{1k} \cup C, \gg \overline{(f_k : t_k)} \vdash e_k^\circ \Leftarrow \prodctx{\Delta_k}{u'_k} \rightsquigarrow C_{2k}, e_k \\
    \gg \Delta_k \vdash u_k \constrain u'_k \rightsquigarrow C_{3k} \\
    C_4 = \textstyle\bigcup_k C_{1k} \cup C_{2k} \cup C_{3k} \cup C \\
    C_5 = \RecCheckLoop{C_4}{\Gamma}{\overline{\getrecvar{t_k}{n_k}}}{\overline{t_k}}{\overline{e_k}}
}{
    C, \gg \vdash \fix{m}{f_k^{n_k}}{t_k^\circ}{e_k^\circ} \rightsquigarrow C_5, \fix{m}{f_k^{n_k}}{|t_k|^*}{e_k} \Rightarrow t_m
}

\inferrule[\defrule{a-cofix}]{
    \textrm{For each $k$:} \\
    C, \gg \vdash t_k^\circ \rightsquigarrow \any, \any \Rightarrow \any \\
    C, \gg \vdash \setcorecstars{t_k^\circ} \rightsquigarrow C_{1k}, t_k \Rightarrow^* U \\
    \prodctx{\Delta_k}{u_k} = \whnf{t_k} \and \prodctx{\Delta'_k}{u'_k} = \shift{\prodctx{\Delta_k}{u_k}} \\
    \textstyle\bigcup_k C_{1k} \cup C, \gg \overline{(f_k : t_k)} \vdash e_k^\circ \Leftarrow \prodctx{\Delta'_k}{u'_k} \rightsquigarrow C_{2k}, e_k \\
    \gg \vdash \Delta'_k \constrain \Delta_k \rightsquigarrow C_{3k} \\
    C_4 = \textstyle\bigcup_k C_{1k} \cup C_{2k} \cup C_{3k} \cup C \\
    C_5 = \RecCheckLoop{C_4}{\Gamma}{\overline{\getcorecvar{t_k}}}{\overline{t_k}}{\overline{e_k}}
}{
    C, \gg \vdash \cofix{m}{f_k}{t_k^\circ}{e_k^\circ} \rightsquigarrow C_5, \cofix{m}{f_k}{|t_k|^*}{e_k} \Rightarrow t_m
}
\end{mathpar}
\caption{Size inference algorithm: Inference (2/2)}
\label{fig:algorithm-coind}
\end{figure*}


\autoref{fig:algorithm-coind} is the second half of inference,
presenting the rules related to \coinductives and \cofixpoints.
A position term from a position-annotated \cofixpoint type can be passed into the algorithm, so we deal with the possibilities separately in Rules \refnorule{a-ind} and \refnorule{a-ind-star}.
In both rules, a bare \coinductive type is annotated with a size variable; in \refrule{a-ind-star}, it is also added to the set of position size variables $\V^*$.
The position annotation of \coinductive types occurs in \refrule{a-fix} or \refrule{a-cofix}, which we discuss shortly.

In \refrule{a-constr}, we generate a single fresh size variable, which gets annotated on the constructor's \coinductive type in the argument types of the constructor type, as well as the return type, which has the successor of that size variable.
All other \coinductive types which aren't the constructor's \coinductive type continue to have $\infty$ annotations.

The key constraint in \refrule{a-case} is generated by \casesize.
Similar to \refrule{a-constr}, we generate a single fresh size variable $\upsilon$ to annotate on $I_k$ in the branches' argument types, which correspond to the constructor arguments of the target.
Then, given the unapplied target type $I_k^s$, \casesize returns $\set{s \sqsubseteq \hat{\upsilon}}$ if $I_k$ is inductive and $\set{\hat{\upsilon} \sqsubseteq s}$ if $I_k$ is coinductive.
This ensures that the target type satisfies $I_k^s ~ \overline{p} ~ \overline{a} \leq I_k^{\hat{\upsilon}_k} ~ \overline{p} ~ \overline{a}$, so that \refrule{case} is satisfied.

The rest of the rule proceeds as we would expect: we infer the sized type of the target and the motive, we check that the motive and the branches have the types we expect given the target type, and we infer that the sized type of the case expression is the annotated motive applied to the target type's annotated indices and the annotated target itself.
We also ensure that the elimination universes are valid using \elim on the motive type's return universe and the target type's universe.
To obtain the motive type's return universe, we use \decompose.
Given a type $t$ and a natural $n$, this metafunction reduces $t$ to a function type $\prodctx{\Delta}{u}$ where $\norm{\Delta} = n$, reduces $u$ to a universe $U$, and returns $U$.
It can fail if $t$ cannot be reduced to a function type, if $\norm{\Delta} < n$, or if $u$ cannot be reduced to a universe.

\subsubsection{Inference: (Co)fixpoints}

Finally, we come to size inference and termination/productivity checking for \cofixpoints.
It uses the following metafunctions:
\begin{itemize}
  \item \setrecstars, given a function type $t$ and an index $n$, decomposes $t$ into arguments and a return type, reduces the $n$th argument type to an inductive type, annotates that inductive type with position annotation $*$, annotates the return type with $*$ if it has the same inductive type, and rebuilds the function type.
    This is how fixpoint types obtain their position annotations without being user-provided; the algorithm will remove other position annotations if size preservation fails.

    Similarly, \setcorecstars annotates the coinductive return type first, then the argument types with the same coinductive type.
    Both of these can fail if the $n$th argument type or the return type respectively are not \coinductive types.
    Note that the decomposition of $t$ may perform reductions using \whnf.
  \item \getrecvar, given a function type $t$ and an index $n$, returns the position size variable of the annotation on the $n$th inductive argument type, while \getcorecvar returns the position size variable of the annotation on the coinductive return type.
    Essentially, they retrieve the position size variable of the annotation on the primary \corecursive type of a \cofixpoint type.
  \item \shift replaces all position size annotations $s$ (\ie $\floor{s} \in \V^*$) by their successor $\hat{s}$.
\end{itemize}

Although the desired \cofixpoint is the $m$th one in the block of mutually-defined \cofixpoints, we must still size-infer and type-check the entire mutual definition.
Rules \refnorule{a-fix} and \refnorule{a-cofix} first run the size inference algorithm on each of the \cofixpoint \emph{types}, ignoring the results, to ensure that any reduction on those types will terminate.
Then we annotate the bare types with position annotations (using \setrecstars/\setcorecstars) and pass these position types through the algorithm to get sized types $\overline{t_k}$.
Next, we check that the \cofixpoint bodies have the successor-sized types of $\overline{t_k}$ when the \cofixpoints have types $\overline{t_k}$ in the local environment.

\begin{figure}
\centering
\def\svgwidth{\columnwidth}
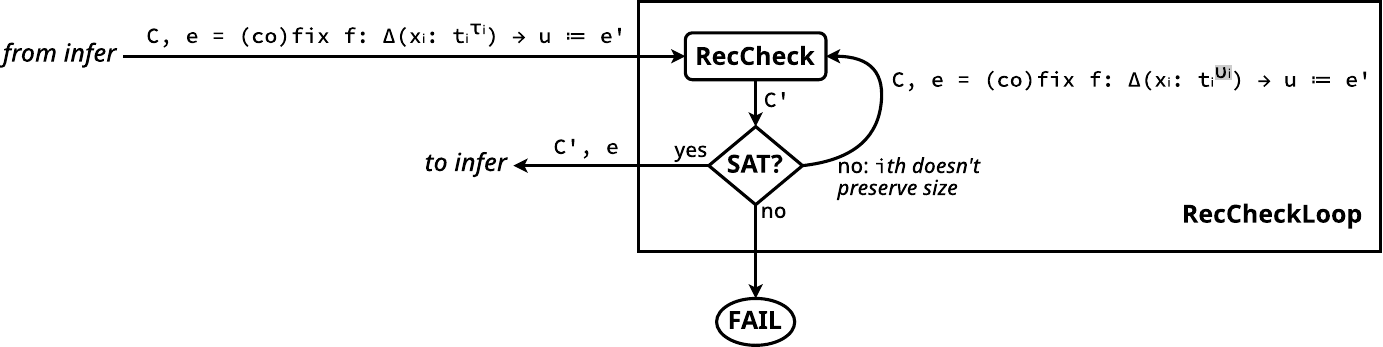
\caption{Illustration of simplification of \RecCheckLoop}
\label{fig:RecCheckLoop}
\end{figure}
\begin{figure}
\centering

\begin{minted}[escapeinside=<>,mathescape=true]{ocaml}
let rec RecCheckLoop <$C$> <$\Gamma$> <$\overline{\tau_k}$> <$\overline{t_k}$> <$\overline{e_k}$> =
  try let <$C'$> = <$\set{}$> in
    let for i = 1 to k do
      let pv<$_i$> = <$\PV$> <${t_i}$> in (* noninfinite $V^{\ast}$ *)
      let sv<$_i$> = (<$\SV$> <$\Gamma$> <$\cup$> <$\SV$> <$t_i$> <$\cup$> <$\SV$> <$e_i$>) <$\setminus$> pv<$_i$> in (* infinite $V^{\neq}$ *)
      <$C'$> := <$C'$> <$\cup$> RecCheck <$C$> <$\tau_i$> pv<$_i$> sv<$_i$>
    done in <$C'$>
  with RecCheckFail <$V$> ->
    if (empty? <$V$>)
    then raise RecCheckLoopFail
    else <$\mathcal{V}^*$> := <$\mathcal{V}^* \setminus V$>; RecCheckLoop <$C$> <$\overline{\tau_k}$> <$\overline{t_k}$> <$\overline{e_k}$>
\end{minted}

\caption{Pseudocode implementation of \RecCheckLoop}
\label{fig:helpers}
\end{figure}


Lastly, we need to check that the constraint set so far is satisfiable.
However, \setrecstars and \setcorecstars optimistically annotate \textit{all} possible \coinductive types in the \cofixpoint type with position annotations, while not all \cofixpoints are size preserving.
Therefore, instead of calling \RecCheck directly to check satisfiability,
\RecCheckLoop iteratively calls \RecCheck and discards incorrect position annotations at each iteration.
A simplification of the algorithm, not handling mutual \cofixpoints and removing single position annotations at a time,
is illustrated in \autoref{fig:RecCheckLoop}.
The size variable $\upsilon_i$ highlighted in grey corresponds to turning the position size variable into a regular size variable.

More precisely, \RecCheck either returns a new constraint set,
or it fails with some set $V$ of position size variables that must be set to infinity
and therefore mark arguments that aren't size preserving.
If $V$ is empty, then \RecCheckLoop fails overall: this indicates that the overall constraint set is unsatisfiable.
If $V$ is not empty, then we can try \RecCheck again after removing the size variables in $V$ from our set of position size variables,
thus no longer enforcing that they must be size preserving.
An OCaml-like pseudocode implementation of \RecCheckLoop is provided by \autoref{fig:helpers}.

\subsection{RecCheck}\label{sec:algorithm:reccheck}

As in previous work on \CChatomega with coinductive streams~\citep{cc-hat-omega} and in \CIChat, we use the same \RecCheck algorithm from \Fhat~\citep{f-hat}.
This algorithm attempts to ensure that the subsizing rules in \autoref{fig:subsizing} can be satisfied within a given set of constraints.
It does so by checking the set of constraints for invalid circular subsizing relations, setting the size variables involved in the cycles to $\infty$, and producing a new set of constraints without these problems; or it fails, indicating nontermination or nonproductivity.
It takes four arguments:

\begin{itemize}
  \item A constraint set $C$, which we treat as a constraint graph.
  \item The size variable $\tau$ of the annotation on the type of the recursive argument (for fixpoints) or on the return type (for cofixpoints).
    While the return type (for fixpoints) or the types of other arguments (for cofixpoints) may optionally be marked as size preserving,
    each \cofixpoint type requires at \textit{least} $\tau$ for the primary \corecursive type.
  \item A set of size variables $V^*$ that must be set to some non-infinite size.
    These are the size annotations in the \cofixpoint type that have position size variables.
    Note that $\tau \in V^*$.
  \item A set of size variables $V^\neq$ that must be set to $\infty$.
    These are all other non-position size annotations, found in the \cofixpoint types and bodies.
\end{itemize}

The key idea of the algorithm is that if there is a negative cycle in $C$,
then for any size variable $\upsilon$ in the cycle,
supposing that the total weight going once around the cycle is $-n$,
by transitivity we have the subsizing relation $\hat{\upsilon}^{n} \sqsubseteq \upsilon$,
This relation can only hold if $\upsilon = \infty$,
since $\succ{\infty} \sqsubseteq \infty$.
The algorithm proceeds as follows:

\begin{enumerate}
  \item \label{item:reccheck:smallest} Let $V^\iota = \bigsqcap V^*$, and add $\tau \sqsubseteq V^\iota$ to $C$.
    This ensures that $\tau$ is the smallest size variable among all the noninfinite size variables.
  \item \label{item:reccheck:neg-cycles} Find all negative cycles in $C$, and let $V^-$ be the set of all size variables in some negative cycle.
  \item Remove all edges with size variables in $V^-$ from $C$, and add $\infty \sqsubseteq V^-$.
  \item \label{item:reccheck:infty} Add $\infty \sqsubseteq \left(\bigsqcup V^\neq \cap \bigsqcup V^\iota\right)$ to $C$.
  \item \label{item:reccheck:bot} Let $V^\bot = \left(\bigsqcup \set{\infty}\right) \cap V^\iota$.
    This is the set of size variables that we have determined to both be infinite and noninfinite.
    If $V^\bot$ is empty, then return $C$.
  \item \label{item:reccheck:fail} Otherwise, let $V = V^\bot \cap (V^* \setminus \set{\tau})$, and fail with $\RecCheckFail{V}$.
    This is the set of contradictory position size variables excluding $\tau$, which we can remove from $\V^*$ in \RecCheckLoop.
    If $V$ is empty, there are no position size variables left to remove, so the check and therefore the size inference algorithm fails.
\end{enumerate}

Disregarding closure operations and set operations like intersection and difference, the time complexity of a single pass is $O(\norm{V}\norm{C})$, where $V$ is the set of size variables appearing in $C$.
This comes from the negative-cycle finding in (\ref{item:reccheck:neg-cycles}) using, for instance, the Bellman--Ford algorithm.

\begin{figure}
\centering

\begin{flushleft}
\fbox{$\Gamma_G^\circ \rightsquigarrow \Gamma_G$}
\end{flushleft}

\vspace{-3ex}

\begin{mathpar}
\inferrule*[right=\defrule{a-global-nil}]{~}{
    \mt \rightsquigarrow \mt
}
\and
\inferrule*[right=\defrule{a-global-assum}]{
    \Gamma_G^\circ \rightsquigarrow \Gamma_G \\
    \set{}, \Gamma_G, \mt \vdash t^\circ \rightsquigarrow C_1, t \Rightarrow^* \varw \\\\
    \rho = \solve{C_1}
}{
    \Gamma_G^\circ(\Assm{x}{t^\circ\!\!}) \rightsquigarrow \Gamma_G(\Assm{x}{\rho t})
}
\and
\inferrule*[right=\defrule{a-global-def}]{
    \Gamma_G^\circ \rightsquigarrow \Gamma_G \\
    \set{}, \Gamma_G, \mt \vdash t^\circ \rightsquigarrow C_1, t \Rightarrow^* \varw \\\\
    \set{}, \Gamma_G, \mt \vdash e^\circ \Leftarrow t \rightsquigarrow C_2, e \\
    \rho = \solve{C_1 \cup C_2}
}{
    \Gamma_G^\circ(\Defn{x}{t^\circ}{e^\circ\!\!}) \rightsquigarrow \Gamma_G(\Defn{x}{\rho t}{\rho e})
}
\end{mathpar}
    
\caption{Size inference algorithm: Well-formedness}
\label{fig:algorithm-wf}
\end{figure}


\subsection{Well-Formedness}\label{sec:algorithm:wf}

A self-contained chunk of code, be it a file or a module, consists of a sequence of \coinductive definitions (signatures) and programs (global declarations).
For our purposes, we assume that there is a singular well-formed signature defined independently.
Then we need to perform size inference on each declaration of $\Gamma_G$ in order,
as given by \autoref{fig:algorithm-wf}.

In \refrule{a-global-assum}, from a bare type $t^\circ$, inference gets us back
a constraint set $C_1$, the annotated type $t$, and \emph{its} type $\varw$,
such that $\Gamma_G, \mt \vdash t : w$ when the constraints in $C_1$ hold.
Similarly, \refrule{a-global-def} gets back from inference a term $e$,
its type $t$, and a constraint sets $C_1, C_2$,
with $\Gamma_G, \mt \vdash e : t$ when the constraints in $C_1 \cup C_2$ hold.
To rid ourselves of the constraint sets, we need to instantiate the size variables involved with size expressions that satisfy those constraints.
This is done by \solve which, when given a constraint set, produces a substitution $\rho$ that performs the instantiation.
Then given $\rho \vDash C_1 \cup C_2$, for instance, $\Gamma_G, \mt \vdash \rho e : \rho t$ holds unconditionally.

\subsubsection{Solving Constraints}

Let $C$ be a constraint graph corresponding to some constraint set for which we want to produce a substitution.
Supposing for now that it contains no negative cycles, for every connected component,
the simplest solution is to assign size expressions to each size variable such that
all of those size expressions have the same base size variable.
For instance, given the constraint set $\set{\upsilon_1 \sqsubseteq \hat{\upsilon}_2, \hat{\upsilon_1} \sqsubseteq \upsilon_3}$,
one solution could be the mapping $\set{\upsilon_1 \mapsto \tau, \upsilon_2 \mapsto \tau, \upsilon_3 \mapsto \hat{\tau}}$.

This kind of problem is a \emph{difference constraint} problem \citep{clrs}.
Generally a solution involves finding a mapping from variables to integers, whereas our solution will map from size variables to size expressions with the same base, but the technique using a single-source shortest-path algorithm still applies.
Given a connected component $C_c$ with no negative cycles or an $\infty$ vertex, our algorithm $\solvecomp$ for finding a solution proceeds as follows:

\begin{enumerate}
  \item Generate a fresh size variable $\tau$.
  \item For every size variable $\upsilon_i$ in $C_c$, add an edge $\tau \sqsubseteq \upsilon_i$ of weight $0$.
  \item \label{item:solvecomp:shortest-path} Find the weights $w_i$ of the shortest paths from $\tau$ to every other size variable $\upsilon_i$ in $C_c$.
    This yields the constraint $\tau \sqsubseteq \hat{\upsilon}_i^{w_i}$.
  \item Na\"ively, we would map each $\upsilon_i$ to the size expression $\hat{\tau}^{-w_i}$
    to trivially satisfy $\tau \sqsubseteq \hat{\upsilon}_i^{w_i}$,
    since this would become $\tau \sqsubseteq \tau$ after substitution.
    However, $-w_i$ may be negative, which would make no sense as the size of a size variable.
    Therefore, we find the largest weight $w_{\max} \coloneqq \max_i w_i$, and shift all the sizes up by $w_{\max}$.
    In other words, we return the map $\rho \coloneqq \set{\upsilon_i \mapsto \hat{\tau}^{w_{\max} - w_i}}$.
\end{enumerate}

Again, the time complexity of a single pass is $O(\norm{V}\norm{C_c})$ (where $V$ is the set of size variables in $C_c$)
due to
finding the single-source shortest paths in (\ref{item:solvecomp:shortest-path}) using,
for instance, the Bellman--Ford algorithm.
(Note that although there are no negative cycles, there are still negative \emph{weights}, so we cannot use, for example, Dijkstra's algorithm.)
The total $\solve$ algorithm, given some constraint graph $C$, is then as follows:

\begin{enumerate}
  \item Initialize an empty size substitution $\rho$.
  \item Find all negative cycles in $C$, and let $V^-$ be all size variables in some negative cycle.
  \item Let $V^\infty = \bigsqcup \set{\infty}$.
  \item Remove all edges with size variables in $V^- \cup V^\infty$ from $C$, and for every $\upsilon_i \in V^- \cup V^\infty$, add $\upsilon_i \mapsto \infty$ to $\rho$.
  \item For every connected component $C_c$ of $C$, add mappings $\solvecomp{C_c}$ to $\rho$.
\end{enumerate}

Since dividing the constraint graph into connected components partitions the size variables and constraints into disjoint sets,
the time complexity of all executions of \solvecomp is $O(\norm{V}\norm{C})$.
This is also the time complexity of negative-cycle finding.
These two dominate the time complexity of finding the connected components,
which is $O(\norm{V} + \norm{C})$.

\subsection{Metatheory}\label{sec:algorithm:metatheory}

In this subsection, we focus on soundness and completeness theorems of various parts of the inference algorithm.
The proof sketches and partial proofs for these and for the intermediate lemmas and theorems can be found in \autoref{sec:proofs}.

We first need soundness and completeness of \RecCheck.
The constraint set returned by \RecCheck ensures that variables that should be infinite are constrained to be so,
and that variables that shouldn't be infinite are not.
Intuitively, soundness then says that if you have a solution to a constraint set returned by \RecCheck,
then there must also be a solution to the original constraint set
that also ensures the same things.
Dually, completeness says that if you have a solution to the original constraint set that ensures these constraints,
then it must also be a solution to the constraint set returned by \RecCheck.

\begin{theorem}[Soundness of \RecCheck (SRC)]
If $\RecCheck(C', \tau, V^*, V^\neq) = C$, for every $\rho$ such that $\rho \vDash C$,
given a fresh size variable $\upsilon$, there exists a $\rho'$ such that the following all hold:
\begin{enumerate}
  \item $\rho' \vDash C'$;
  \item $\rho'\tau = \upsilon$;
  \item $\floor{\rho' V^*} = \upsilon$;
  \item $\floor{\rho' V^\neq} \neq \upsilon$;
  \item For all $\upsilon' \in V^\neq$, $(\set{\upsilon \mapsto \rho\tau} \circ \rho')(\upsilon') = \rho\upsilon'$; and
  \item For all $\tau' \in V^*$, $(\set{\upsilon \mapsto \rho\tau} \circ \rho')(\tau') \sqsubseteq \rho\tau'$.
\end{enumerate}
\end{theorem}

\begin{theorem}[Completeness of \RecCheck (CRC)]~\\
Suppose the following all hold:
\begin{itemize}
  \item $\rho \vDash C'$;
  \item $\rho\tau = \upsilon$;
  \item $\floor{\rho V^*} = \upsilon$; and
  \item $\floor{\rho V^\neq} \neq \upsilon$.
\end{itemize}
Then $\rho \vDash \RecCheck(C', \tau, V^*, V^\neq)$.
\end{theorem}

\RecCheck returns a constraint set for a single \cofixpoint definition with a fixed set of position variables;
\RecCheckLoop, on the other hand, returns a constraint set for an entire mutual \cofixpoint definition, finding a suitable set of position variables.
There are then two properties we want to ensure.

\begin{theorem}[Correctness of RecCheckLoop]~\\[-4ex]
\begin{enumerate}
  \item \RecCheckLoop terminates on all inputs.
  \item If $\RecCheckLoop{C', \Gamma, \vec{\tau_k}, \vec{t_k}, \vec{e_k}} = C$ with an initial position variable set $\mathcal{V}^*$,
  then for every $i \in \vec{k}$, $\RecCheck{C', \tau_i, \PV(t_i), \SV(\Gamma, t_i, e_i) \setminus \PV(t_i)} \subseteq C$ with a final position variable set $\mathcal{V}^*_\subseteq \subseteq \mathcal{V}^*$.
\end{enumerate}
\end{theorem}

We also want to ensure that \solvecomp and \solve actually return solutions of the constraint sets they are solving.

\begin{theorem}[Correctness of \solve and \solvecomp]~\\[-4ex]
\begin{enumerate}
  \item If the constraint set $C_c$ contains no negative cycles, then $\solvecomp{C_c} \vDash C_c$ and
  \item $\solve{C} \vDash C$.
\end{enumerate}
\end{theorem}

Before proceeding onto the main soundness theorems, we need a few lemmas ensuring that the positivity/negativity judgements
and algorithmic subtyping are sound and complete with respect to subtyping.

\begin{lemma}[Soundness of positivity/negativity]
Suppose that $\forall \upsilon \in \SV{t}, \rho_1 \upsilon \sqsubseteq \rho_2 \upsilon$.
\begin{enumerate}
  \item If $\gg \vdash \upsilon \pos t$, then $\gg \vdash \rho_1 t \leq \rho_2 t$; and
  \item If $\gg \vdash \upsilon \neg t$, then $\gg \vdash \rho_2 t \leq \rho_1 t$.
\end{enumerate}
\end{lemma}

\begin{lemma}[Completeness of positivity/negativity]~\\[-4ex]
\begin{enumerate}
  \item If $\Gamma_G, \Gamma \vdash t \leq \subst{t}{\upsilon}{\hat{\upsilon}}$ then $\gg \vdash \upsilon \pos t$.
  \item If $\Gamma_G, \Gamma \vdash \subst{t}{\upsilon}{\hat{\upsilon}} \leq t$ then $\gg \vdash \upsilon \neg t$.
\end{enumerate}
\end{lemma}

\begin{lemma}[Soundness of algorithmic subtyping]
Let $\Gamma_G, \Gamma \vdash t \constrain u \rightsquigarrow C$,
and suppose that $\rho \vDash C$.
Then $\Gamma_G, \rho \Gamma \vdash \rho t \leq \rho u$.
\end{lemma}

Now we are ready to tackle the main theorems, in particular soundness of checking, inference, and well-formedness
with respect to the typing rules.
We leave completeness of checking and inference as a conjecture,
but show that if it holds, then completeness of well-formedness will hold as well.

\begin{theorem}[Soundness (check/infer)]
Let $\Sigma$ be a fixed, well-formed signature, $\Gamma_G$ a global environment, $\Gamma$ a local environment, and $C$ a constraint set.
Suppose we have the following:
\begin{enumerate}[label=\alph*)]
  \item $\forall \rho \vDash C, \WF{\Gamma_G, \rho \Gamma}$.
  \item If $\Gamma \equiv \Gamma_1 (\defn{x}{t}{e}) \Gamma_2$ then $\forall \upsilon \in \SV(e, t), \upsilon \notin \SV(\Gamma_G, \Gamma_1)$.
\end{enumerate}
Then the following hold:
\begin{enumerate}
  \item If $C, \gg \vdash e^\circ \Leftarrow t \rightsquigarrow C', e$,
  then $\forall \rho' \vDash C \cup C'$,
  we have $\Gamma_G, \rho\Gamma \vdash \rho e : \rho t$.
  \item If $C, \gg \vdash e^\circ \rightsquigarrow C', e \Rightarrow t$,
  then $\forall \rho \vDash C \cup C'$,
  we have $\Gamma_G, \rho\Gamma \vdash \rho e : \rho t$.
\end{enumerate}
\end{theorem}

\begin{conjecture}[Completeness (check/infer)]
Let $\Sigma$ be a fixed, well-formed signature, $\Gamma_G$ a global environment, $\Gamma$ a local environment, $C$ a constraint set, and $\rho \vDash C$ a solution to the constraint set.
\begin{enumerate}
  \item If $\Gamma_G, \rho\Gamma \vdash e : \rho t$,
    then there exist $C', \rho'$ such that:
    \begin{itemize}
      \item $\rho' \vDash C'$;
      \item $\forall \upsilon \in \SV(\Gamma, t), \rho \upsilon = \rho' \upsilon$; and
      \item $C, \Gamma_G, \Gamma \vdash \erase{e} \Leftarrow t \rightsquigarrow C', e'$ where $\Gamma_G, \Gamma \vdash \rho' e' \conv* e$.
    \end{itemize}
  \item If $\Gamma_G, \rho\Gamma \vdash e : t$,
    then there exist $C, \rho'$ such that:
    \begin{itemize}
      \item $\rho' \vDash C'$;
      \item $\forall \upsilon \in \SV(\Gamma, t), \rho \upsilon = \rho' \upsilon$; and
      \item $C, \Gamma_G, \Gamma \vdash \erase{e} \rightsquigarrow C', e' \Rightarrow t'$ where $\Gamma_G, \Gamma \vdash \rho' e' \conv* e$ and $\Gamma_G, \Gamma \vdash \rho' t \leq t$.
    \end{itemize}
\end{enumerate}
\end{conjecture}

\begin{theorem}[Soundness (well-formedness)]
If $\Gamma_G^\circ \rightsquigarrow \Gamma_G$ then $\WF{\Gamma_G, \mt}$.
\end{theorem}

The completeness theorem for well-formedness is slightly different than expected:
a well-formed global environment, when erased, should successfully have sizes inferred,
but we don't place any requirements on the newly size-inferred environment,
just that there should be \emph{some} answer.

\begin{theorem}[Completeness (well-formedness)]
If $\WF{\Gamma_G, \mt}$ then $\erase{\Gamma_G} \rightsquigarrow \Gamma'_G$.
\end{theorem}


\section{Prototype Implementation and Evaluation} \label{sec:impl}

We have implemented the sized typing algorithm in a beta version of Coq 8.12~\citep{impl},
which can also be found in the supplementary materials.
Naturally, the full core language of Coq has more features (irrelevant for sized typing) than \lang,
and the implementation has to interact with these as well as other proof assistant features such as elaboration and the vernacular.
Furthermore, many details of the sized typing algorithm are left underspecified.
In this section, we take a take a brief look at these details,
verify the time complexity of the implementation against the theoretical complexities from the previous section,
determine its impact on performance as a whole,
and discuss the practical feasibility of the implementation as well as the problems encountered.

\subsection{Architecture of the Coq Kernel}

The core type checking/inference algorithm is found in Coq's \emph{kernel}.
Before reaching the kernel, terms go through a round of \emph{pretyping}
where existential metavariables (essentially typed holes) are solved for
and the recursive indices of fixpoints are determined.
Size inference is implemented as an augmentation of the existing type checking/inference algorithm,
making use of the recursive indices.

\autoref{fig:kernel} summarizes the relevant file/module structure.
Most of the added code specifically for size inference is in the new \texttt{Sized} and \texttt{Subsizing} modules;
the remaining structure remains the same as that of Coq 8.13's codebase~\citep{coq}.
(\texttt{Subsizing} is only separate from \texttt{Sized} to break circular dependencies: it relies on the global environment, while the environment depends on \texttt{Sized}.)

\begin{figure}
\dirtree{%
.1 coq.
.2 lib.
.3 WeightedDigraph\DTcomment{graph data structure and algorithms for constraints}.
.2 pretyping.
.2 kernel.
.3 Constr\DTcomment{core AST and traversals}.
.3 Environ\DTcomment{environments and lookups}.
.3 Reduction\DTcomment{reduction and convertibility}.
.3 Inductive\DTcomment{functions on (co)fixpoints (guard checking)}.
.3 Typeops\DTcomment{entrypoint to type checker for terms}.
.3 Term\_typing\DTcomment{entrypoint to type checker for declarations}.
.3 Sized\DTcomment{constructs and functions for sized types}.
.3 Subsizing\DTcomment{producing subsizing constraints}.
}
\caption{Selected excerpts of the Coq codebase structure}
\label{fig:kernel}
\end{figure}

The \texttt{Sized} module contains several submodules, four of which are relevant to our performance discussion:
\begin{itemize}
  \item \texttt{State} keeps track of the (position) size variables that have been used;
  \item \texttt{SMap} defines the data structure for and operations on size substitutions;
  \item \texttt{Constraints} defines the data structure for and operations on constraint sets; and
  \item \texttt{RecCheck} implements the \RecCheck and \solve algorithms.
\end{itemize}

Sized typing is implemented as a vernacular flag that can be set and unset, just like guard checking.
By default, the flag is off; the commands
$$\coqinline{Set Sized Typing. Unset Guard Checking.}$$
will enable sized typing only.
If both are set, then guard checking will only occur if sized typing fails.
When sized typing is not set, size annotations are still added, but constraints aren't collected,
meaning that global definitions checked in this state will never be marked as size preserving.

\subsection{Analysis of Performance Degradation}

When compiling parts of the Coq standard library with sized typing on, we noticed some severe performance degradation.
This is bad news if we hope to replace guard checking with sized typing,
or even if we simply wish to use sized typing as the primary method of termination or productivity checking throughout.
In particular, we examine compilation of the \texttt{Coq.MSets.MSetList} library%
\footnote{This file can be found in the artifact at \texttt{coq/theories/MSets/MSetList.v}.},
which is an implementation of finite sets using ordered lists
that contains a fair amount of both fixpoints and proof terms
and that happens to compile successfully with sized typing on.
In this file alone, we find a $5.5\times$ increase in compilation time with \texttt{coqc}.
Other files may have even worse degradation; for an earlier version of the algorithm,
there was a $15\times$ increase in compilation time for \texttt{Coq.setoid\_ring.Field\_theory}%
\footnote{This file can be found in the artifact at \texttt{coq/theories/setoid\_ring/Field\_theory.v}.},
which is about twice as large as \msetlist and contains mostly proofs.
We investigate possible causes of the performance degradation and discuss potential solutions.

\subsubsection{Profiling \texttt{Sized} Functions}

To measure the performance degradation, we compare compiling \msetlist against itself with sized typing on and guard checking off, which we refer to as \msetlistsized.
Both compilations are run five times each.
The compilation times are significantly different ($t = 463.94$, $p \ll 0.001$),
with \msetlist's compilation time at $15.122 \pm 0.073$ seconds and \msetlistsized's at $82.660 \pm 0.286$ seconds.

\begin{table}
\centering
\begin{tabular}{| l | r | r | r | r |}
\hline
\textbf{Function(s)} & \textbf{Unsized time (s)} & \textbf{Sized time (s)} & \textbf{$t$} & \textbf{Sized time \%} \\
\hline
\textbf{\solve}               & $ 0.029 \pm 0.002$ & $ 62.397 \pm 0.414$ &  337 &  74.6  \\
\textbf{\RecCheck}            & $ 0.000 \pm 0.000$ & $  2.203 \pm 0.023$ &  219 &   2.63  \\
\textbf{\texttt{Constraints}} & $ 0.186 \pm 0.005$ & $  2.899 \pm 0.028$ &  217 &   3.46 \\
\textbf{\texttt{SMap}}        & $ 0.011 \pm 0.001$ & $  0.281 \pm 0.003$ &  215 &   0.34 \\
\textbf{\texttt{State}}       & $ 0.047 \pm 0.002$ & $  0.104 \pm 0.002$ &   55 &   0.12 \\
\textbf{\texttt{foldmap}}     & $ 0.163 \pm 0.004$ & $  0.266 \pm 0.004$ &   46 &   0.32 \\
\hline
\textbf{Total of above}       & $ 0.436 \pm 0.014$ & $ 68.150 \pm 0.474$ &      &  81.5  \\
\textbf{Total compilation}    & $15.122 \pm 0.073$ & $ 83.660 \pm 0.286$ &      & 100    \\
\hline
\end{tabular}
\caption{Relevant function runtimes when compiling \msetlist vs. \msetlistsized}
\label{table:timing}
\end{table}

To identify the source of the slowdown and test our hypothesis that the majority of it
is intrinsically due to size inference,
we first profile the performance of functions relevant to the \texttt{Sized} module during the compilation.
We divide these functions into five groups: the \solve and \RecCheck functions,
the \texttt{foldmap}%
\footnote{This is the \texttt{foldmap\_annots} function in \texttt{coq/kernel/Constr.ml}.}
function common to all operations manipulating size annotations on the AST (such as applying size substitutions),
the functions in \texttt{State}, the functions in \texttt{SMap}, and the functions in \texttt{Constraints}.
\autoref{table:timing} summarizes the results, as well as the relative time spent in the functions in \msetlistsized.
The differences in execution times of the functions in each group are all statistically significant ($p \ll 0.001$ for all of the $t$-statistics).

$77.2\%$ of the total compilation time in \msetlistsized is taken up by \solve and \RecCheck.
Other \texttt{Sized}-related overhead is smaller, although not insignificant,
especially \texttt{Constraint} operations, which form a proportion slightly larger than that of \RecCheck.
We conjecture that some of this other overhead can be reduced with more clever implementations.
For instance, instead of explicitly performing size substitutions, the sizes can be looked up as needed;
or instead of explicitly passing around a size state, we could use a state monad or a mutable global state;
or constraints could be stored in a data structure incrementally checked for negative cycles,
similar to the current implementation of universe level constraints.
We therefore focus our attention on \solve and \RecCheck, where performance degradation may not be limited to mere implementational details.

\begin{figure}
\begin{subfigure}{0.475\textwidth}
\includegraphics[width=0.9\textwidth]{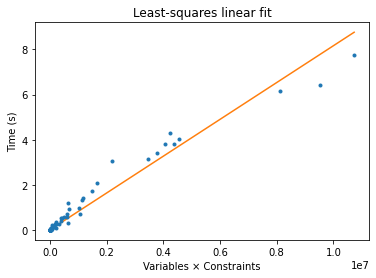}
\caption{\solve execution time vs. $\norm{V}\norm{C}$ \\ (blue dots), linear model (orange line)}
\label{fig:stats:linregress-solve}
\end{subfigure}
\hfill
\begin{subfigure}{0.475\textwidth}
\includegraphics[width=0.9\textwidth]{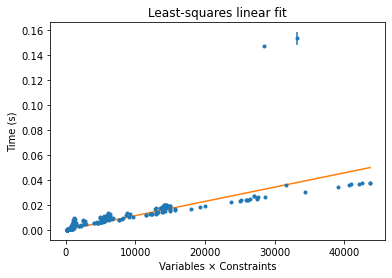}
\caption{\RecCheck execution time vs. $\norm{V}\norm{C}$ \\ (blue dots), linear model (orange line)}
\label{fig:stats:linregress-reccheck}
\end{subfigure}
\\[3ex]
\begin{subfigure}{0.475\textwidth}
\includegraphics[width=0.9\textwidth]{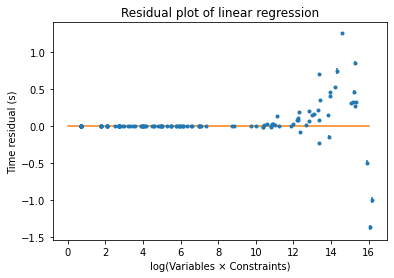}
\caption{\solve model residuals plot (log scale)}
\label{fig:stats:residuals-solve}
\end{subfigure}
\hfill
\begin{subfigure}{0.475\textwidth}
\includegraphics[width=0.9\textwidth]{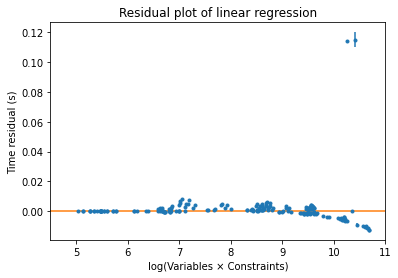}
\caption{\RecCheck model residuals plot (log scale)}
\label{fig:stats:residuals-reccheck}
\end{subfigure}
\\[3ex]
\begin{subfigure}{0.475\textwidth}
\includegraphics[width=0.9\textwidth]{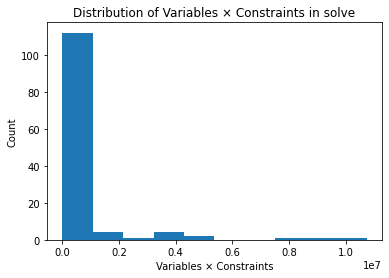}
\caption{$\norm{V}\norm{C}$ distribution in \solve, 10 bins}
\label{fig:stats:distr-solve}
\end{subfigure}
\hfill
\begin{subfigure}{0.475\textwidth}
\includegraphics[width=0.9\textwidth]{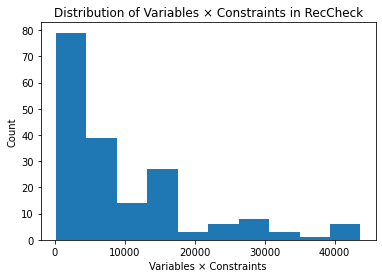}
\caption{$\norm{V}\norm{C}$ distribution in \RecCheck, 10 bins}
\label{fig:stats:distr-reccheck}
\end{subfigure}
\\[2ex]
\caption{Execution vs. $\norm{V}\norm{C}$, residuals, $\norm{V}\norm{C}$ distributions for \solve and \RecCheck}
\label{fig:stats}
\end{figure}

\subsubsection{Time Complexity of \titlesolve and \titleRecCheck}

As shown in \autoref{sec:algorithm}, given a constraint graph from constraint graph $C$ with size variables $V$,
the time complexities of \solve and \RecCheck are $O(\norm{V}\norm{C})$.
Indeed, in \autoref{fig:stats:linregress-solve},
plotting the mean execution times of each of the 155 calls to \solve against $\norm{V}\norm{C}$ for that call (shown as blue dots),
we see a strong positive correlation ($r = 0.983$).
Doing the same for the 186 calls to \RecCheck in \autoref{fig:stats:linregress-reccheck},
we have a weaker positive correlation ($r = 0.698$),
likely due to the two visible outliers.
(Without the outliers, we have $r = 0.831$.)

To verify that the execution times are dominated by linear relationships to $\norm{V}\norm{C}$,
we fit the data to linear models using least-square regressions (shown as orange lines),
and examine the residuals plots in \autoref{fig:stats:residuals-solve} and \autoref{fig:stats:residuals-reccheck}
for \solve and \RecCheck, respectively.
(Note the logarithmic horizontal scale used for clarity, as there are more calls with fewer variables and constraints).

For \solve, the model appears to be a good fit at first,
but residuals increase in magnitude as $\norm{V}\norm{C}$ increases,
indicating some additional behaviour unexplained by the model.
Similarly, for \RecCheck, the model also appears to be a good fit at first,
but then follow a downward curving pattern,
also indicating additional behaviour unexplained by the model (such as the two outliers).

Although there might be additional smaller factors influencing the time complexities of \solve and \RecCheck beyond the number of variables and constraints,
we can at least reasonably conclude that execution time increases with $\norm{V}\norm{C}$.
Since this time complexity is intrinsic to the algorithms and is not due to mere implementational details,
and more than three-fourths of the total compilation time is due to $\solve$ and $\RecCheck$,
the majority of the slowdown is therefore intrinsic and unavoidable.

\subsubsection{An Explosion of Size Variables}

\begin{figure}
\begin{minted}{coq}
Unset Guard Checking.
Set Sized Typing.
Time Definition nats1  := (nat, nat, nat, nat, nat, nat, nat, nat).
Time Definition nats2  := (nats1, nats1, nats1, nats1).
Time Definition nats3  := (nats2, nats2, nats2, nats2).
Time Definition nats4  := (nats3, nats3, nats3, nats3).
Time Definition nats5  := (nats4, nats4, nats4, nats4).
Time Definition nats6  := (nats5, nats5, nats5, nats5).
\end{minted}
\caption{Coq definitions with an explosion in size variables and elapsed time}
\label{fig:nats}
\end{figure}

As $\solve$ and $\RecCheck$ contribute so much to the compilation time
and their complexities depend on $\norm{V}\norm{C}$,
there must be a significant number of calls involving large numbers of variables and constraints.
Indeed, \autoref{fig:stats:distr-solve} and \autoref{fig:stats:distr-reccheck}
show that despite most calls during compilation of \msetlist involving small values of $\norm{V}\norm{C}$,
the number of calls with large values of $\norm{V}\norm{C}$ is not trivial.

Of course, the presence of large numbers of variables and constraints of only \msetlist with sized typing
doesn't tell us whether this is a common property of Coq files on average,
but the fact that this occurs in the standard library indicates that it is absolutely possible
for an explosion in size variables and constraints to be a significant barrier to the adoption of sized typing for general-purpose use.
In fact, this behaviour is easily reproducible with the artificial but comparatively small and simple example in \autoref{fig:nats}.
\autoref{table:nats-stats} lists the number of size variables present in the types and bodies of these definitions,
along with the elapsed type checking time reported by Coq.

\begin{table}
\centering
\begin{tabular}{| l | r | r | r | r |}
\hline
\textbf{Definition} & \textbf{$\norm{V}$ (body)} & \textbf{$\norm{V}$ (type)} & \textbf{Time (s)} \\
\hline
\coqinline{nats1}  &  14    & 7    &  0.004 \\
\coqinline{nats2}  &  62    & 31   &  0.020 \\
\coqinline{nats3}  &  254   & 127  &  0.177 \\
\coqinline{nats4}  &  1022  & 511  &  2.299 \\
\coqinline{nats5}  &  4094  & 2047 & 35.385 \\
\coqinline{nats6}  &  16382 & 8191 & $>120$ \\
\hline
\end{tabular}
\caption{Size variables and time elapsed for definitions in \autoref{fig:nats}}
\label{table:nats-stats}
\end{table}

There are two things we can do to reduce the execution time of \solve and \RecCheck:
eliminate constraints when possible, or reduce the number of size variables in definitions.
One way to accomplish the first option would be to turn on sized typing only for certain definitions,
in particular the ones involving \cofixpoints where they will be most useful.
Then for all other definitions, since no constraints are collected,
calls to \solve will be trivial regardless of how many size variables they contain.

However, a \cofixpoint might require that some non-\corecursive definition with many size variables be size preserving,
which means that that definition also needs to be checked with sized typing on,
or the \cofixpoint itself may have a large number of size variables.
Furthermore, there's no clear indication of which definitions might have a large number of size variables and which don't,
leaving it up to a lot of guesswork and experimentation.
Using sized typing as a targeted tool for particular programs is not viable
if we cannot directly tell \emph{which} particular programs will benefit the most
in terms of tradeoffs between non-structural \corecursion and performance.

The second option to reduce the number of size variables can be done by allowing manual annotation of \coinductive types with the infinite size,
reducing the number of free size variables that need to be substituted for
and that propagate throughout subsequent definitions.
For example, the tuple types in \autoref{fig:nats} can have infinite size annotations
without affecting the sizes of the \coqinline{nat} arguments.
In other words, we allow size annotations in the surface language that users write,
which would no longer be plain CIC;
as this solution deviates from the philosophy of being wholly backward compatible,
it is beyond the scope of this paper.

\subsection{Inferring Recursive Indices}\label{sec:impl:recind}

As previously mentioned, users aren't obligated to indicate which fixpoint argument is the one on which we recur,
and its position index is inferred during pretyping in the kernel.
In Coq, for a mutual fixpoint, this is done by trying the guard predicate on every combination of indices%
\footnote{This is the \texttt{search\_guard} function in \texttt{coq/kernel/Pretyping.ml}.}.
This is possible because the guard predicate is a syntactic check, requiring nothing but the elaborated fixpoint term.

Unfortunately, we run into problems when attempting to apply the same strategy to inferring recursive indices through sized typing alone.
Because termination checking is inextricably tied to type checking,
the entire fixpoint needs to be type checked to verify whether the current set of indices is correct,
and this type checking in the kernel can fail if the fixpoint still contains unsolved metavariables.
Furthermore, because we only have access to the bare environments (\ie with no sizes inferred),
local definitions in scope at the fixpoint may not yet known to be size preserving,
thus causing the check to fail.
As an example, in the following Coq term, \coqinline{id} has no size annotations and is therefore treated as \emph{not} size preserving,
even though it ought to be, which causes the recursive call on the smaller argument wrapped in \coqinline{id} to not pass type checking.
\begin{minted}{coq}
let id (x : nat) := x in
  fix f (n : nat) :=
    match n with
    | O => O
    | S k => f (id k)
    end
\end{minted}
This suggests that size inference should be done during the pretyping phase:
size inference could be viewed as part of the elaboration step from the surface CIC to the core \lang.
This, too, is beyond the scope of this paper, especially as there is no past work on the interaction between size inference and elaboration to build on.

\section{Related Work}\label{sec:related}

The history of sized types is vast and varied.
Extensive prior accounts are given in dissertations by \citet{lambda-hat-diss} and \citet{abel-diss}.
Here, we focus on two lineages towards sized dependent type theories: first, the more-or-less direct ancestry of \lang, and second, a contrasting line of work on type systems with explicit higher-order sized types.

\subsection{Ancestry of \titlelang}

Perhaps the most well-known work on sized types is by \citet{hughes},
who introduce sized types for a Hindley--Milner type system with \coinductives and a size inference algorithm,
as well as the term ``sized types''.
Their size algebra is more expressive than ours, with size addition $s_1 + s_2$ and constant multiplication $n \times s$.
Independently, \citet{ccr} introduces \CCR, a Calculus of Constructions (CoC) with \textit{guarded types},
a type-based termination checking alternative to the earlier syntactic guard condition~\citep{guard-condition}.
There, types are guarded with a type operator $\widehat{\ph}$,
similar to the later modality $\rhd \ph$ in modern guarded type theories.
Based on a semantic interpretation of \CCR,
\citet{acg} introduce a simply-typed lambda calculus (STLC) with coinductives with \textit{type labels},
corresponding roughly to size annotations with successor sizes.

Following this, \citet{lambda-hat} and \citet{lambda-hat-diss} introduce \lambdahat,
another STLC with inductives and size annotations with the same size algebra we use,
although they are instead called \textit{stages}.
It improves upon the work of \citet{acg} by adding an implicit form of size polymorphism:
the position size variable of fixpoint types are substituted by an arbitrary size expression,
just as in \refrule{fix}.
\citet{f-hat} extend \lambdahat to System F with \Fhat, and introduce and prove correct a size inference algorithm.
This includes the \RecCheck algorithm that we use.
They continue on to extend \Fhat with \textit{sized products} (that is, pairs with size annotations) in \FXhat~\citep{fx-hat}, whose size expressions include size addition, and to CIC in \CIChat~\citep{cic-hat}.
Our size inference algorithm is based directly on that of \CIChat.
We add to it global and local definitions and explicitly treat mutually-defined \coinductives and \cofixpoints,
while removing polarities and subtyping based on these polarities.

However, normalization of \CIChat is only a conjecture; it is later proven for the restricted language \CIChatminus by \citet{cic-hat-minus-nat} (with only naturals) and by \citet{cic-hat-minus} (with inductive types).
The restrictions include disallowing size variables in the bodies of functions, in the arguments of applications, in the branches of case expressions, and in the indices of inductives; erasing the parameters to constructors; and disallowing strong elimination to types with size variables.
We remove these restrictions to allow using sized definitions and for backward compatibility with Coq.

Our typing rules and inference algorithm for coinductives and cofixpoints are based on \CChatomega~\citep{cc-hat-omega},
which extends CoC with sized coinductive streams.
Further extensions to the size algebra are linear sized types in \CIChatl~\citep{cic-hat-l},
which adds constant multiplication to a sized CoC with naturals and streams;
and well-founded sized types in \CIChatsub~\citep{wellfounded},
which changes the premise type checking the \cofixpoint body in Rules \refnorule{fix} and \refnorule{cofix}
to the recursive reference having \emph{any} smaller size according to the subsizing rules,
rather than the direct predecessor.
All three include size inference algorithms similar to that of \CIChat.

There are prototype implementations of \CIChatminus~\citep{cicminus} and \CIChatsub~\citep{cic-wf}.
It appears that there were also plans to implement sized types in Coq by \citet{coq-hat}, but seem to be abandoned.

\subsection{Higher-Order Sized Types}

For the purposes of size inferrability from unannotated code,
the type systems from \lambdahat up to \CIChat and its variations treat sizes as merely annotations
and feature only what can be considered as prenex size polymorphism.
On the other hand, \citet{abel-diss} introduces \Fomegahat, a sized type system for System F$_\omega$ that treats size as a kind,
which therefore allows for higher-order size polymorphism via explicit quantification.
While \Fomegahat subsumes \Fhat and uses the same size algebra,
it uses recursive and corecursive type constructors ($\mu$- and $\nu$-types) rather than inductive (and coinductive) type definitions.

Higher-order sized types of the same flavour are implemented in a dependently-typed setting in MiniAgda~\citep{miniagda}.
To avoid inconsistencies introduced by the interplay between sized types and pattern matching,
it also introduces bounded size patterns \mbox{$\upsilon_1 < \upsilon_2$}.
\citet{flationary} expands upon the theoretical side with bounded size quantification \mbox{$\forall \upsilon \mathrel{<} s\mathpunct{.} t$} and well-founded recursion on sizes,
which are also implemented in MiniAgda.
\citet{f-omega-cop} combine well-founded sized types and copatterns for System F$_\omega$ with \corecursive type constructors in \Fomegacop (which was cited as the inspiration for \CIChatsub).

\cite{nbe} prove normalization of a higher-order sized dependent type theory with naturals, but without bounded size quantification.
To our knowledge, this is the only formalization of higher-order sized dependent types in the literature.

Sized types with higher-order bounded size quantification are implemented in Agda%
\footnote{For Agda's documentation on sized types, see \url{https://agda.readthedocs.io/en/latest/language/sized-types.html}.};
however, it is known to be inconsistent%
\footnote{For a detailed discussion on the issue, see \url{https://github.com/agda/agda/issues/2820}.}.
In short, it is possible to express the well-foundedness of sizes within Agda,
but the infinite size $\infty$ itself is \emph{not} well founded,
as $\infty + 1 = \infty$ and $\infty < \infty$ hold,
making it possible to derive a contradiction.


\section{Perspectives and the Future of Sized Types}
\label{sec:conclusion}

We have introduced \lang, a sized type system based on CIC and made to be compatible with Coq,
more than a decade since (prenex, fully-inferrable) sized types for CIC were first introduced in \CIChat and \CIChatminus.
And yet, despite good metatheoretical properties having been established for \CIChatminus,
no functioning attempt at implementing sized types in Coq has previously been made.
This we have done, finding significant performance problems caused by size inference for definitions yielding an explosion in size variables.

This doesn't necessarily spell doom for \lang.
The seasoned type system implementor may employ implementational tricks to improve performance in practice.
Perhaps with some program analysis of how definitions are used, certain size variables known to be irrelevant could immediately be instantiated to the infinite size;
maybe a dependency analysis would reveal which definitions need to be checked with the sized typing flag turned on.
Our na\"ive implementation tries to be as general as possible to accept as many programs as possible,
and heuristics could be used to guess where and why the user wants to use sized types,
whittling down the number of open possibilities for size-inferred programs.

But all of these feel like arbitrary and potentially fragile hacks---and perhaps it's because they \emph{are}.
We have discussed some more sensible solutions to not only the performance problems but also the theoretical ones:
Why don't we explicitly quantify over the size variables of a definition to specify which ones are actually relevant?
Why don't we require that all recursive arguments be marked?
Why not solve the problem of nested inductives using polarities?,
but we immediately shoot them down because they require additional work from the user's perspective and therefore violate the philosophy of backward compatibility.
Perhaps this philosophy maintained for more than a decade of past work from \lambdahat to \CIChatsub is \emph{wrong}.

So far, size inference seems to work for programs because the notion of programs were merely single terms.
Inference was merely extracting hidden information already present in the term.
The moment we introduce a little modularity with definitions,
we don't have concrete information on how these definitions will be used,
and by being as general as possible to accommodate all usages,
we end up with terrible performance.
Inference becomes a guessing game we are losing.

If we make size quantification, abstraction, and application explicit,
then there won't be any more size variables involved than are strictly necessary.
To ease the tedious burden of all the extra annotations from the user,
sizes that can be deduced could be marked as implicit and filled in by the elaborator, as is done for terms.
In contrast to inference, elaboration merely deduces information already available,
and fails as soon as more information is required,
rather than trying to heuristically fill in the gaps itself.
Another benefit of explicit sized types is that it can easily be extended to higher-order size quantification.
This appears to be the best future direction for sized types;
after all, Agda is still to date the only practical proof assistant with sized types.

So is sized typing for Coq practical?
Our answer is that it might be---but only if we allow ourselves to ask users to put in a little work as well.


\section*{Acknowledgements}

We gratefully thank Bruno Barras, Amin Timany, and Andreas Abel for helpful discussions on the metatheory,
in particular on strong normalization,
and Felipe Ba\~nados Schwerter for testing the implementation and finding bugs.

We acknowledge the support of the Natural Sciences and Engineering Research Council of Canada (NSERC), funding reference number RGPIN-2019-04207,
and the Canada Graduate Scholarships -- Master’s (CGS M) programme.
Cette recherche a \'et\'e financ\'ee par le Conseil de recherches en sciences naturelles et en g\'enie du Canada (CRSNG), num\'ero de r\'ef\'erence RGPIN-2019-04207,
et le Programme de bourses d’\'etudes sup\'erieures du Canada au niveau de la maitrise (BESC M).

\paragraph*{Conflicts of Interest.} None.
\bibliographystyle{JFPlike}
\bibliography{main.bib}

\clearpage
\appendix

\section{Well-Formedness of (Co)Inductive Definitions}\label{sec:wf-ind}

In this section we define what it means for a \coinductive definition to be \emph{well-formed}, including some required auxilliary definitions.
A signature is then well formed if each of its \coinductive definitions are well-formed.
Note that although we prove subject reduction for \lang without nested inductive types, we include their definitions for completeness.

\begin{definition}[Strict Positivity]
  Given some existing sigature $\Sigma$, the variable $x$ occurs \emph{strictly positively} in the term $t$, written $x \POS t$, if any of the following holds:

  \begin{itemize}
    \item $x \notin \FV(t)$
    \item $t \conv x ~ \vec{e}$ and $x \notin \FV(\vec{e})$
    \item $t \conv \prod{x}{u}{v}$ and $x \notin \FV(u)$ and $x \POS v$
  \end{itemize}

  If nested inductive types are permitted, then $x \POS t$ may hold if the following also holds:

  \begin{itemize}
    \item $t \conv I_k^\infty ~ \vec{p} ~ \vec{a}$ where
      $\defn{\langle \app{I_i}{\Delta_p}}{\any \rangle}{\seq{\assm{c_j}{\prodctx{\Delta_j}{\app{I_j}{\dom{\Delta_p}}{\vec{t}_j}}}}} \in \Sigma$
      for some $k \in \vec{\imath}$ and all of the following hold:
      \begin{itemize}
        \item $\norm{\vec{p}} = \norm{\Delta_p}$
        \item $x \notin \FV(\vec{a})$
        \item For every $j$, if $I_j = I_k$, then $x \nPOS_{I_k} \subst{(\prodctx{\Delta_j}{\app{I_j}{\vec{p}}{\vec{t}_j}})}{\dom{\Delta_p}}{\vec{p}}$
      \end{itemize}
  \end{itemize}
\end{definition}

\begin{definition}[Nested Positivity]
  Given some existing signature $\Sigma$, the variable $x$ is \emph{nested positive} in $t$ of $I_k$, written $x \nPOS_{I_k} t$, if
  $\defn{\langle \app{I_i}{\Delta_p}}{\any \rangle}{\any} \in \Sigma$
  for some $k \in \vec{\imath}$ and any of the following holds:

  \begin{itemize}
    \item $t \conv \app{I_k^\infty}{\vec{p}}{\vec{a}}$ and $\norm{\vec{p}} = \norm{\Delta_p}$ and $x \notin \FV(\vec{a})$
    \item $t \conv \prod{x}{u}{v}$ and $x \POS u$ and $x \nPOS_{I_k} v$
  \end{itemize}

  In short, $x \nPOS_I t$ if $t \conv \prodctx{\Delta}{\app{I}{\vec{p}}{\vec{a}}}$ and $x \POS \Delta$ and $x \notin \FV(\vec{a})$.
\end{definition}

\begin{definition}[Constructor Type]
  The term $t$ is a constructor type for $I$ when:

  \begin{itemize}
    \item $t = I ~ \vec{e}$; or
    \item $t = \prod{x}{u}{v}$ where $x \notin \FV{u}$ and $v$ is a constructor type for $I$; or
    \item $t = u \to v$ where $x \POS u$ and $v$ is a constructor type for $I$.
  \end{itemize}

  Note that in particular, this means that $t = \prodctx{\Delta}{I ~ \vec{e}}$ such that $I \POS u$
  for every $u \in \codom{\Delta}$, and the recursive arguments of $t$ are not dependent.
\end{definition}

\begin{definition}[Well-formedness of \Coinductive Definitions]
  Suppose we have some signature $\Sigma$ and some global environment $\Gamma_G$. Consider the following \coinductive definition, where $\vec{p} = \dom{\Delta_p}$.
  \begin{displaymath}
    \defn{\langle \app{I_i}{\Delta_p}}{\prodctx{\Delta_i}{U_i} \rangle}{\seq{\assm{c_j}{\prodctx{\Delta_j}{\app{I_j}{\vec{p}}{\vec{t}_j}}}}}
  \end{displaymath}
  This \coinductive definition is \emph{well-formed} if the following all hold:

  \begin{enumerate}[label = \textbf{(I\arabic*)}.]
    \item For every $i$, there is some $U'_i$ such that $\Sigma, \Gamma_G, \Delta_p \vdash \prodctx{\Delta_i}{U_i} : U'_i$ holds.
    \item For every $j$, there is some $U_j$ such that $\Sigma, \Gamma_G, \Delta_p(I_j^\infty: \prodctx{\Delta_p}{\prodctx{\Delta_i}{U_i}}) \vdash \prodctx{\Delta_j}{I_j^\infty ~ \vec{p} ~ \vec{t}_j} : U_j$ holds.
    \item For every $j$, $\prodctx{\Delta_j}{\app{I_j}{\vec{p}}{\vec{t}_j}}$ is a constructor type for $I_j$. Note that this implies $I_j \POS \codom{\Delta_j}$.
    \item For every $i, j$, all \coinductive types in the terms $\codom{\Delta_p}, \codom{\Delta_i}, \codom{\Delta_j}$ are annotated with $\infty$.
  \end{enumerate}
\end{definition}


\section{Inference Soundness and Completeness Proofs}\label{sec:proofs}

Here we provide some more detailed proof sketches for the various soundness and completeness theorems
found in \autoref{sec:algorithm:metatheory}.
Further details when not specified can be found in \citet{f-hat}, \citet{cic-hat}, and \citet{cc-hat-omega}.

\begin{theorem}[Soundness of \RecCheck (SRC)]\label{thm:src}
If $\RecCheck(C', \tau, V^*, V^\neq) = C$, for every $\rho$ such that $\rho \vDash C$,
given a fresh size variable $\upsilon$, there exists a $\rho'$ such that the following all hold:
\begin{enumerate}
  \item $\rho' \vDash C'$;
  \item $\rho'\tau = \upsilon$;
  \item $\floor{\rho' V^*} = \upsilon$;
  \item $\floor{\rho' V^\neq} \neq \upsilon$;
  \item For all $\upsilon' \in V^\neq$, $(\set{\upsilon \mapsto \rho\tau} \circ \rho')(\upsilon') = \rho\upsilon'$; and
  \item For all $\tau' \in V^*$, $(\set{\upsilon \mapsto \rho\tau} \circ \rho')(\tau') \sqsubseteq \rho\tau'$.
\end{enumerate}
\end{theorem}

\begin{proof}[{[Partial]}]
Let $C^\iota$ be $C$ with all vertices in $\bigsqcup \set{\infty}$ removed.
By the definition of \RecCheck, since all negative cycles in $C'$ are removed and the only constraints that are added are of the form $\infty \sqsubseteq s$,
$C^\iota$ has no negative cycles either.
Let $V^\iota = \bigsqcap V^*$.
Note that the constraints $\tau \sqsubseteq V^\iota$ are in $C^\iota$.
Then we are able to compute the weights $w_i$ of the shortest paths from $\tau$ to $\bigsqcup V^\iota$ with respect to $C^\iota$.
According to \citet{f-hat}, these weights are nonnegative.
Then we can define $\rho' \coloneqq \rho \circ \set{\upsilon_i \mapsto \hat{\upsilon}^{w_i} \mid \upsilon_i \in \bigsqcup V^\iota, \rho\upsilon_i \neq \infty}$.

\begin{enumerate}
  \item The proof is more involved; see \citet{f-hat}.
  \item The shortest path from $\tau$ to itself is no path at all, so $\rho'\tau = \upsilon$.
  \item Since $V^* \subseteq V^\iota \subseteq \bigsqcup V^\iota$,
    for every $\upsilon_i \in V^*$, $\rho'\upsilon_i = \hat{\upsilon}^{w_i}$ where $w_i$ is the weight of the shortest path from $\upsilon$ to $\upsilon_i$,
    and its size variable is obviously $\upsilon$.
  \item Let $\upsilon' \in V^\neq$.
    If $\upsilon' \in \bigsqcup V^\iota$, then $\infty \sqsubseteq \upsilon'$ must be in $C$, and therefore $\rho\upsilon' = \infty$, so $\rho'\upsilon' = \rho\upsilon'$.
    Otherwise, if $\upsilon' \notin \bigsqcup V^\iota$, we again have $\rho'\upsilon' = \rho\upsilon'$.
    Since $\upsilon$ is fresh, it could not be mapped to by $\rho$, so the size variable of $\rho\upsilon'$ cannot be $\upsilon$.
  \item Let $\upsilon' \in V^\neq$.
    If $\upsilon' \in \bigsqcup V^\iota$,
    then we must have the constraint $\infty \sqsubseteq \upsilon'$ in $C$, so $\rho\upsilon' = \infty$.
    Therefore, $(\set{\upsilon \mapsto \rho\tau} \circ \rho')\upsilon' = (\set{\upsilon \mapsto \rho\tau} \circ \rho)\upsilon' = \rho\upsilon'$.
  \item Let $\tau' \in V^*$. Note that $V^* \subseteq V^\iota \subseteq \bigsqcup V^\iota$.
    Then letting $w'$ be the weight of the shortest path from $\upsilon$ to $\tau'$, we have $\rho'\tau' = \hat{\upsilon}^{w'}$,
    and $(\set{\upsilon \mapsto \rho\tau} \circ \rho')\tau' = \succ{\rho\tau}^{w'}\!$.
    Since $\rho \vDash C$ and there is a path of weight $w'$ from $\tau$ to $\tau'$ in $C$,
    we have $\succ{\rho\tau}^{w'}\! \sqsubseteq \rho\tau'$. \qed
\end{enumerate}
\end{proof}

\begin{theorem}[Completeness of \RecCheck (CRC)]~\\
Suppose the following all hold:
\begin{itemize}
  \item $\rho \vDash C'$;
  \item $\rho\tau = \upsilon$;
  \item $\floor{\rho V^*} = \upsilon$; and
  \item $\floor{\rho V^\neq} \neq \upsilon$.
\end{itemize}
Then $\rho \vDash \RecCheck(C', \tau, V^*, V^\neq)$.
\end{theorem}

\begin{proof}
Let $C = \RecCheck(C', \tau, V^*, V^\neq)$.
To show that $\rho \vDash C$, we need to show that for every constraint $s_1 \sqsubseteq s_2$ in $C$,
$\rho s_1 \sqsubseteq \rho s_2$ holds.
Since $\rho \vDash C'$, this means we need to show that $\rho$ satisfies every constraint added to $C'$ in \RecCheck.
We handle them step by step.
Let $V^\iota \coloneqq \bigsqcap V^*$, and let $V^-$ be the set of size variables involved in some negative cycle in $C'$.
\begin{itemize}
  \item \autoref{item:reccheck:smallest}: $\tau \sqsubseteq V^\iota$. Since we have $\rho \tau = \upsilon$ and $\rho V^* = \hat{\upsilon}^{n}$ for some $n$ by assumption,
  $\rho \tau \sqsubseteq \rho V^\iota$ holds.
  \item \autoref{item:reccheck:neg-cycles}: $\infty \sqsubseteq V^-$. For all size variables $\upsilon' \in V^-$,
  since being in a negative cycle transitively implies a subsizing relation $\hat{\upsilon}'^{n} \sqsubseteq \upsilon'$ for some $n$,
  the only way for $\rho \hat{\upsilon}'^{n} \sqsubseteq \rho \upsilon'$ to hold is if $\rho \upsilon' = \infty$,
  which satisfies $\infty \sqsubseteq \rho \upsilon'$.
  \item \autoref{item:reccheck:infty}: $\infty \sqsubseteq (\bigsqcup V^\neq \cap \bigsqcup V^\iota)$. Since $\rho V^\neq$ and $\rho V^\iota$ have different size variables by assumption,
  if a size variable $\upsilon'$ is in both $\bigsqcup V^\neq$ and $\bigsqcup V^\iota$,
  it must be set to $\infty$, which satisfies $\infty \sqsubseteq \upsilon'$. \qed
\end{itemize}
\end{proof}

\begin{theorem}[Correctness of RecCheckLoop]\label{thm:reccheckloop}~\\[-4ex]
\begin{enumerate}
  \item \RecCheckLoop terminates on all inputs.
  \item If $\RecCheckLoop{C', \Gamma, \vec{\tau_k}, \vec{t_k}, \vec{e_k}} = C$ with an initial position variable set $\mathcal{V}^*$,
  then for every $i \in \vec{k}$, $\RecCheck{C', \tau_i, \PV(t_i), \SV(\Gamma, t_i, e_i) \setminus \PV(t_i)} \subseteq C$ with a final position variable set $\mathcal{V}^*_\subseteq \subseteq \mathcal{V}^*$.
\end{enumerate}
\end{theorem}

\begin{proof}[{[Sketch]}]
\begin{enumerate}
  \item \RecCheckLoop does a recursive call only when \RecCheck fails with a size variable set $V$, which by definition is a subset of $\PV(t_i)$ for some $t_i$.
  Since $V$ is removed from $\mathcal{V}^*$ every time, $\PV(\vec{t_k})$ is the decreasing measure of \RecCheckLoop.
  \item Again, $\mathcal{V}^*$ is only removed from, not added to, so the final set must be a subset of the initial set.
  By inspection, $C$ is a union of the constraints returned by \RecCheck when they all succeed. \qed
\end{enumerate}
\end{proof}

\begin{theorem}[Correctness of \solve and \solvecomp]\label{thm:solve}~\\[-4ex]
\begin{enumerate}
  \item If the constraint set $C_c$ contains no negative cycles, then $\solvecomp{C_c} \vDash C_c$ and
  \item $\solve{C} \vDash C$.
\end{enumerate}
\end{theorem}

\begin{proof}[{[Sketch]}]
\begin{enumerate}
  \item By \citet{clrs}, any constant shift ($w_{\max}$, in our case) of a shortest-path solution is a valid solution to the difference constraint problem.
  \item By the same reasoning for \RecCheck, any variables involved in negative cycles must be set to $\infty$ in a solution.
  Remaining constraints are solved by \solvecomp. \qed
\end{enumerate}
\end{proof}

Before proceeding, we need a few lemmas ensuring that the positivity/negativity judgements
and algorithmic subtyping are sound and complete with respect to subtyping.

\begin{lemma}[Soundness of positivity/negativity]\label{lem:posneg-subtyping}
Suppose that $\forall \upsilon \in \SV{t}, \rho_1 \upsilon \sqsubseteq \rho_2 \upsilon$.
\begin{enumerate}
  \item If $\gg \vdash \upsilon \pos t$, then $\gg \vdash \rho_1 t \leq \rho_2 t$; and
  \item If $\gg \vdash \upsilon \neg t$, then $\gg \vdash \rho_2 t \leq \rho_1 t$.
\end{enumerate}
\end{lemma}

\begin{proof}[{[Sketch]}]
By mutual induction on the positivity and negativity rules in \autoref{fig:posneg}. \qed
\end{proof}

\begin{lemma}[Completeness of positivity/negativity]\label{lem:subtyping-posneg}~\\[-4ex]
\begin{enumerate}
  \item If $\Gamma_G, \Gamma \vdash t \leq \subst{t}{\upsilon}{\hat{\upsilon}}$ then $\gg \vdash \upsilon \pos t$.
  \item If $\Gamma_G, \Gamma \vdash \subst{t}{\upsilon}{\hat{\upsilon}} \leq t$ then $\gg \vdash \upsilon \neg t$.
\end{enumerate}
\end{lemma}

\begin{proof}[{[Sketch]}]
By induction on the subtyping rules in \autoref{fig:subtyping}. \qed
\end{proof}

\begin{lemma}[Soundness of algorithmic subtyping]\label{lem:subtyping}
Let $\Gamma_G, \Gamma \vdash t \constrain u \rightsquigarrow C$,
and suppose that $\rho \vDash C$.
Then $\Gamma_G, \rho \Gamma \vdash \rho t \leq \rho u$.
\end{lemma}

\begin{proof}[{[Sketch]}]
By induction on the algorithmic subtyping rules in \autoref{fig:algorithm-subtyping}. \qed
\end{proof}

The following lemma and corollary asserting the absence of certain size variables
will later let us commute some substitutions.

\begin{lemma}\label{lem:fresh-vars}~\\[-4ex]
\begin{enumerate}
  \item If $\Gamma_G, \Gamma \vdash e^\circ \Leftarrow t \rightsquigarrow C, e$, then $\forall \upsilon \in \SV{e}, \upsilon \notin \SV{\Gamma_G, \Gamma}$.
  \item If $\Gamma_G, \Gamma \vdash e^\circ \rightsquigarrow C, e \Rightarrow t$, then $\forall \upsilon \in \SV{e}, \upsilon \notin \SV{\Gamma_G, \Gamma}$.
\end{enumerate}
\end{lemma}

\begin{proof}[{[Sketch]}]
By mutual induction on the checking and inference rules of the algorithm.
For checking, it follows by the induction hypothesis on the inference premise.
For inference, most cases are straightforward applications of the induction hypothesis;
new size annotations are only introduced in $e$ in Rules \refnorule{a-ind} and \refnorule{a-ind-star},
which introduce fresh size variables that are by definition not in $\SV{\Gamma_G, \Gamma}$. \qed
\end{proof}

\begin{corollary}\label{lem:global-fresh-vars}
  \item If $\Gamma^\circ_G D^\circ \rightsquigarrow \Gamma_G D$ for bare and sized declarations $D^\circ, D$, then $\forall \upsilon \in \SV(D), \upsilon \notin \Gamma_G$.
\end{corollary}

Finally, we can proceed with the main theorems.

\begin{theorem}[Soundness (check/infer)]\label{thm:soundness}
Let $\Sigma$ be a fixed, well-formed signature, $\Gamma_G$ a global environment, $\Gamma$ a local environment, and $C$ a constraint set.
Suppose we have the following:
\begin{enumerate}[label=\alph*)]
  \item \label{item:soundness:wf} $\forall \rho \vDash C, \WF{\Gamma_G, \rho \Gamma}$.
  \item \label{item:soundness:sv} If $\Gamma \equiv \Gamma_1 (\defn{x}{t}{e}) \Gamma_2$ then $\forall \upsilon \in \SV(e, t), \upsilon \notin \SV(\Gamma_G, \Gamma_1)$.
\end{enumerate}
Then the following hold:
\begin{enumerate}
  \item If $C, \gg \vdash e^\circ \Leftarrow t \rightsquigarrow C', e$,
  then $\forall \rho' \vDash C \cup C'$,
  we have $\Gamma_G, \rho\Gamma \vdash \rho e : \rho t$.
  \item If $C, \gg \vdash e^\circ \rightsquigarrow C', e \Rightarrow t$,
  then $\forall \rho \vDash C \cup C'$,
  we have $\Gamma_G, \rho\Gamma \vdash \rho e : \rho t$.
\end{enumerate}
\end{theorem}

\begin{proof}[{[Partial]}]
By mutual induction on the checking and inference rules of the algorithm.
Suppose \ref{item:soundness:wf} and \ref{item:soundness:sv} hold.
\begin{enumerate}
  \item By \refrule{a-check}, we have
  \begin{displaymath}
    \inferrule[]{
      C, \gg \vdash e^\circ \rightsquigarrow C_1, e \Rightarrow t \\
      \Gamma_G, \Gamma \vdash t \constrain u \rightsquigarrow C_2
    }{
      C, \gg \vdash e^\circ \Leftarrow u \rightsquigarrow C_1 \cup C_2, e
    }
  \end{displaymath}
  Let $\rho \vDash C \cup C_1 \cup C_2$.
  By the induction hypotheses on the premise, we have $C, \Gamma_G, \rho \Gamma \vdash \rho e : \rho t$.
  By \autoref{lem:subtyping}, we have $\Gamma_G, \rho \Gamma \vdash \rho t \leq \rho u$.
  Then by \refrule{conv}, we have $\Gamma_G, \rho \Gamma \vdash \rho e : \rho u$.
  \item We will prove the cases for definitions, let expressions, case expressions, and fixpoints;
  the case for cofixpoints is similar to that of fixpoints, and the remaining cases are straightforward.
  \begin{itemize}
    \item \refrule{a-var-def}.
    \begin{displaymath}
      \inferrule[]{
        (\defn{x}{t}{e}) \in \Gamma \and
        \overline{\upsilon'_i} = \SV(e, t) \setminus \SV(C) \and
        \overline{\upsilon_i} = \fresh{\norm{\overline{\upsilon'_i}}} \and
        \rho = \set{\vec{\upsilon'_i \mapsto \upsilon_i}}
      }{
        C, \gg \vdash x \rightsquigarrow \set{}, x^\rho \Rightarrow \rho t
      }
    \end{displaymath}
    Let $\rho' \vDash C$.
    We must show that $\Gamma_G, \rho' \Gamma \vdash \rho' x^\rho : \rho' (\rho t)$ holds.
    By well-formedness of $\rho' \Gamma$, we have that $\Gamma_G, \rho' \Gamma_1 \vdash \rho' e : \rho' t$,
    where $\Gamma \equiv \Gamma_1 (\defn{x}{t}{e}) \Gamma_2$.
    Since $\rho$ only does a size variable renaming, we also have $\Gamma_G, \rho (\rho' \Gamma_1) \vdash \rho (\rho' e) : \rho (\rho' t)$.
    Furthermore, since the size variables in $\rho$ and $\rho'$ are fresh,
    and $\rho$ only affects size variables in $\SV(e, t) \setminus \SV(C)$
    while $\rho'$ only affects size variables in $\SV(C)$,
    the two substitutions commute, giving us
    $\Gamma_G, \rho' (\rho \Gamma_1) \vdash \rho' (\rho e) : \rho' (\rho t)$.
    Finally, since $\vec{\upsilon'_i} \notin \Gamma_1$, the substitution $\rho$ on $\Gamma_1$ has no effect, yielding
    $\Gamma_G, \rho' \Gamma_1 \vdash \rho' (\rho e) : \rho' (\rho t)$.
    Then we can use \refrule{var-def} to obtain our goal.
    \item \refrule{a-const-def}.
    Similar to \refrule{a-var-def}, but using \autoref{lem:global-fresh-vars} instead of \ref{item:soundness:sv}.
    \item \refrule{a-let-in}.
    \begin{displaymath}
      \inferrule[]{
        C, \gg \vdash t^\circ \rightsquigarrow C_1, t \Rightarrow^* U \\
        C, \gg \vdash e_1^\circ \Leftarrow t \rightsquigarrow C_2, e_1 \\
        C \cup C_1 \cup C_2, \gg (\defn{x}{t}{e_1}) \vdash e_2^\circ \rightsquigarrow C_3, e_2 \Rightarrow u
      }{
        C, \gg \vdash \letin{x}{t^\circ}{e_1^\circ}{e_2^\circ} \rightsquigarrow C_1 \cup C_2 \cup C_3, \letin{x}{|t|}{e_1}{e_2} \Rightarrow u[x \coloneqq e_1]
      }
    \end{displaymath}
    Let $\rho \vDash C \cup C_1 \cup C_2 \cup C_3$.
    We must show that $\Gamma_G, \Gamma \vdash \rho (\letin{x}{|t|}{e_1}{e_2}) : \rho (u[x \coloneqq e_1])$.
    The induction hypotheses on the first two premises tell us the following:
    \begin{itemize}
      \item $\forall \rho_1 \vDash C \cup C_1$,
      $\Gamma_G, \rho_1 \Gamma \vdash \rho_1 t : U$; and
      \item $\forall \rho_2 \vDash C \cup C_2$
      $\Gamma_G, \rho_2 \Gamma \vdash \rho_2 e_1 : \rho_2 t$.

    \end{itemize}
    To obtain the third induction hypothesis, we need to first show that
    $\forall \rho' \vDash C \cup C_1 \cup C_2, \WF{\Gamma_G, \rho' (\Gamma (\defn{x}{t}{e_1}))}$ holds.
    Letting $\rho' \vDash C \cup C_1 \cup C_2$, by \ref{item:soundness:wf}, we have that $\WF{\Gamma_G, \rho' \Gamma}$.
    Applying $\rho'$ to the second induction hypothesis, we have that $\Gamma_G, \rho' \Gamma \vdash \rho' e_1 : \rho t$.
    Then using \refrule{wf-local-def}, we have $\WF{\Gamma_G, \rho' \Gamma (\defn{x}{\rho' t}{\rho' e_1})}$ as desired.
    Furthermore, by \autoref{lem:fresh-vars}, we know that $\forall \upsilon \in \SV{e, t}, \upsilon \notin \SV{\Gamma}$
    Finally, we have the third induction hypothesis:
    \begin{itemize}
      \item $\forall \rho_3 \vDash C \cup C_1 \cup C_2 \cup C_3$,
      $\Gamma_G, \rho_3 \Gamma (\defn{x}{t}{e_1}) \vdash \rho_3 e_2 : \rho_3 u$.
    \end{itemize}
    Applying $\rho$ to all three induction hypotheses and using \refrule{let} yields our goal.
    \item \refrule{a-case}.
      \begin{displaymath}
      \inferrule[]{
        C, \gg \vdash e^\circ \rightsquigarrow C_1, e \Rightarrow^* \app{I_k^s}{\vec{p}}{\vec{a}} \\
        C, \gg \vdash P^\circ \rightsquigarrow C_2, P \Rightarrow t_p \\
        \prodctx{\any}{\prodctx{\Delta_k}{U_k}} = \indtype{I_k} \\
        U = \decompose{t_p}{\norm{\Delta_k} + 1} \\
        \elim{U_k}{U}{I_k} \\
        \upsilon = \fresh{1} \\
        \Gamma_G, \Gamma \vdash t_p \constrain \motivetype{\overline{p}}{U}{I_k^{\hat{\upsilon}}} \rightsquigarrow C_3 \\
        \textrm{For each $j$:} \\
        C, \gg \vdash e^\circ_j \Leftarrow \branchtype{\overline{p}}{c_j}{\upsilon}{P} \rightsquigarrow C_{4j}, e_j \\
        C_5 = \casesize{I_k^s}{\hat{\upsilon}} \cup C_1 \cup C_2 \cup C_3 \cup (\textstyle\bigcup_j C_{4j})
      }{
        C, \gg \vdash \caseof{P^\circ}{e^\circ}{c_j}{e_j^\circ} \rightsquigarrow C_5, \caseof{|P|}{e}{c_j}{e_j} \Rightarrow \app{P}{\vec{a}}{e}
      }
      \end{displaymath}
      Let $\rho \vDash C \cup C_5$.
      We must show that $\Gamma_G, \rho \Gamma \vdash \rho (\caseof{|P|}{e}{c_j}{e_j}) : \rho (\app{P}{\vec{a}}{e})$.
      The induction hypotheses and \autoref{lem:subtyping} tell us the following:
      \begin{itemize}
        \item $\forall \rho_1 \vDash C \cup C_1, \Gamma_G, \rho_1 \Gamma \vdash \rho_1 e : \rho_1 (\app{I^s_k}{\vec{p}}{\vec{a}})$;
        \item $\forall \rho_2 \vDash C \cup C_2, \Gamma_G, \rho_2 \Gamma \vdash \rho_2 P : \rho_2 t_p$;
        \item $\forall \rho_3 \vDash C_3, \Gamma_G, \rho_3 \Gamma \vdash \rho_3 t_p \leq \rho_3 (\motivetype{\vec{p}}{U}{I_k^{\hat{\upsilon}}})$; and
        \item $\forall \rho_{4j} \vDash C \cup C_{4j}, \Gamma_G, \rho_{4j} \Gamma \vdash \rho_{4j} e_j : \rho_{4j} (\branchtype{\vec{p}}{c_j}{\upsilon}{P})$.
      \end{itemize}
      We can apply $\rho$ to all four of these.
      By \refrule{conv}, we have that $\Gamma_G, \rho \Gamma \vdash \rho P : \rho (\motivetype{\vec{p}}{U}{I_k^{\hat{\upsilon}}})$.
      Because $\rho \vDash \casesize{I_k^s, \hat{\upsilon}}$,
      $\rho s \sqsubseteq \rho \hat{\upsilon}$ if $I_k$ is inductive and
      $\rho \hat{\upsilon} \sqsubseteq s$ if $I_k$ is coinductive.
      Then by Rules \refnorule{st-ind} or \refnorule{st-coind} respectively,
      we have $\Gamma_G, \rho \Gamma \vdash \rho I_k^s \leq \rho I_k^{\hat{\upsilon}}$,
      and by \refrule{conv}, we have $\Gamma_G, \rho \Gamma \vdash \rho e : \rho (\app{I_k^{\hat{\upsilon}}}{\vec{p}}{\vec{a}})$.
      Finally, using \refrule{case}, we have our goal.
    \item \refrule{a-fix}.
      \begin{displaymath}
        \inferrule[]{
          \textrm{For each $k$:} \\
          C, \gg \vdash t_k^\circ \rightsquigarrow \any, \any \Rightarrow \any \\
          C, \gg \vdash \setrecstars{t_k^\circ}{n_k} \rightsquigarrow C_{1k}, t_k \Rightarrow^* U \\
          \prodctx{\Delta_k}{u_k} = \whnf{t_k} \and \prodctx{\Delta_k}{u'_k} = \shift{\prodctx{\Delta_k}{u_k}} \\
          \textstyle\bigcup_k C_{1k} \cup C, \gg \overline{(f_k : t_k)} \vdash e_k^\circ \Leftarrow \prodctx{\Delta_k}{u'_k} \rightsquigarrow C_{2k}, e_k \\
          \gg \Delta_k \vdash u_k \constrain u'_k \rightsquigarrow C_{3k} \\
          C_4 = \textstyle\bigcup_k C_{1k} \cup C_{2k} \cup C_{3k} \cup C \\
          C_5 = \RecCheckLoop{C_4}{\overline{\getrecvar{t_k}{n_k}}}{\overline{t_k}}{\overline{e_k}}
        }{
          C, \gg \vdash \fix{m}{f_k^{n_k}}{t_k^\circ}{e_k^\circ} \rightsquigarrow C_5, \fix{m}{f_k^{n_k}}{|t_k|^*}{e_k} \Rightarrow t_m
        }
      \end{displaymath}
      Let $\rho \vDash C \cup C_5$.
      We must show that $\Gamma_G, \rho \Gamma \vdash \rho (\fix{m}{f_k^{n_k}}{|t_k|^*}{e_k}) : \rho t_m$.
      The induction hypotheses and \autoref{lem:subtyping} tell us the following:
      \begin{itemize}
        \item $\forall \rho_{1k} \vDash C \cup C_{1k}, \Gamma_G, \rho_{1k} \Gamma \vdash \rho_{1k} t_k : U$;
        \item $\forall \rho_{2k} \vDash C \cup (\textstyle\bigcup_k C_{1k}) \cup C_{2k}, \Gamma_G, \rho_{2k} (\Gamma \vec{(f_k : t_k)}) \vdash \rho_{2k} e_k : \rho_{2k} (\prodctx{\Delta_k}{u'_k})$;
        \item $\forall \rho_{3k} \vDash C_{3k}, \Gamma_G, \rho_{3k} (\Gamma \Delta_k) \vdash \rho_{3k} u_k \leq \rho_{3k} u'_k$.
      \end{itemize}
      By \autoref{thm:reccheckloop}, from $\rho \vDash C_5$,
      we also have that for every $i \in \vec{k}$,
      $\rho \vDash \RecCheck{C_4, \tau_i, \PV{t_i}, \SV{\Gamma, t_i, e_i} \setminus \PV{t_i}}$,
      where $\tau_i = \getrecvar{t_i, n_i}$.
      Then applying \autoref{thm:src}, letting $\upsilon_i$ be a fresh size variable, there exists a $\rho'$ such that the following hold:
      \begin{enumerate}
        \item \label{item:soundness:fix:1} $\rho' \vDash C_4$;
        \item \label{item:soundness:fix:2} $\rho' \tau_i = \upsilon_i$
        \item \label{item:soundness:fix:3} $\floor{\rho' \PV{t_i}} = \upsilon_i$
        \item \label{item:soundness:fix:4} $\floor{\rho' (\SV{\Gamma, t_i, e_i} \setminus \PV{t_i})} \neq \upsilon_i$;
        \item \label{item:soundness:fix:5} $\forall \upsilon' \in \SV{\Gamma, t_i, e_i} \setminus \PV{t_i}, (\set{\upsilon_i \mapsto \rho \tau_i} \circ \rho') \upsilon' = \rho \upsilon'$; and
        \item \label{item:soundness:fix:6} $\forall \tau' \in \PV{t_i}, (\set{\upsilon_i \mapsto \rho \tau_i} \circ \rho') \tau' \sqsubseteq \rho \tau'$.
      \end{enumerate}
      By \ref{item:soundness:fix:4} and \ref{item:soundness:fix:5} together, we can conclude that
      $\forall \upsilon' \in \SV{\Gamma, t_i, e_i} \setminus \PV{t_i}, \rho' \upsilon' = \rho \upsilon'$, so $\rho' \Gamma = \rho \Gamma$ and $\rho' e_k = \rho e_k$.
      Then by \ref{item:soundness:fix:1}, we can apply $\rho'$ to each the inductive hypotheses and simplify to yield:
      \begin{itemize}
        \item $\Gamma_G, \rho \Gamma \vdash \rho' t_k : U$;
        \item $\Gamma_G, (\rho \Gamma)\vec{(f_k : \rho' t_k)} \vdash \rho e_k : \rho' (\prodctx{\Delta_k}{u'_k})$; and
        \item $\Gamma_G, (\rho \Gamma)(\rho'\Delta_k) \vdash \rho' u_k \leq \rho' u'_k$.
      \end{itemize}
      Notice that \shift only shifts position variables up by one, which means that by \ref{item:soundness:fix:2}, $\rho'u'_k = \set{\upsilon_i \mapsto \hat{\upsilon}_i}(\rho'u_k)$.
      Then by \autoref{lem:subtyping-posneg}, the last subtyping judgement implies that $\upsilon_k$ is positive in $\rho' u_k$.
      At last, we are able to apply \refrule{fix}, picking $s = \rho \tau_m$:
      \begin{equation}\label{eqn:soundness:fix}
        \Gamma_G, \rho \Gamma \vdash \fix{m}{f_k^{n_k}}{|\rho' t_k|^{\upsilon_k}}{\rho e_k} : \subst{(\rho' t_m)}{\upsilon_m}{\rho \tau_m}
      \end{equation}
      By \ref{item:soundness:fix:3} and \ref{item:soundness:fix:4}, we have $\erase{\rho' t_i}^{\upsilon_i} = \erase{t_i}^*$,
      as all position variables in $t_i$ are mapped to $\upsilon_i$ by $\rho'$.
      Finally, by \ref{item:soundness:fix:5}, $\set{\upsilon_m \mapsto \rho \tau_m} \circ \rho' = \rho$ when applied to non-position variables,
      while $\set{\upsilon_m \mapsto \rho \tau_m} \circ \rho' \sqsubseteq \rho$ when applied to position variables.
      Since $\Delta_m$ contains no position variables, and all position variables appear positively in $u_m$, by \autoref{lem:posneg-subtyping},
      $\Gamma_G, \rho \Gamma \vdash (\set{\upsilon_m \mapsto \rho \tau_m} \circ \rho') t_m \leq \rho t_m$.
      The goal then follows by \refrule{conv} on \autoref{eqn:soundness:fix}. \qed
  \end{itemize}
\end{enumerate}
\end{proof}

\begin{conjecture}[Completeness (check/infer)]\label{thm:completeness}
Let $\Sigma$ be a fixed, well-formed signature, $\Gamma_G$ a global environment, $\Gamma$ a local environment, $C$ a constraint set, and $\rho \vDash C$ a solution to the constraint set.
\begin{enumerate}
  \item \label{thm:completeness:check} If $\Gamma_G, \rho\Gamma \vdash e : \rho t$,
    then there exist $C', \rho'$ such that:
    \begin{itemize}
      \item $\rho' \vDash C'$;
      \item $\forall \upsilon \in \SV(\Gamma, t), \rho \upsilon = \rho' \upsilon$; and
      \item $C, \Gamma_G, \Gamma \vdash \erase{e} \Leftarrow t \rightsquigarrow C', e'$ where $\Gamma_G, \Gamma \vdash \rho' e' \conv* e$.
    \end{itemize}
  \item \label{thm:completeness:infer} If $\Gamma_G, \rho\Gamma \vdash e : t$,
    then there exist $C, \rho'$ such that:
    \begin{itemize}
      \item $\rho' \vDash C'$;
      \item $\forall \upsilon \in \SV(\Gamma, t), \rho \upsilon = \rho' \upsilon$; and
      \item $C, \Gamma_G, \Gamma \vdash \erase{e} \rightsquigarrow C', e' \Rightarrow t'$ where $\Gamma_G, \Gamma \vdash \rho' e' \conv* e$ and $\Gamma_G, \Gamma \vdash \rho' t \leq t$.
    \end{itemize}
\end{enumerate}
\end{conjecture}

\begin{theorem}[Soundness (well-formedness)]
If $\Gamma_G^\circ \rightsquigarrow \Gamma_G$ then $\WF{\Gamma_G, \mt}$.
\end{theorem}

\begin{proof}
By cases on the size inference rules for global declarations.
\begin{itemize}
  \item \refrule{a-global-nil}: Trivial.
  \item \refrule{a-global-assum}.
    \begin{displaymath}
      \inferrule*[]{
        \Gamma_G^\circ \rightsquigarrow \Gamma_G \\
        \set{}, \Gamma_G, \mt \vdash t^\circ \rightsquigarrow C_1, t \Rightarrow^* U \\
        \rho = \solve{C_1}
      }{
        \Gamma_G^\circ(\Assm{x}{t^\circ\!\!}) \rightsquigarrow \Gamma_G(\Assm{x}{\rho t})
      }
    \end{displaymath}
    By \autoref{thm:solve}, we have that $\rho \vDash C_1$.
    By the induction hypothesis, we have that $\WF{\Gamma_G, \mt}$.
    Then by \autoref{thm:soundness}, we have that $\Gamma_G, \mt \vdash \rho t : U$,
    and by \refrule{wf-global-assum}, we conclude that $\WF{\Gamma_G(\Assm{x}{\rho t}), \mt}$.
  \item \refrule{a-global-def}.
    \begin{displaymath}
      \inferrule*[]{
        \Gamma_G^\circ \rightsquigarrow \Gamma_G \\
        \set{}, \Gamma_G, \mt \vdash t^\circ \rightsquigarrow C_1, t \Rightarrow^* U \\\\
        \set{}, \Gamma_G, \mt \vdash e^\circ \Leftarrow t \rightsquigarrow C_2, e \\
        \rho = \solve{C_1 \cup C_2}
      }{
        \Gamma_G^\circ(\Defn{x}{t^\circ}{e^\circ\!\!}) \rightsquigarrow \Gamma_G(\Defn{x}{\rho t}{\rho e})
      }
    \end{displaymath}
    By \autoref{thm:solve}, we have that $\rho \vDash C_1 \cup C_2$.
    By the induction hypothesis, we have that $\WF{\Gamma_G, \mt}$.
    Then by \autoref{thm:soundness}, we have that $\Gamma_G, \mt \vdash \rho t : U$ and $\Gamma_G, \mt \vdash \rho e : \rho t$.
    Finally, by \refrule{wf-global-def}, we conclude $\WF{\Gamma_G(\Defn{x}{\rho t}{\rho e}), \mt}$.
    \qed
\end{itemize}
\end{proof}

\begin{theorem}[Completeness (well-formedness)]
If $\WF{\Gamma_G, \mt}$ then $\erase{\Gamma_G} \rightsquigarrow \Gamma'_G$.
\end{theorem}

\begin{proof}
By cases on the well-formedness rules for global declarations.
\begin{itemize}
  \item \refrule{wf-nil}: Trivial.
  \item \refrule{wf-global-assum}.
    \begin{displaymath}
      \inferrule[]
        {\Gamma, \mt \vdash t: U \and x \notin \Gamma_G}
        {\WF{\Gamma_G (\Assm{x}{t}), \mt}}
    \end{displaymath}
    Follows from \autoref{thm:completeness} on the premise.
  \item \refrule{wf-global-def}.
    \begin{displaymath}
      \inferrule[]
        {\Gamma, \mt \vdash e: t \and x \notin \Gamma_G}
        {\WF{\Gamma_G (\Defn{x}{t}{e}), \mt}}
    \end{displaymath}
    Follows from \autoref{thm:completeness} on the premise. \qed
\end{itemize}
\end{proof}

\label{lastpage}

\end{document}